\documentclass[preprint,aps,prd]{revtex4-1}
\usepackage{graphicx}
\usepackage{slashed}
\usepackage{verbatim} 
\newcommand{\be}{\begin{eqnarray}}
\newcommand{\ee}{\end{eqnarray}}
\usepackage[colorlinks]{hyperref}
\usepackage{cancel,bm} 
\usepackage{amsmath,amssymb}
\usepackage{mathtools}
\usepackage{float}
\usepackage{color}
\usepackage{leftidx}
\usepackage{booktabs}
\usepackage{latexsym}
\usepackage{revsymb}
\usepackage{multirow}
\usepackage{hypcap}
\usepackage{array}
\hypersetup{
    colorlinks=red,
    citecolor=blue,
    filecolor=magenta,      
    urlcolor=red,
}

\begin{document}

\title{Sivers effect in Inelastic $J/\psi$ Photoproduction in $ep^\uparrow$ Collision in Color Octet Model}

\author{Sangem Rajesh, Raj Kishore and Asmita Mukherjee}
 %\email{rajeshphy@phy.iitb.ac.in}   
\affiliation{ Department of Physics,
Indian Institute of Technology Bombay, Mumbai-400076,
India.}
\date{\today}

\begin{abstract}
The prediction of single-spin asymmetry  in inelastic photoproduction of $J/\psi$ in $ep^\uparrow$ collision 
is
presented. At next-to-leading order, the dominating process is  photon-gluon fusion, $\gamma+g\rightarrow 
J/\psi+g$ for the production  of $J/\psi$ in $e+p^\uparrow\rightarrow  J/\psi+X$, which directly 
probes the gluon Sivers function. Using the non-relativistic QCD based color octet model, the color octet 
states
$\leftidx{^{3}}{S}{_1}^{(8)}$, $\leftidx{^{1}}{S}{_0}^{(8)}$ and $\leftidx{^{3}}{P}{_{J(0,1,2)}}^{(8)}$
contribution to $J/\psi$ production is calculated. Sizable  asymmetry is estimated as a 
function of transverse momentum $P_T$ and energy fraction $z$ of $J/\psi$ in the range $0<P_T\leq 1$ GeV and 
$0.3<z\leq 0.9$. The unpolarized differential cross section of inelastic $J/\psi$ photoproduction is
found to be in good agreement with H1 and ZEUS data.
\end{abstract}

\maketitle
\raggedbottom 
%=================================================================================
\section{Introduction}\label{sec1}
Among the transverse momentum dependent pdfs (TMDs), Sivers function has attracted
considerable interest in the scientific community in recent days, largely
because of a large amount of experimental results coming in. The Sivers function
gives the asymmetric distribution of unpolarized quarks/gluons inside a
transversely polarized nucleon. The non-zero Sivers function gives a
coupling between the intrinsic transverse momentum of the parton
(quark/gluon) and the transverse spin of the nucleon \cite{Sivers:1989cc,Sivers:1990fh}, this gives an 
azimuthal
asymmetry in the distribution of the final state particle in $ep^\uparrow$
and $pp^\uparrow$ collision that has been measured at HERMES 
\cite{Airapetian:2004tw,Airapetian:2009ae,Airapetian:2013bim}, 
COMPASS  \cite{Adolph:2012sp,Adolph:2014fjw,Adolph:2017pgv,Aghasyan:2017jop},
JLAB \cite{Qian:2011py,Zhao:2014qvx} and RHIC \cite{Adare:2010bd,Adamczyk:2015gyk} 
respectively. Sivers 
function is a time reversal odd (T-odd) object 
\cite{Collins:2002kn}. 
The initial and final state interactions (gauge links) play an important role in
the Sivers asymmetry. This gives a dependence on the specific process in
which the Sivers function is studied. For example, Sivers function probed in
semi-inclusive deep inelastic scattering (SIDIS) is expected to be the same
in magnitude but opposite in sign compared to the one probed in the
Drell-Yan (DY) process. More complex processes have complex gauge links \cite{Boer:2003cm}.
Experimental data on the Sivers asymmetry have now made it possible for the
extraction of $u$ and $d$ quark Sivers function \cite{Anselmino:2016uie}, but the 
gluon Sivers function (GSF) is still unknown. There is no constraint on GSF except a
positivity bound \cite{Mulders:2000sh}. The GSF contains two gauge links, and
the process dependence is more involved. It has been shown \cite{Buffing:2013kca} that the GSF in any process can
be written in terms of two independent
Sivers functions, an f-type GSF (this contains $[++]$ gauge link and
also called WW gluon distributions) and a d-type GSF 
(this contains $[+-]$ gauge link and are called dipole distributions) \cite{Buffing:2013kca}. The
operator structures in these two Sivers function have different charge
conjugation properties. \par
%--------------------------------------------------------------------------
Heavy quarkonium production in $ep$ 
\cite{Godbole:2012bx,Mukherjee:2016qxa,Boer:2016bfj,Anselmino:2016fhz,Boer:2016fqd} and
$pp$ \cite{Anselmino:2004nk,DAlesio:2017rzj} collision 
has been studied theoretically quite extensively for probing the gluon TMDs, in particular
the GSF and linearly polarized gluon distribution 
\cite{Mukherjee:2015smo,Mukherjee:2016cjw}. 
This is because the heavy quarkonium is produced at leading order (LO)
through photon-gluon fusion ($ep$) or two gluon fusion ($pp$) channel.
Although the production mechanism of heavy quarkonium is still not
well established, the most widely used theoretical approach is based on
non-relativistic QCD (NRQCD) \cite{Bodwin:1994jh}. This gives  systematic way to separate the
high energy and low energy effects of the production mechanism. 
In this approach, the heavy quark pair is produced at a short distance in
color singlet (CS) \cite{Carlson:1976cd,Berger:1980ni,Baier:1981uk,Baier:1981zz} or in color octet (CO) 
\cite{Braaten:1994vv,Cho:1995vh,Cho:1995ce} configuration and then they hadronize to
form a quarkonium state of given quantum numbers through a soft process. The
short distance coefficients are calculated perturbatively for each process
and the long distance matrix elements (LDMEs) are extracted from the
experimental data. The LDMEs are categorized by performing an expansion in
terms of the relative velocity of the heavy quark $v$ in the limit $v <<1$ \cite{Lepage:1992tx}. 
The theoretical predictions are arranged as double expansions in terms of
$v$ as well as $\alpha_s$.  The heavy quark pair may be produced in CO state which then form the CS quarkonium by emitting a soft
gluon. NRQCD  has been successful to explain the
$J/\psi$ hadroproduction at Tevatron \cite{Abe:1997jz,Acosta:2004yw}, also data from  $J/\psi$ photoproducton at
HERA \cite{Adloff:2002ex,Aaron:2010gz,Chekanov:2002at,Abramowicz:2012dh} suggests substantial contribution 
from CO states 
\cite{Butenschoen:2010rq,Ma:2010jj,Chao:2012iv,Butenschoen:2011yh,Zhang:2014ybe,Butenschoen:2012px}. In the
single-spin asymmetry (SSA) in $ep$ collision, when the $J/\psi$ is produced in the CS
state, the two final state interactions  with quark and anti-quark lines
cancel each other, and the final state interaction with unobserved particles
cancel between diagrams having different cuts. As a result, SSA in $J/\psi$
production in $ep$ collision is zero when the heavy quark pair 
is produced in the CS state, and non-zero asymmetry can be observed 
when the pair is produced in CO state \cite{Yuan:2008vn}. The final state interactions
are more involved  for $pp$ collision processes, and there, non-zero SSA is
expected when the heavy quark pair is produced in a CS state. 
In the study of TMDs in SSA in heavy quarkonium production, one assumes that
TMD factorization holds for such processes. \par
 In our previous work \cite{Mukherjee:2016qxa}, we calculated the Sivers asymmetry in $J/\psi$ 
electroproduction at LO, which is a
photon-gluon $2 \rightarrow 1$ process, in color octet model (COM). We showed
that the calculated asymmetry at $z=1$ agrees within the error bar of the
recent COMPASS \cite{Matousek:2016xbl} measurement . Here we extend the analysis to estimate the SSA in 
photoproduction of $J/\psi$ at next-to-leading order (NLO). This allows to calculate
the asymmetry over a wider kinematical region accessible to the present
experiments at  COMPASS and at the planned EIC in the future. We will
use NRQCD based COM in our calculation for estimating the asymmetry.\par
%----------------------
The paper is organized into five sections including the introduction in Sec.\ref{sec1}.  The 
SSA and $J/\psi$ production framework are presented in Sec.\ref{sec2} and 
Sec.\ref{sec3} respectively. Sec.\ref{sec4}
discusses about the numerical results. The conclusion of the paper is given in Sec.\ref{sec5}. 
A few details of calculation are given in the appendices.                             
%==================================================================================
\section{Single-spin asymmetry}\label{sec2}
\begin{figure}[h]
\begin{center} 
\includegraphics[height=5cm,width=14cm]{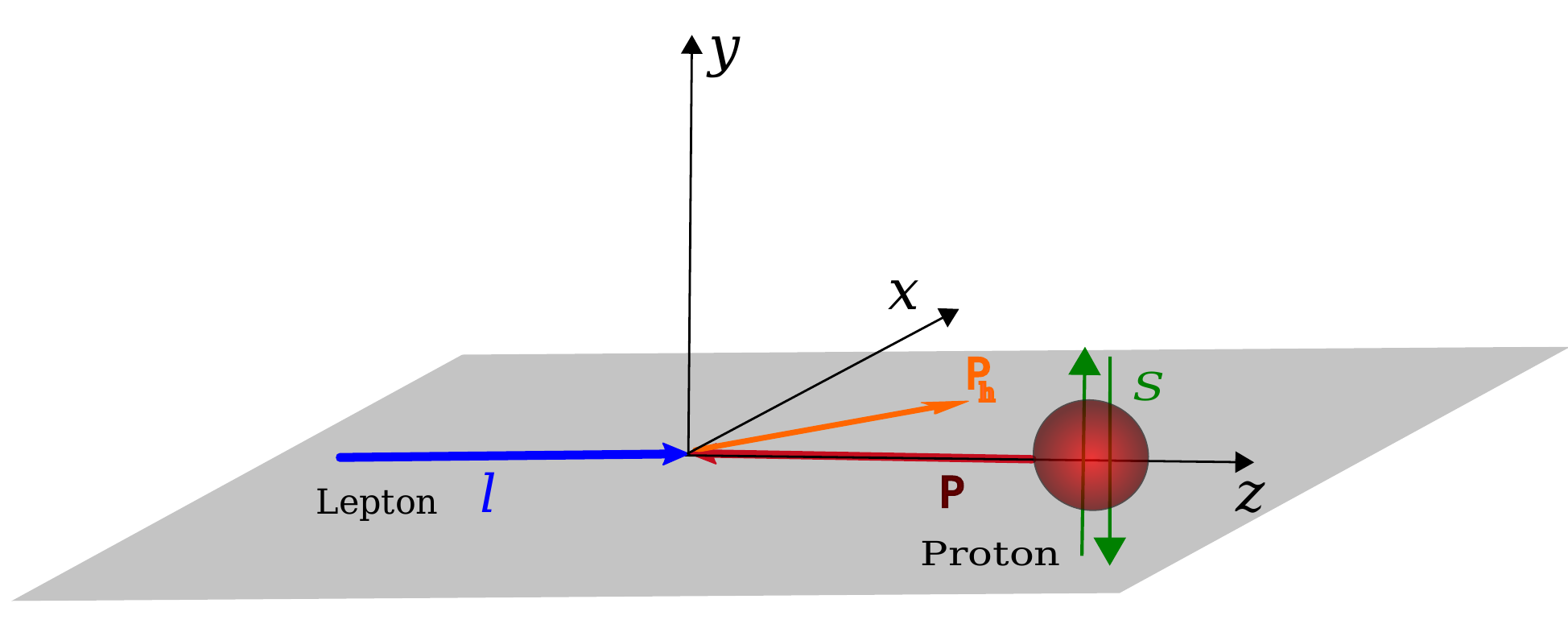}
\end{center}
\caption{\label{fig1}Kinematical configuration for  $ep\rightarrow J/\psi+X$ process.}
\end{figure}

In general the transverse  single-spin asymmetry (SSA) is  defined as following
\begin{eqnarray}\label{asy}
 A_N=\frac{d \sigma^{\uparrow}-d 
\sigma^{\downarrow}}{d \sigma^{\uparrow}+d \sigma^{\downarrow}},
\end{eqnarray}
where $d\sigma^{\uparrow}$ and $d\sigma^{\downarrow}$ are respectively the  differential cross-sections 
measured when one of the particle is transversely polarized up ($\uparrow$) and down ($\downarrow$) with 
respect to the scattering plane. Here $\uparrow(\downarrow)$ direction is  the proton  polarization
direction along the +y (-y) axis with momentum along -z axis and the final hadron is produced in the xz plane 
as shown in \figurename{\ref{fig1}}.
%===========================================================================================
%===============================================================================================
We consider the inclusive process   $e(l)+p^{\uparrow}(P)\rightarrow 
J/\psi(P_h)+X$. The virtual photon radiated by the initial electron scattering will interact with the proton.
The virtual photon carries the momentum $q$ such that $q^2\approx -2EE^\prime(1-\cos\theta)$ with 
$E$ and $E^\prime$ are energies of the initial and final electron respectively.
In the forward scattering limit, however, the four momentum of virtual photon $q^2=-Q^2\rightarrow 0$ as a 
result the virtual photon becomes the real  photon. The dominating subprocess at NLO for quarkonium production 
in $ep$ collision is photon-gluon fusion process, i.e., $\gamma(q)+g(k)\rightarrow J/\psi(P_h) +g(p_g)$.
The letters within the round brackets represent the four momentum of each particle.
There are two types of $J/\psi$ photoproductions. One is the direct photoproduction in which the 
photon electromagnetically interacts with the partons of the proton. The 
second, resolved photoproduction wherein 
the photon acts as a source of partons and then they strongly interact with partons of the proton.\par
In this paper we have not considered the resolved photoproduction channel which basically contributes at low 
$z$ region ($z\leq0.3$) \cite{Kniehl:1998qy}, where $z=\frac{P.P_h}{P.q}$ is the energy 
fraction  transferred from the photon to $J/\psi$ in the proton rest frame. The LO photon-gluon fusion 
subprocess ($\gamma +g\rightarrow J/\psi$) contributes to 
elastic photoproduction at $z=1$ \cite{Mukherjee:2016qxa}. The process of a colorless exchanged particle 
between
quasi-real photon and proton, diffractive process, contributes to $J/\psi$ production in the elastic region, 
i.e., $z\approx1$ and $P_T\approx 0$ GeV \cite{Ryskin:1992ui,Butenschoen:2009zy}. $P_T$ is the transverse 
momentum of $J/\psi$. Moreover, gluon and heavy quark fragmentation also contribute for quarkonium production 
significantly at $P_T>4$ GeV \cite{Li:1996jk}, which are excluded by imposing $P_T$ cut. The feed-down contribution 
from an excited state $\psi(2S)$ and 
the decay of $\chi_c$ states contribution to $J/\psi$ are $15\%$ \cite{Chekanov:2002at} and $1\%$ 
\cite{Butenschoen:2009zy,Artoisenet:2009xh} respectively, are not considered in this work.
Therefore, we impose the following kinematical cut $0.3<z\leq0.9$  to account for 
inelastic photoproduction \cite{Artoisenet:2009xh,Kramer:1995nb} events only.
 For true inelastic $J/\psi$ production, one has to impose low $P_T$ cut as in 
\cite{Artoisenet:2009xh,Kramer:1995nb}, however, to validate asymmetry calculation in the TMD framework, we 
have considered $0<P_T\leq1$ GeV and low  $P_T$ cut is not imposed.
The softening of final gluon, i.e., $z\to 1$, leads to infrared singularity in the inelastic photoproduction 
as shown in Eq.\eqref{soft}.
Hence,  $z\leq0.9$ kinematical cut is motivated to keep the final gluon  hard and the perturbative calculation
is under good control. At the same order in $\alpha_s$, another channel $\gamma+q\rightarrow J/\psi+q$ also 
gives the CO contribution to $J/\psi$ production. Since the process is initiated by light quarks, 
the contribution is expected to be negligible compared to the photon-gluon fusion process \cite{Ko:1996xw}. For the dominating channel of $J/\psi$ production through $\gamma g$ fusion, the contribution to the numerator of $A_N$ comes mainly from the gluon Sivers distribution \cite{DAlesio:2017nrd}. As the heavy quark pair in the final state is produced unpolarized, there is no contribution from Collins function \cite{Anselmino:2004nk}. Also the linearly polarized gluons  do not contribute to the denominator as long as the lepton is unpolarized \cite{DAlesio:2017nrd}. Within the generalized parton model formalism, the differential cross section for an 
unpolarized process is given by 

\begin{equation}\label{d1}
 \begin{aligned}
E_h\frac{d\sigma}{d^3{\bm P}_h}={}&\frac{1}{2(2\pi)^2}\int dx_\gamma dx_g 
d^2{\bm k}_{\perp g}
f_{\gamma/e}(x_\gamma)f_{g/p}(x_g,{\bm k}_{\perp 
g})\delta(\hat{s}+\hat{t}+\hat{u}-M^2)\\
&\times\frac{1}{2\hat{s}}|\mathcal{M}_{\gamma+g\rightarrow J/\psi +g}|^2.
\end{aligned}
\end{equation}
Here $x_\gamma$ and $x_g$ are the light-cone momentum fractions of photon and gluon respectively.  The 
Weizs$\ddot{a}$ker-Williams distribution function, $f_{\gamma/e}(x_\gamma)$, describes the density of photons inside the electron
which is given by \cite{Frixione:1993yw}
\begin{eqnarray}  \label{flux}
f_{\gamma/e}(x_{\gamma})=\frac{\alpha}{2\pi }\left[
2m_e^2x_\gamma \left(\frac{1}{Q^2_{min}}-\frac{1}{Q^2_{max}}\right)+\frac{1+(1-x_\gamma)^2}{x_\gamma}\ln\frac{Q^2_{max}}{Q^2_{min}}\right ]
\end{eqnarray}
where $\alpha$ is the electromagnetic coupling and $Q^2_{min}=m_e^2\frac{x_\gamma^2}{1-x_\gamma}$, $m_e$ 
being the electron mass.
We have considered $Q^2_{max}=1$ GeV$^2$ for estimating the SSA. For photoproduction of $J/\psi$
at HERA, we have taken two different values of  $Q^2_{max}=2.5$ GeV$^2$ and 1 GeV$^2$ in line with  H1 
\cite{Adloff:2002ex,Aaron:2010gz} and ZEUS \cite{Chekanov:2002at,Abramowicz:2012dh} data respectively.
 The unpolarized gluon TMD, $f_{g/p}$, represents the density of gluons inside an unpolarized proton. The 
$\hat{s}$, $\hat{t}$ and $\hat{u}$ are the Mandelstam variables whose definitions are given in appendix 
\ref{ap2}. $\mathcal{M}_{\gamma+g\rightarrow J/\psi +g}$ is the  amplitude of photon-gluon 
fusion process which will be discussed in Sec.\ref{sec3} and its square is given in appendix \ref{ap1}. 
The mass of $J/\psi$ is represented with $M$. Now, we are in a position to write down the expression of 
numerator and denominator terms of Eq.\eqref{asy} when the target proton is polarized and are given by
\begin{equation}\label{d3}
 \begin{aligned}
 d\sigma^{\uparrow}-d\sigma^{\downarrow}={}&\frac{d\sigma^{ep^{\uparrow}\rightarrow J/\psi X}}{dzd^2{\bm 
P}_T}-\frac{d\sigma^{ep^{\downarrow}\rightarrow J/\psi X}}{dzd^2{\bm P}_T}\\
={}&\frac{1}{2z(2\pi)^2}\int dx_\gamma dx_g  
d^2{\bm k}_{\perp g}
f_{\gamma/e}(x_\gamma)\Delta^N f_{g/p^{\uparrow}}(x_g,{\bm k}_{\perp 
g})\\
&\times\delta(\hat{s}+\hat{t}+\hat{u}-M^2)\frac{1}{2\hat{s}}|\mathcal{M}_{\gamma+g\rightarrow J/\psi +g}|^2,
\end{aligned}
\end{equation}
and 
\begin{equation}\label{d4}
 \begin{aligned}
 d\sigma^{\uparrow}+d\sigma^{\downarrow}={}&\frac{d\sigma^{ep^{\uparrow}\rightarrow  J/\psi X}}{dzd^2{\bm 
P}_T}+\frac{d\sigma^{ep^{\downarrow}\rightarrow  J/\psi X}}{dzd^2{\bm P}_T}=2\frac{d\sigma}{dzd^2{\bm P}_T}\\
={}&\frac{2}{2z(2\pi)^2}\int dx_{\gamma} dx_g 
d^2{\bm k}_{\perp g}
f_{\gamma/e}(x_\gamma) f_{g/p}(x_g,{\bm k}_{\perp 
g})\\
&\times\delta(\hat{s}+\hat{t}+\hat{u}-M^2)\frac{1}{2\hat{s}}|\mathcal{M}_{\gamma+g\rightarrow J/\psi +g}|^2.
\end{aligned}
\end{equation} 
where $\Delta^N f_{g/p^{\uparrow}}(x_g,{\bm k}_{\perp g})$, GSF, describes the density of unpolarized gluons 
inside the transversely polarized proton and is defined as below
\be
\Delta^Nf_{g/p^{\uparrow}}(x_g,{\bm k}_{\perp g})
&=& f_{g/p^{\uparrow}}(x_g,{\bm k}_{\perp g})- f_{g/p^{\downarrow}}(x_g,{\bm k}_{\perp g})\nonumber\\
&=&\Delta^Nf_{g/p^{\uparrow}}(x_g,k_{\perp g})~{\hat{\bm S}}.(\hat{\bm P}\times\hat{\bm 
k}_{\perp g})
\ee
For estimating the SSA numerically, we have to discuss about the parameterization of TMDs.
Generally, it is assumed that the unpolarized  gluon TMDs follow the Gaussian distribution. The 
Gaussian parameterization of unpolarized TMD is
\be \label{unp}
 f_{g/p}(x_g,{\bm k}^2_{\perp g},\mu)=f_{g/p}(x_g,\mu)\frac{1}{\pi \langle k^2_{\perp g}\rangle}
 e^{-{\bm k}^2_{\perp g}/\langle k^2_{\perp g}\rangle}.
\ee
Here, $x_g$ and $k_{\perp g}$ dependencies of the TMD are factorized. The collinear PDF is  denoted with 
$f_{g/p}(x_g,\mu)$ which is measured at the scale $\mu=\sqrt{M^2+P_T^2}$. The collinear PDF obeys the 
Dokshitzer-Gribov-Lipatov-Altarelli-Parisi (DGLAP) scale evolution.
We choose a frame (shown in \figurename{\ref{fig1}}) as discussed in appendix \ref{ap2} wherein the polarized 
proton is moving along $-z$ axis with momentum ${\bm P}$, is 
transversely polarized $\hat{\bm S}=(\cos\phi_s,\sin\phi_s,0)$. The transverse momentum of the initial 
gluon is 
${\bm k}_{\perp g}=k_{\perp g}(\cos\phi,\sin\phi,0)$,
\be
 \hat{{\bm S}}.(\hat{\bm P}\times\hat{\bm k}_{\perp g})=\sin(\phi-\phi_s).
\ee
For numerical estimation we have taken $\phi_s=\pi/2$.
 The parameterization of GSF is given by \cite{alesio,Anselmino:2016uie}
\be
\Delta^{N}f_{g/p^{\uparrow}}(x_g, k_{\perp g},\mu)=2\mathcal{N}_g(x_g)f_{g/p}(x_g,\mu)h(k_{\perp g})
\frac{e^{-{\bm k}^2_{\perp g}/\langle k^2_{\perp g}\rangle}}{\pi\langle k^2_{\perp g}\rangle},
\ee
here $f_{g/p}(x_g,\mu)$ is the usual collinear gluon PDF and 
\be\label{ngx}
\mathcal{N}_g(x_g)=N_gx_g^\alpha(1-x_g)^\beta\frac{(\alpha+\beta)^{(\alpha+\beta)}}{
\alpha^\alpha\beta^\beta}.
\ee
The definition of $h(k_{\perp g})$ is given by
\be
h(k_{\perp g})=\sqrt{2e}\frac{k_{\perp g}}{M_1}e^{-{\bm k}^2_{\perp g}/M^2_1}.
\ee
The $k_{\perp g}$ dependent part of Sivers function can be written as
\be
h(k_{\perp g})\frac{e^{-{\bm k}^2_{\perp g}/\langle k^2_{\perp g}\rangle}}{\pi\langle k^2_{\perp g}\rangle}
=\frac{\sqrt{2e}}{\pi}\sqrt{\frac{1-\rho}{\rho}}k_{\perp g}
\frac{e^{-{\bm k}^2_{\perp g}/\rho\langle k^2_{\perp g}\rangle}}{\langle k^2_{\perp g}\rangle^{3/2}},
\ee
where we defined 
\be
\rho=\frac{M^2_1}{\langle k^2_{\perp g}\rangle + M^2_1}.
\ee
D'Alesio et al. \cite{alesio} have extracted the GSF from  pion production data at RHIC \cite{rhic1}  first 
time and  
two sets of best fit parameters were presented which are  denoted with SIDIS1 and 
SIDIS2. Moreover, using the latest SIDIS data  Anselmino et al. \cite{Anselmino:2016uie} have  extracted the 
quark and 
anti-quark Sivers function. However, GSF has not been extracted yet from SIDIS 
data. Therefore, in order to estimate the asymmetry,  best fit 
parameters of Sivers function corresponding to $u$ and $d$ quark  will be used in 
the following  parameterizations \cite{Boer:2003tx} : 
\be \label{ab}
(a)~~\mathcal{N}_g(x_g)&=&\left(\mathcal{N}_u(x_g)+\mathcal{N}_d(x_g)\right)/2, \nonumber\\
(b)~~\mathcal{N}_g(x_g)&=&\mathcal{N}_d(x_g).
\ee
We call the  parameterization (a) and (b) as BV-a and BV-b respectively. The best fit parameters are 
tabulated in \tablename{ \ref{table1}}.
%=====================================================

%===============================================================================================
\begin{table}[H]
\centering
\caption{\label{table1}Best fit parameters of Sivers function.}
\begin{tabular}{ | >{\centering\arraybackslash}m{2cm}| >{\centering\arraybackslash}m{1.2cm}| 
>{\centering\arraybackslash}m{1.2cm}| >{\centering\arraybackslash}m{1.2cm}| 
>{\centering\arraybackslash}m{1.2cm}| >{\centering\arraybackslash}m{1.2cm}| 
>{\centering\arraybackslash}m{2cm}| >{\centering\arraybackslash}m{2cm}| 
>{\centering\arraybackslash}m{1.5cm}| }
\hline
\multicolumn{9}{ |c| }{Best fit parameters} \\
\cline{1-9}
Evolution & $a$ & $N_a$ & $\alpha$ & $\beta$ &$\rho$ &$M_1^2$ GeV$^2$& $\langle k^2_\perp\rangle$ 
GeV$^2$ & 
Notation  \\ \cline{1-9}
\multicolumn{1}{ |c  } {\multirow{4}{*}{DGLAP}} &
\multicolumn{1}{ |c| } {$g$ \cite{alesio}} & 0.65 & 2.8 & 2.8 & 0.687 & & 0.25 & SIDIS1 \\
  \cline{2-9}
  \multicolumn{1}{ |c  }{} &
  \multicolumn{1}{ |c| } {$g$ \cite{alesio}} & 0.05 & 0.8 & 1.4 & 0.576 & & 0.25 & SIDIS2 \\
  \cline{2-9}
  \multicolumn{1}{ |c  }{} &
  \multicolumn{1}{ |c| } {$u$ \cite{Anselmino:2016uie}} & 0.18 & 1.0 & 6.6 &  & 0.8 & 0.57 & BV-a \\
  \cline{2-8}
  \multicolumn{1}{ |c  }{} &
  \multicolumn{1}{ |c| } {$d$ \cite{Anselmino:2016uie}} & -0.52 & 1.9 & 10.0 &  & 0.8 & 0.57 & BV-b \\
  \cline{1-9}
\end{tabular}
\end{table}
%========================================================================================
The final expressions of  numerator and denominator terms of Eq.\eqref{asy} within DGLAP evolution 
approach are given by
\begin{equation}\label{d3}
 \begin{aligned}
 d\sigma^{\uparrow}-d\sigma^{\downarrow}
={}&\frac{1}{2z(2\pi)^2}\int dx_\gamma dx_g  
d^2{\bm k}_{\perp g}
f_{\gamma/e}(x_\gamma)2\mathcal{N}_g(x_g)f_{g/p}(x_g,\mu)
\frac{\sqrt{2e}}{\pi}\sqrt{\frac{1-\rho}{\rho}}k_{\perp g}
\frac{e^{-{\bm k}^2_{\perp g}/\rho\langle k^2_{\perp g}\rangle}}{\langle k^2_{\perp g}\rangle^{3/2}}
\\
&\times\delta(\hat{s}+\hat{t}+\hat{u}-M^2)\frac{1}{2\hat{s}}|
\mathcal{M}_{\gamma+g\rightarrow J/\psi +g}|^2\sin(\phi-\phi_s),
\end{aligned}
\end{equation}
and 
\begin{equation}\label{d4}
 \begin{aligned}
 d\sigma^{\uparrow}+d\sigma^{\downarrow}={}&\frac{2}{2z(2\pi)^2}\int dx_{\gamma} dx_g 
d^2{\bm k}_{\perp g}
f_{\gamma/e}(x_\gamma) f_{g/p}(x_g,\mu)\frac{1}{\pi \langle k^2_{\perp g}\rangle}
 e^{-{\bm k}^2_{\perp g}/\langle k^2_{\perp g}\rangle}\\
&\times\delta(\hat{s}+\hat{t}+\hat{u}-M^2)\frac{1}{2\hat{s}}|\mathcal{M}_{\gamma+g\rightarrow J/\psi +g}|^2.
\end{aligned}
\end{equation}

%==============================================================================================
\section{$J/\psi$ production in COM framework}\label{sec3}
%=============================================================================================

\begin{figure}[h]
\begin{center} 
\includegraphics[height=10cm,width=17cm]{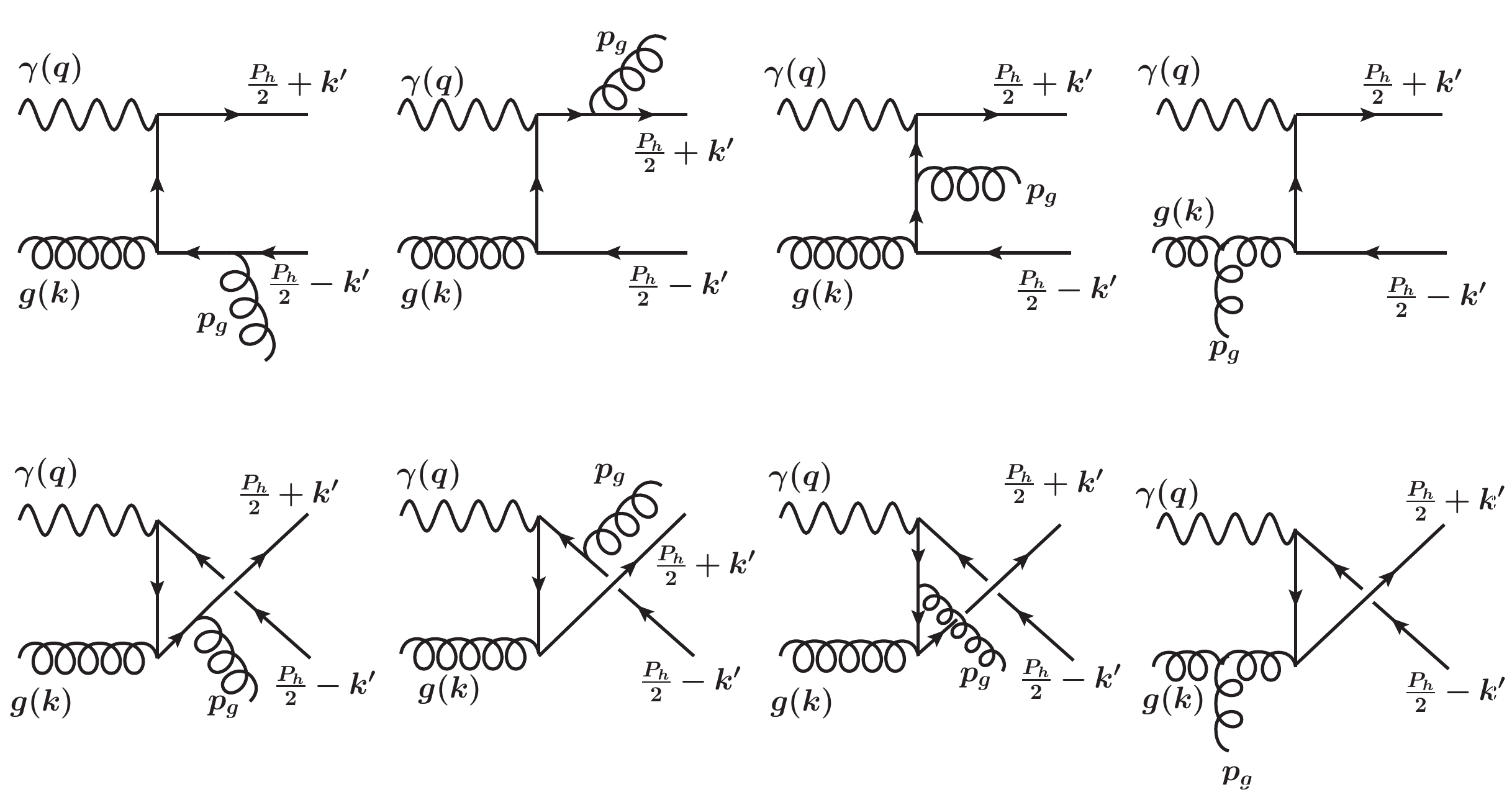}
\end{center}
\caption{\label{fig2}Feynman diagrams for  $\gamma+g\rightarrow J/\psi+g$ process.}
\end{figure}

Let us consider the $J/\psi$ production in  $e+p\rightarrow J/\psi+X$ process. The NLO subprocess 
 is $\gamma+g\rightarrow J/\psi +g$ and the related Feynman diagrams to this process are shown in 
\figurename{\ref{fig2}}. The amplitude expression for bound state production in NRQCD framework can be 
written as below \cite{Baier:1983va,Boer:2012bt}

%===========================================================================================
%===============================================================================================

 \begin{equation}\label{e3}
 \begin{aligned}
 \mathcal{M}\left(\gamma g\rightarrow  Q\bar{Q}[\leftidx{^{2S+1}}{L}{_J}^{(1,8)}](P_h)+g\right)=\sum_{L_zS_z}
 \int \frac{d^3\bm{k}^{\prime}}{(2\pi)^3}\Psi_{LL_z}(\bm{k}^\prime) \langle LL_z;SS_z|JJ_z\rangle \\
 \times\mathrm{Tr}[O(q,k,P_h,k^\prime)\mathcal{P}_{SS_z}(P_h,k^\prime)],
 \end{aligned}
\end{equation}
where $k^\prime$ is the relative momentum of the heavy quark in the quarkonium rest frame.
In Eq.\eqref{e3}, $O(q,k,P_h,k^\prime)$ represents the amplitude of $Q\bar{Q}$ pair  without considering the 
external heavy quark and 
anti-quark legs, which is given by
\begin{equation}
 O(q,k,P_h,k^\prime)=\sum_{m=1}^8 \mathcal{C}_m O_m(q,k,P_h,k^\prime).
\end{equation}
From \figurename{\ref{fig1}}, the amplitude expression of individual Feynman diagram is given below 
\begin{equation}\label{ee1}
 \begin{aligned}
O_1= 
4g^2_s(ee_c)\varepsilon^\mu_{\lambda_a}(k)\varepsilon^\nu_{\lambda_b}(q)\varepsilon^{\rho\ast}_{\lambda_g}
(p_g)
\gamma_\nu\frac{\slashed{P_h}+2\slashed{k}^\prime-2\slashed{q}+M}{(P_h+2k^\prime-2q)^2-M^2}\gamma_\mu
\frac{-\slashed{P_h}+2\slashed{k}^\prime-2\slashed{p}_g+M}{(P_h-2k^\prime+2p_g)^2-M^2}\gamma_\rho,
\end{aligned}
\end{equation}
\begin{equation}\label{ee2}
 \begin{aligned}
O_2= 
4g^2_s(ee_c)\varepsilon^\mu_{\lambda_a}(k)\varepsilon^\nu_{\lambda_b}(q)\varepsilon^{\rho\ast}_{\lambda_g}
(p_g)
\gamma_\rho\frac{\slashed{P_h}+2\slashed{k}^\prime+2\slashed{p}_g+M}{(P_h+2k^\prime+2p_g)^2-M^2}\gamma_\nu
\frac{-\slashed{P_h}+2\slashed{k}^\prime+2\slashed{k}+M}{(P_h-2k^\prime-2k)^2-M^2}\gamma_\mu,
\end{aligned}
\end{equation}
\begin{equation}\label{ee3}
 \begin{aligned}
O_3= 
4g^2_s(ee_c)\varepsilon^\mu_{\lambda_a}(k)\varepsilon^\nu_{\lambda_b}(q)\varepsilon^{\rho\ast}_{\lambda_g}
(p_g)
\gamma_\nu\frac{\slashed{P_h}+2\slashed{k}^\prime-2\slashed{q}+M}{(P_h+2k^\prime-2q)^2-M^2}\gamma_\rho
\frac{-\slashed{P_h}+2\slashed{k}^\prime+2\slashed{k}+M}{(P_h-2k^\prime-2k)^2-M^2}\gamma_\mu,
\end{aligned}
\end{equation}
\begin{equation}\label{ee4}
 \begin{aligned}
O_4= 
2g^2_s(ee_c)\varepsilon^\mu_{\lambda_a}(k)\varepsilon^\nu_{\lambda_b}(q)\varepsilon^{\rho\ast}_{
\lambda_g}(p_g)
\gamma_\nu\frac{\slashed{P_h}+2\slashed{k}^\prime-2\slashed{q}+M}{(P_h+2k^\prime-2q)^2-M^2}\gamma^\sigma\frac{
1}{(k-p_g)^2}\\
\times\left[g_{\mu\rho}(k+p_g)_\sigma+g_{\rho\sigma}(k-2p_g)_\mu+g_{\sigma\mu}(p_g-2k)_\rho\right].
\end{aligned}
\end{equation}
Here $M=2m_c$, $m_c$ being the charm quark mass.
Charge conjugation invariance implies that all the eight Feynman diagrams are symmetric by reversing the 
fermion flow. The amplitude expressions of $O_5$, $O_6$, $O_7$ and $O_8$ can be obtained by reversing the 
fermion 
flow and
replacing $k^\prime \to -k^\prime$.
The color factor of each diagram is given by 
\begin{equation}
\begin{aligned}
\mathcal{C}_1=\mathcal{C}_6=\mathcal{C}_7= \sum_{ij}\langle 
3i;\bar{3}j|8c\rangle(t_at_b)_{ij},~~~\mathcal{C}_2=\mathcal{C}_3=\mathcal{C}_5= \sum_{ij}\langle
3i;\bar{3}j|8c\rangle(t_bt_a)_{ij}
\end{aligned}
\end{equation}
\begin{equation}
\begin{aligned}
\mathcal{C}_4=\mathcal{C}_8= \sum_{ij}\langle 3i;\bar{3}j|8c\rangle 
if_{abd}(t_d)_{ij} \nonumber
\end{aligned}
\end{equation}
here the summation is over the colors of the outgoing quark and anti-quark.
The  SU(3) Clebsch-Gordan coefficients for CS and CO states respectively are given by
\be\label{e2}
\langle 3i;\bar{3}j|1\rangle=\frac{\delta^{ij}}{\sqrt{N_c}}~,~~~~~~~~~~~
\langle 3i;\bar{3}j|8a\rangle=\sqrt{2}(t^a)^{ij}
\ee
and they project out  the color state of $Q\bar{Q}$ pair either it is in CS or CO state, where 
$N_c$ is 
the number of colors. The generators of SU(3) group in 
fundamental representation is denoted by $t_a$ which fallows $\mathrm{Tr}(t_at_b)=\delta_{ab}/2$ and 
$\mathrm{Tr}(t_at_bt_c)=
\frac14(d_{abc}+if_{abc})$.
Using Eq.\eqref{e2}, we have the following color factors for the production of initial $Q\bar{Q}$ in CO state
\begin{equation} 
\mathcal{C}_1=\mathcal{C}_6=\mathcal{C}_7=\frac{\sqrt{2}}{4}(d_{abc}+if_{abc}),~\mathcal{C}_2=\mathcal{C}
_3=\mathcal{C}_5=\frac{\sqrt{2 } }{4}(d_{abc}
-if_{abc}),~ \mathcal{C}_4=\mathcal{C}_8=\frac{\sqrt{2}}{2}if_{abc}.
\end{equation}
The excluded heavy quark and anti-quark spinors are absorbed in the definition of spin projection operator 
which
is given by \cite{Baier:1983va,Boer:2012bt}
\be \label{e6}
\mathcal{P}_{SS_z}(P_h,k^\prime)&=&\sum_{s_1s_2}\langle\frac12s_1;\frac12s_2|SS_z\rangle v(\frac{P_h}{2}-
k^\prime,s_1)\bar{u}(\frac{P_h}{2}+k^\prime,s_2)\nonumber\\
&=&\frac{1}{4M^{3/2}}(-\slashed{P}_h+2\slashed{k}^\prime+M)\Pi_{SS_z}(\slashed{P}_h+2\slashed{k}^\prime+M)
+\mathcal{O}(k^{\prime 2}),
\ee
bearing $\Pi_{SS_z}=\gamma^{5}$ for singlet ($S=0$) state and $\Pi_{SS_z}=\slashed{\varepsilon}_{s_z}(P_h)$
for triplet ($S=1$) state. Here spin polarization vector of the $Q\bar{Q}$ system is denoted with
$\varepsilon_{s_z}(P_h)$. Since the relative momentum $k^\prime$ is very small w.r.t $P_h$, Taylor expansion 
can be performed around $k^\prime=0$ in Eq.\eqref{e3}. The first term in the expansion gives the $S$-wave 
amplitude. Since the radial wavefunction $R_1(0)=0$ for $P-$wave ($L=1,~ J=0,1,2$), one has to consider the 
second term in the Taylor expansion to calculate  $P$-wave amplitude. By following Ref.\cite{Boer:2012bt}, 
one obtains the  $S$ and $P$ state amplitude expressions which are given by 
\begin{eqnarray}\label{e8}
 \mathcal{M}[\leftidx{^{2S+1}}{S}{_J}^{(8)}](P_h,k)&=&\frac{1}{\sqrt{4\pi}}R_0(0)\mathrm{Tr}[O(q,k,P_h,
k^\prime) \mathcal{P}_{SS_z}(P_h,k^\prime)]\Big\rvert_{k^\prime=0}\nonumber\\
&=&\frac{1}{\sqrt{4\pi}}R_0(0)\mathrm{Tr}[O(0) \mathcal{P}_{SS_z}(0)],
\end{eqnarray}
\begin{eqnarray}\label{e9} 
 \mathcal{M}[\leftidx{^{2S+1}}{P}{_J}^{(8)}]&=&-i
\sqrt{\frac{3}{4\pi}}R^\prime_1(0) \sum_{L_zS_z}\varepsilon^\alpha_{L_z}(P_h)\langle LL_z;SS_z|JJ_z\rangle 
\frac{\partial}{\partial k^{\prime \alpha}}\mathrm{Tr}[O(q,k,P_h,k^\prime)
 \mathcal{P}_{SS_z}(P_h,k^\prime)]\Big\rvert_{k^\prime=0}\nonumber \\
 &=&-i\sqrt{\frac{3}{4\pi}}R^\prime_1(0)\sum_{L_zS_z}\varepsilon^\alpha_{L_z}(P_h)\langle 
LL_z;SS_z|JJ_z\rangle \mathrm{Tr}[O_\alpha(0)\mathcal{P}_{SS_z}(0)+O(0)
 \mathcal{P}_{SS_z\alpha}(0)]
\end{eqnarray}
The following shorthand notations are defined in the above expressions
\be
O(0)=O(q,k,P_h,k^\prime)\Big\rvert_{k^\prime=0}~,~~~~~~~~  
\mathcal{P}_{SS_z}(0)=\mathcal{P}_{SS_z}(P_h,k^\prime)\Big\rvert_{k^\prime=0}
\ee
\be
O_\alpha(0)=\frac{\partial}{\partial k^{\prime \alpha}}O(q,k,P_h,k^\prime)\Big\rvert_{k^\prime=0}~,~~~~
\mathcal{P}_{SS_z\alpha}(0)=\frac{\partial}{\partial k^{\prime 
\alpha}}\mathcal{P}_{SS_z}(P_h,k^\prime)\Big\rvert_{k^\prime=0}.
\ee
For $P-$ wave amplitude calculation, we use the Clebsch-Gordan coefficients  as defined in 
Ref.\cite{Kuhn:1979bb,Guberina:1980dc}
\be\label{s0}
\sum_{L_zS_z} \langle 1L_z;SS_z|00\rangle \varepsilon^\alpha_{ s_z }(P_h)\varepsilon^\beta_{ L_z }(P_h)
=\sqrt{\frac13}\left(g^{\alpha\beta}-\frac{1}{M^2}P_h^\alpha P_h^\beta\right),
\ee
\be\label{s1}
\sum_{L_zS_z} \langle 1L_z;1S_z|1J_z\rangle \varepsilon^\alpha_{ s_z }(P_h)\varepsilon^\beta_{ L_z }(P_h)
=-\frac{i}{M}\sqrt{\frac12}\epsilon_{\delta\lambda\rho\sigma}g^{\rho\alpha}g^{\sigma\beta}
P_h^\delta\varepsilon^\lambda_{ J_z }(P_h),
\ee
\be\label{s2}
\sum_{L_zS_z} \langle 1L_z;1S_z|2J_z\rangle \varepsilon^\alpha_{ s_z }(P_h)\varepsilon^\beta_{ L_z }(P_h)
=\varepsilon^{\alpha\beta}_{ J_z }(P_h).
\ee
Here $\varepsilon^\alpha_{ J_z }(P_h)$ is the polarization vector of bound state with $J=1$ and it obeys the 
following relations 
\be
\varepsilon^\alpha_{ J_z }(P_h)P_{h\alpha}&=&0, \nonumber\\
\sum_{L_z}\varepsilon^\alpha_{ J_z }(P_h)\varepsilon^{\ast\beta}_{ J_z }(P_h)&=&
-g^{\alpha\beta}+\frac{P_h^\alpha P_h^\beta}{M^2}\equiv\mathcal{Q}^{\alpha\beta}.
\ee
The $\varepsilon^{\alpha\beta}_{ J_z }(P_h)$ represents the polarization tensor for $J=2$ bound state and 
obeys the below relation \cite{Kuhn:1979bb,Guberina:1980dc}
\be
\varepsilon^{\alpha\beta}_{ J_z }(P_h)=\varepsilon^{\beta\alpha}_{ J_z }(P_h), ~~~~
\varepsilon^{\alpha}_{ J_z\alpha }(P_h)=0,~~~
P_{h\alpha}\varepsilon^{\alpha}_{ J_z}(P_h)=0,~~~\nonumber\\
\varepsilon^{\mu\nu}_{ J_z }(P_h)\varepsilon^{\ast\alpha\beta}_{ J_z }(P_h)=\frac12[\mathcal{Q}^{\mu\alpha}
\mathcal{Q}^{\nu\beta}+\mathcal{Q}^{\mu\beta}\mathcal{Q}^{\nu\alpha}]-\frac13\mathcal{Q}^{\mu\nu}
\mathcal{Q}^{\alpha\beta}.
\ee
 The $R_0(0)$ and $R_1^\prime(0)$ are the radial wave function and its 
derivative at the origin, and have 
the 
following relation with LDME \cite{Ko:1996xw}
\be\label{a111}
\langle 0\mid \mathcal{O}_1^{J/\psi}(\leftidx{^{2S+1}}{S}{_J})\mid 0\rangle=\frac{N_c}{2\pi}(2J+1)|R_0(0)|^2,
\ee
\be\label{a11}
\langle 0\mid \mathcal{O}_8^{J/\psi}(\leftidx{^{2S+1}}{S}{_J})\mid 0\rangle=\frac{2}{\pi}(2J+1)|R_0(0)|^2,
\ee
\be\label{a12}
\langle 0\mid \mathcal{O}_8^{J/\psi}(\leftidx{^3}{P}{_J})\mid 0\rangle=\frac{2N_c}{\pi}(2J+1)|R^\prime_1(0)|^2.
\ee
The numerical values of LDMEs are given in \tablename{~\ref{table2}}.
Now, let's  discuss about the each CO state (${^3}{S}{_1}$, ${^1}{S}{_0}$
${^3}{P}{_J}$) amplitudes in detail. 
%%%%%%%%%%%%%%%%%%%%%%%%%%%%%%%%%%%%%%%%%%%%%%%%%%%%%%%%%%%%%%%%%%%%%%%%%%%%%%%%%%%%%%%%%%%%%%%%%%%%%%
\subsection{  ${^3}{S}{_1}$ Amplitude}

We have the following symmetry relations for ${^3}{S}{_1}$ state
\begin{eqnarray}\label{e15}
 \mathrm{Tr}[O_1(0)(-\slashed{P}_h+M)\slashed{\varepsilon}_{s_z}]&=&
 \mathrm{Tr}[O_5(0)(-\slashed{P}_h+M)\slashed{\varepsilon}_{s_z}]\nonumber\\
 \mathrm{Tr}[O_2(0)(-\slashed{P}_h+M)\slashed{\varepsilon}_{s_z}]&=&
 \mathrm{Tr}[O_6(0)(-\slashed{P}_h+M)\slashed{\varepsilon}_{s_z}]\nonumber\\
 \mathrm{Tr}[O_3(0)(-\slashed{P}_h+M)\slashed{\varepsilon}_{s_z}]&=&
 \mathrm{Tr}[O_7(0)(-\slashed{P}_h+M)\slashed{\varepsilon}_{s_z}]\nonumber\\
 \mathrm{Tr}[O_4(0)(-\slashed{P}_h+M)\slashed{\varepsilon}_{s_z}]&=&
 -\mathrm{Tr}[O_8(0)(-\slashed{P}_h+M)\slashed{\varepsilon}_{s_z}].
\end{eqnarray}

Using Eq.\eqref{e15}, we can sum the color factors and we have
\begin{equation}
 \mathcal{C}_1+\mathcal{C}_5=\mathcal{C}_2+\mathcal{C}_6=\mathcal{C}_3+\mathcal{C}_7=\frac{\sqrt{2}}{2}d_{abc}
\end{equation}
The diagrams 4 and 8 do not contribute to ${^3}{S}{_1}$ state as from Eq.\eqref{e15}.
The final amplitude expression for ${^3}{S}{_1}$ state can be obtained by using Eq.\eqref{e8} and is given by
\begin{eqnarray}\label{e16}
\begin{aligned}
 \mathcal{M}[\leftidx{^{3}}{S}{_1}^{(8)}](P_h,k)=&{}\frac{1}{4\sqrt{\pi M}}R_0(0)\frac{\sqrt{2}}{2}d_{abc}
 \mathrm{Tr}\left[\sum_{m=1}^3O_m(0)(-\slashed{P}_h+M)\slashed{\varepsilon}_{s_z}\right],
\end{aligned}
\end{eqnarray}
where
\begin{eqnarray}\label{e17}
\begin{aligned}
\sum_{m=1}^3O_m(0)=&{}g^2_s(ee_c)\varepsilon^\mu_{\lambda_a}(k)\varepsilon^\nu_{\lambda_b}(q)\varepsilon^{
\rho\ast}_{
\lambda_g}(p_g)\Bigg[\frac{\gamma_\nu(\slashed{P_h}-2\slashed{q}+M)\gamma_\mu
(-\slashed{P_h}-2\slashed{p}_g+M)\gamma_\rho}{(\hat{s}-M^2)(\hat{u}-M^2)}\\
+&\frac{\gamma_\rho(\slashed{P_h}+2\slashed{p}_g+M)\gamma_\nu
(-\slashed{P_h}+2\slashed{k}+M)\gamma_\mu}{(\hat{s}-M^2)(\hat{t}-M^2)}+
\frac{\gamma_\nu(\slashed{P_h}-2\slashed{q}+M)\gamma_\rho(-\slashed{P_h}+2\slashed{k}+M)\gamma_\mu
}{(\hat{t}-M^2)(\hat{u}-M^2)}
\Bigg].
\end{aligned}
\end{eqnarray}
\subsection{  ${^1}{S}{_0}$ Amplitude}
The symmetry relations for ${^1}{S}{_0}$ state are given by 
\begin{eqnarray}\label{e20}
 \mathrm{Tr}[O_1(0)(-\slashed{P}_h+M)\gamma^5]&=&-
 \mathrm{Tr}[O_5(0)(-\slashed{P}_h+M)\gamma^5]\nonumber\\
 \mathrm{Tr}[O_2(0)(-\slashed{P}_h+M)\gamma^5]&=&-
 \mathrm{Tr}[O_6(0)(-\slashed{P}_h+M)\gamma^5]\nonumber\\
 \mathrm{Tr}[O_3(0)(-\slashed{P}_h+M)\gamma^5]&=&-
 \mathrm{Tr}[O_7(0)(-\slashed{P}_h+M)\gamma^5]\nonumber\\
 \mathrm{Tr}[O_4(0)(-\slashed{P}_h+M)\gamma^5]&=&
 \mathrm{Tr}[O_8(0)(-\slashed{P}_h+M)\gamma^5]
\end{eqnarray}

One can sum the color factors using  Eq.\eqref{e20} and we have the below relation
\begin{equation} \label{colr}
\mathcal{C}_1-\mathcal{C}_5=-\mathcal{C}_2+\mathcal{C}_6=-\mathcal{C}_3+\mathcal{C}_7=\frac{\sqrt{2}}{2}if_{
abc }, ~~~\mathcal{C}_4+\mathcal{C}_8=\sqrt{2}if_{abc }
 \end{equation}
Using Eq.\eqref{e8} the final amplitude expression for ${^1}{S}{_0}$ state  is given by
\begin{eqnarray}\label{e21}
\begin{aligned}
 \mathcal{M}[\leftidx{^{1}}{S}{_0}^{(8)}](P_h,k)=&{}\frac{1}{4\sqrt{\pi M}}R_0(0)\frac{\sqrt{2}}{2}if_{abc}
 \mathrm{Tr}\big[\left(O_1(0)-O_2(0)-O_3(0)+2O_4(0)\right)\\
 &\times(-\slashed{P}_h+M)\gamma^5\big]
\end{aligned}
\end{eqnarray}
where $O_1(0),O_2(0)$ and $O_3(0)$ are given in Eq.\eqref{e17} and 
\begin{equation}\label{ee4}
 \begin{aligned}
O_4(0)= 
g^2_s(ee_c)\varepsilon^\mu_{\lambda_a}(k)\varepsilon^\nu_{\lambda_b}(q)\varepsilon^{\rho\ast}_{
\lambda_g}(p_g)
\frac{\gamma_\nu(\slashed{P_h}-2\slashed{q}+M)\gamma^\sigma}{\hat{u}(\hat{u}-M^2)}\\
\times\left[g_{\mu\rho}(k+p_g)_\sigma+g_{\rho\sigma}(k-2p_g)_\mu+g_{\sigma\mu}(p_g-2k)_\rho\right]
\end{aligned}
\end{equation}
\subsection{  ${^3}{P}{_J}$ Amplitude}
The symmetry relations for $P$-state ($J=0,1,2$) are given by
\begin{eqnarray}\label{3pj1} 
\begin{aligned}
\mathrm{Tr}\big[O_{1\alpha}(0)\mathcal{P}_{1S_z}(0)+ O_1(0)\mathcal{P}_{1\alpha S_z}(0) \big]=
 -\mathrm{Tr}\big[O_{5\alpha}(0)\mathcal{P}_{1S_z}(0)+ O_5(0)\mathcal{P}_{1\alpha S_z}(0) \big]\nonumber
\end{aligned}
\end{eqnarray}
\begin{eqnarray}\label{3pj2} 
\begin{aligned}
\mathrm{Tr}\big[O_{2\alpha}(0)\mathcal{P}_{1S_z}(0)+ O_2(0)\mathcal{P}_{1\alpha S_z}(0) \big]=
 -\mathrm{Tr}\big[O_{6\alpha}(0)\mathcal{P}_{1S_z}(0)+ O_6(0)\mathcal{P}_{1\alpha S_z}(0) \big]\nonumber
\end{aligned}
\end{eqnarray}
\begin{eqnarray}\label{3pj3} 
\begin{aligned}
\mathrm{Tr}\big[O_{3\alpha}(0)\mathcal{P}_{1S_z}(0)+ O_3(0)\mathcal{P}_{1\alpha S_z}(0) \big]=
 -\mathrm{Tr}\big[O_{7\alpha}(0)\mathcal{P}_{1S_z}(0)+ O_7(0)\mathcal{P}_{1\alpha S_z}(0) \big]\nonumber
\end{aligned}
\end{eqnarray}
\begin{eqnarray}\label{3pj4} 
\begin{aligned}
\mathrm{Tr}\big[O_{4\alpha}(0)\mathcal{P}_{1S_z}(0)+ O_4(0)\mathcal{P}_{1\alpha S_z}(0) \big]=
 \mathrm{Tr}\big[O_{8\alpha}(0)\mathcal{P}_{1S_z}(0)+ O_8(0)\mathcal{P}_{1\alpha S_z}(0) \big].
\end{aligned}
\end{eqnarray}
From above equations, we get  the  color factors as given in Eq.\eqref{colr}.
Using these color factors, the Eq.\eqref{e9} can be further simplified as below
\begin{multline}\label{3pj8} 
\mathcal{M}[\leftidx{^{3}}{P}{_J}^{(8)}](P_h,k)=\frac{\sqrt{2}}{2}f_{
abc }\sqrt{\frac{3}{4\pi}}R^\prime_1(0)\sum_{L_zS_z}
\varepsilon^\alpha_{L_z}(P_h)
\langle 1L_z;1S_z|JJ_z\rangle\\
\mathrm{Tr}\bigg[\left(O_{1\alpha}(0)-O_{2\alpha}(0)-O_{3\alpha}(0)+2O_{4\alpha}(0)\right)\mathcal{P}_{SS_z}
(0)+\left( O_1(0)-O_2(0)-O_3(0)+2O_4(0)\right)
\mathcal{P}_{SS_z\alpha}(0)
 \bigg].
 \end{multline}
 In order to calculate the amplitude expression for $J=0,1$ and 2, we have used the Clebsch-Gordan 
coefficients 
as defined in 
 Eq.\eqref{s0}, \eqref{s1} and \eqref{s2}. After summing and averaging over the colors and spins, the amplitude square of each state is given 
 in appendix \ref{ap1}.

%======================================================================================================
\section{Numerical Results}\label{sec4} 
%======================================================================================================
In this section, we discuss the numerical results of SSA  and inelastic photoproduction
of $J/\psi$ in polarized and unpolarized $ep$ collision respectively.
For numerical estimation of SSA, best fit parameters of GSF from \cite{alesio}
and up and down quark Sivers function parameters from \cite{Anselmino:2016uie} are considered. MSTW2008 
\cite{Martin:2009iq} is used for PDF which is probed at the scale $\mu=\sqrt{M^2+P^2_T}$. Mass of $J/\psi$, 
M=3.096 GeV is taken. The NLO subprocess  
$\gamma+g\rightarrow J/\psi+g$ is considered for $J/\psi$ production in $ep^\uparrow\rightarrow J/\psi+X$ 
process. The COM is employed for calculating production rate of $J/\psi$. The 
$\leftidx{^{3}}{S}{_1}^{(8)}$, 
$\leftidx{^{1}}{S}{_0}^{(8)}$, $\leftidx{^{3}}{P}{_0}^{(8)}$, $\leftidx{^{3}}{P}{_1}^{(8)}$ and 
$\leftidx{^{3}}{P}{_2}^{(8)}$ states amplitudes  are calculated using FORM package 
\cite{Kuipers:2012rf}, and are given in appendix \ref{ap1}. For comparison, we have considered three sets of 
LDMEs  from the References \cite{Chao:2012iv,Butenschoen:2011yh,Zhang:2014ybe}, which are tabulated in 
  \tablename{~\ref{table2}}. The LDMEs for $J=1,2$ are obtained by using the relations $\langle 
\mathcal{O}_8^{J/\psi}(\leftidx{^3}{P}{_1})\rangle=3\langle 
\mathcal{O}_8^{J/\psi}(\leftidx{^3}{P}{_0})\rangle$ and $\langle 
\mathcal{O}_8^{J/\psi}(\leftidx{^3}{P}{_2})\rangle=5\langle 
\mathcal{O}_8^{J/\psi}(\leftidx{^3}{P}{_0})\rangle$. The transverse momentum of the initial gluon 
$k_{\perp g}$ in Eq.\eqref{d1} is 
integrated within the limits $0<k_{\perp g}<3$ GeV. We have noticed that the higher values of $k_{\perp 
g~\mathrm{max}}$ (upper limit of the $k_{\perp g}$ integration) do not affect the SSA and 
unpolarized differential cross section.
  \par
\begin{figure}[H]
\begin{minipage}[c]{0.99\textwidth}
\small{(a)}\includegraphics[width=8cm,height=6.5cm,clip]{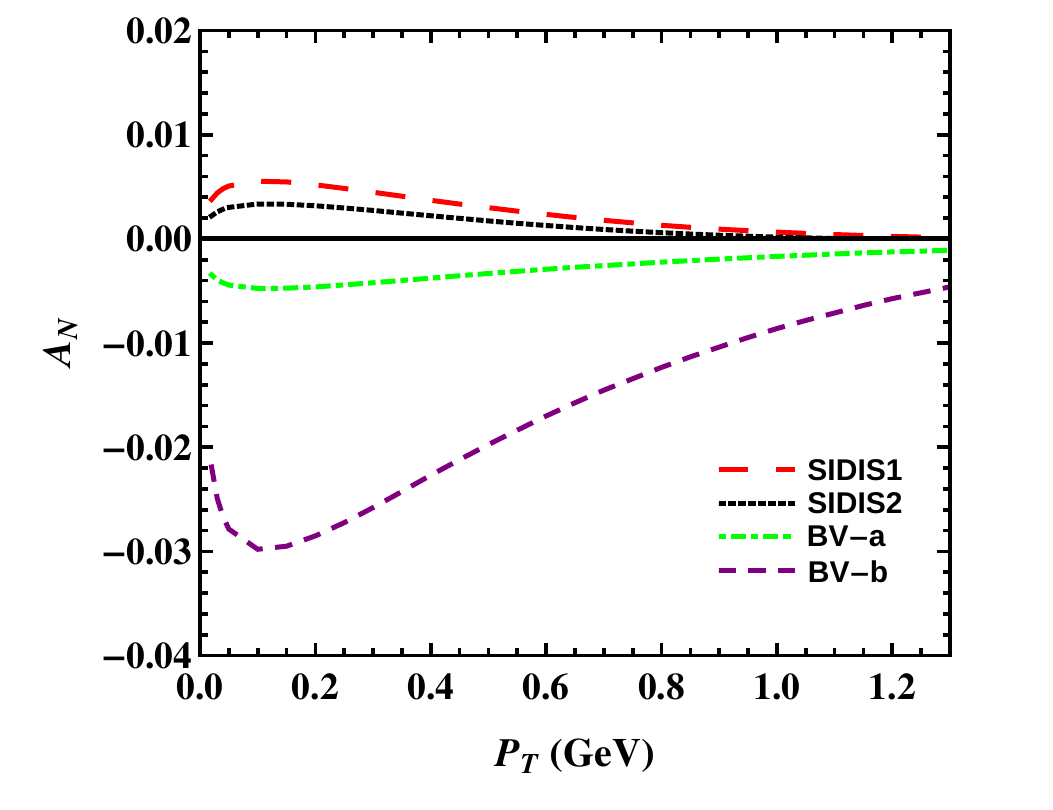}
\hspace{0.1cm}
\small{(b)}\includegraphics[width=8cm,height=6.5cm,clip]{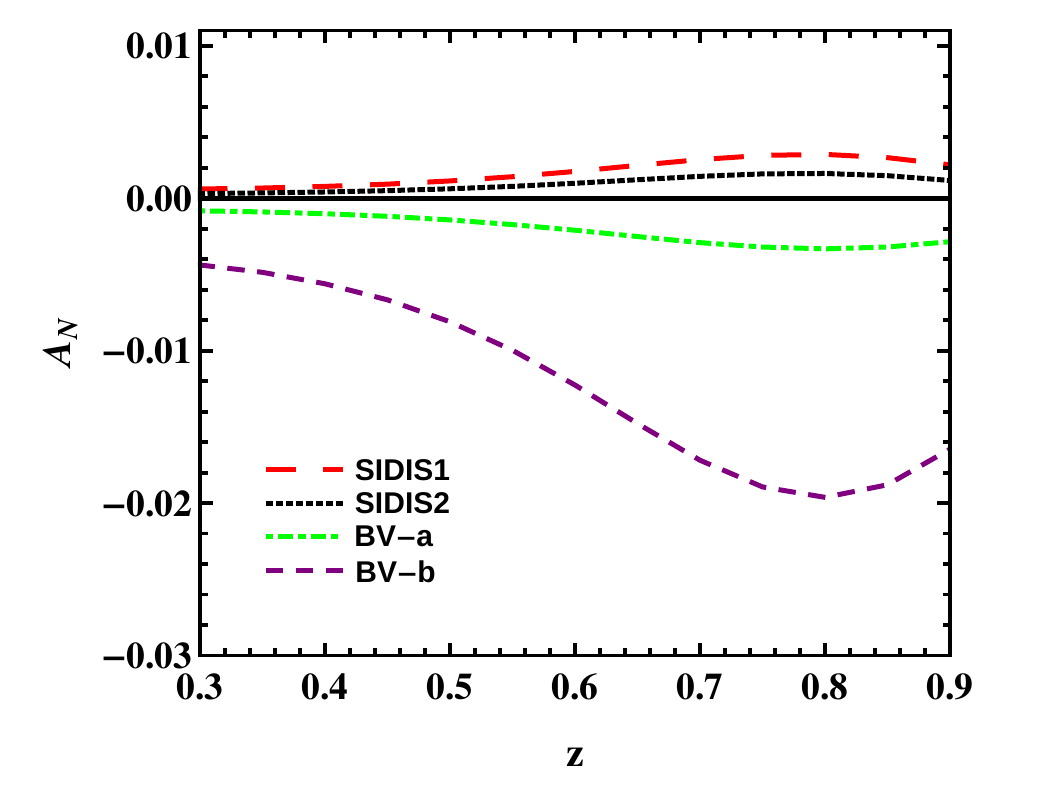}
\end{minipage}
\caption{\label{fig3}Single spin asymmetry  in $e+p^{\uparrow}\rightarrow J/\psi +X$
process as function of 
(a) $P_T$ (left panel) and  (b) $z$ (right panel) at $\sqrt{s}=100$ GeV (EIC). The integration ranges are 
$0<P_{T}\leq1$ GeV and $0.3<z<0.9$. For convention of lines see the text.}
\end{figure}
\begin{figure}[H]
\begin{minipage}[c]{0.99\textwidth}
\small{(a)}\includegraphics[width=8cm,height=6.5cm,clip]{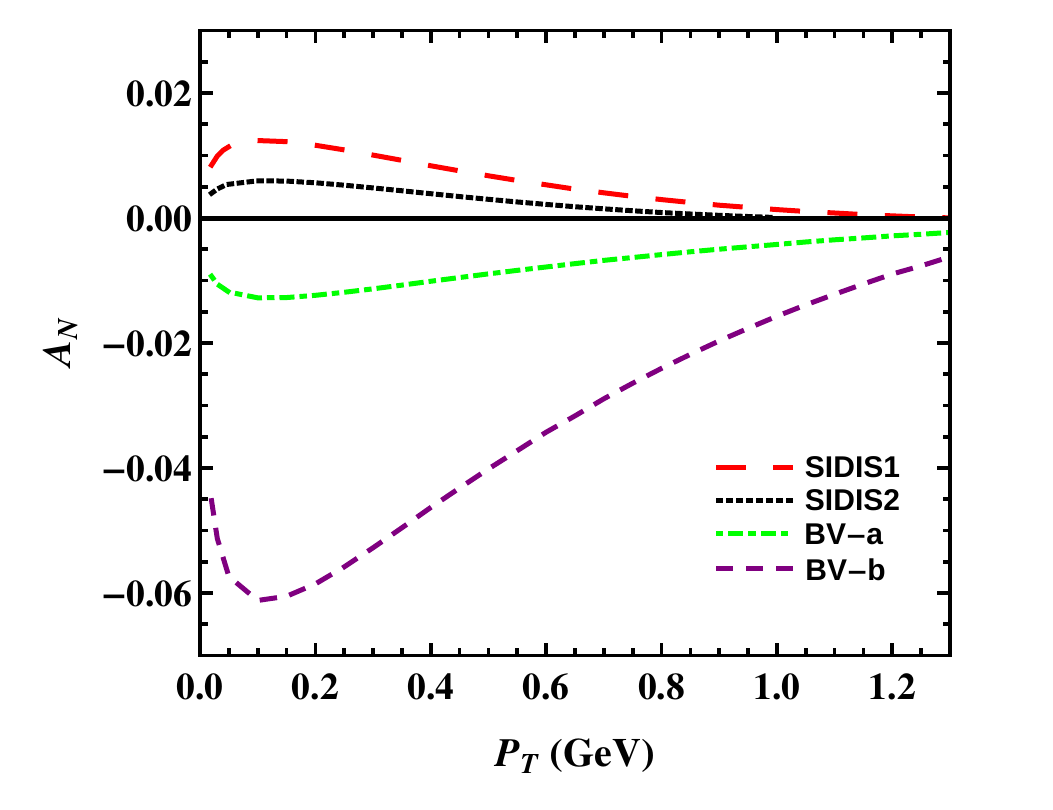}
\hspace{0.1cm}
\small{(b)}\includegraphics[width=8cm,height=6.5cm,clip]{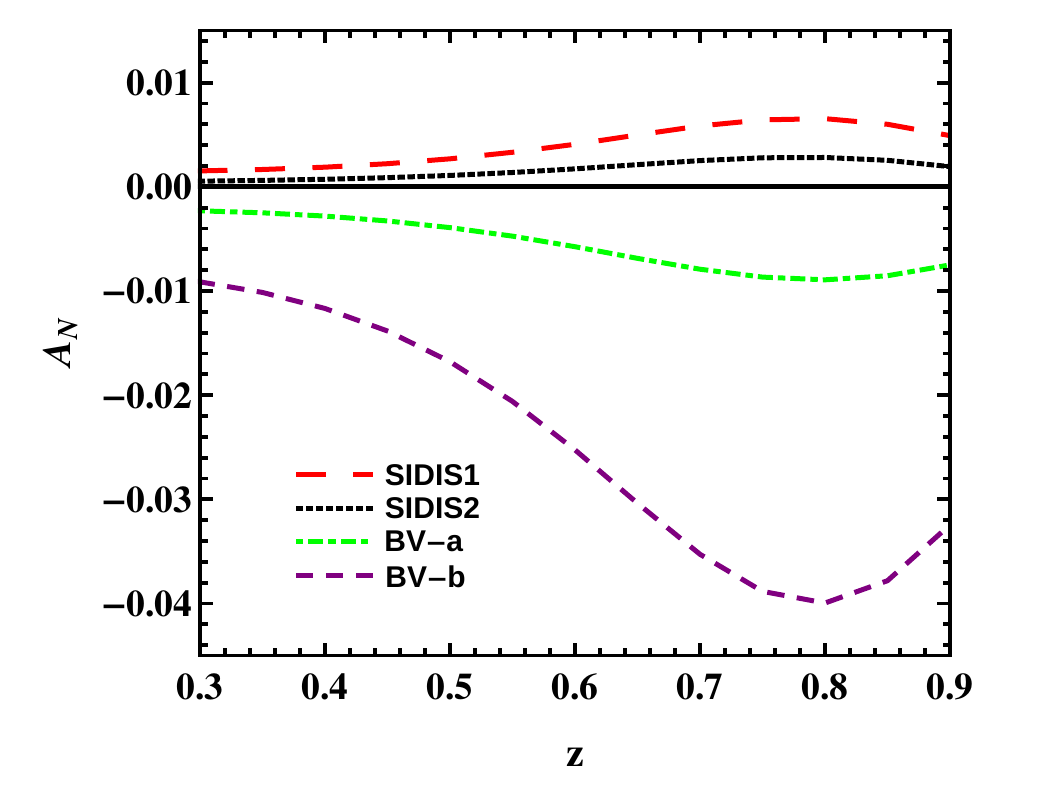}
\end{minipage}
\caption{\label{fig4}Single spin asymmetry  in $e+p^{\uparrow}\rightarrow J/\psi +X$
process as function of 
(a) $P_T$ (left panel) and  (b) $z$ (right panel) at $\sqrt{s}=45$ GeV (EIC). The integration ranges are 
$0<P_{T}\leq1$ GeV  and $0.3<z<0.9$.  For convention of lines see the text.}
\end{figure}
\begin{figure}[H]
\begin{minipage}[c]{0.99\textwidth}
\small{(a)}\includegraphics[width=8cm,height=6.5cm,clip]{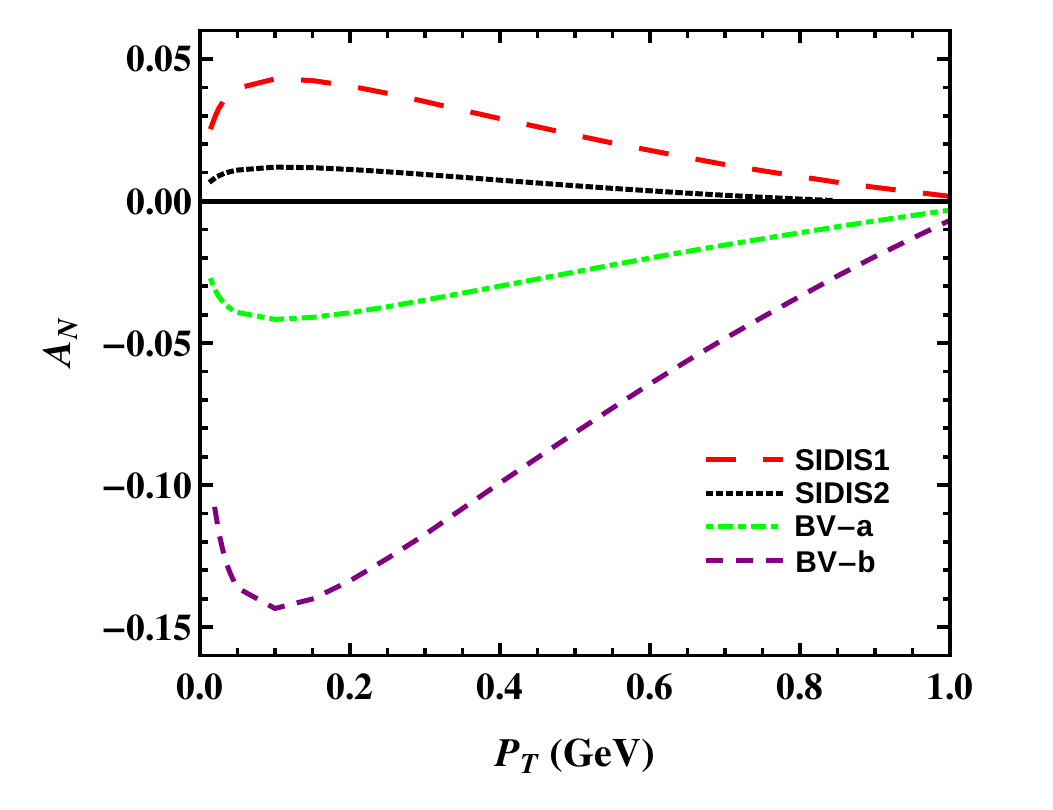}
\hspace{0.1cm}
\small{(b)}\includegraphics[width=8cm,height=6.5cm,clip]{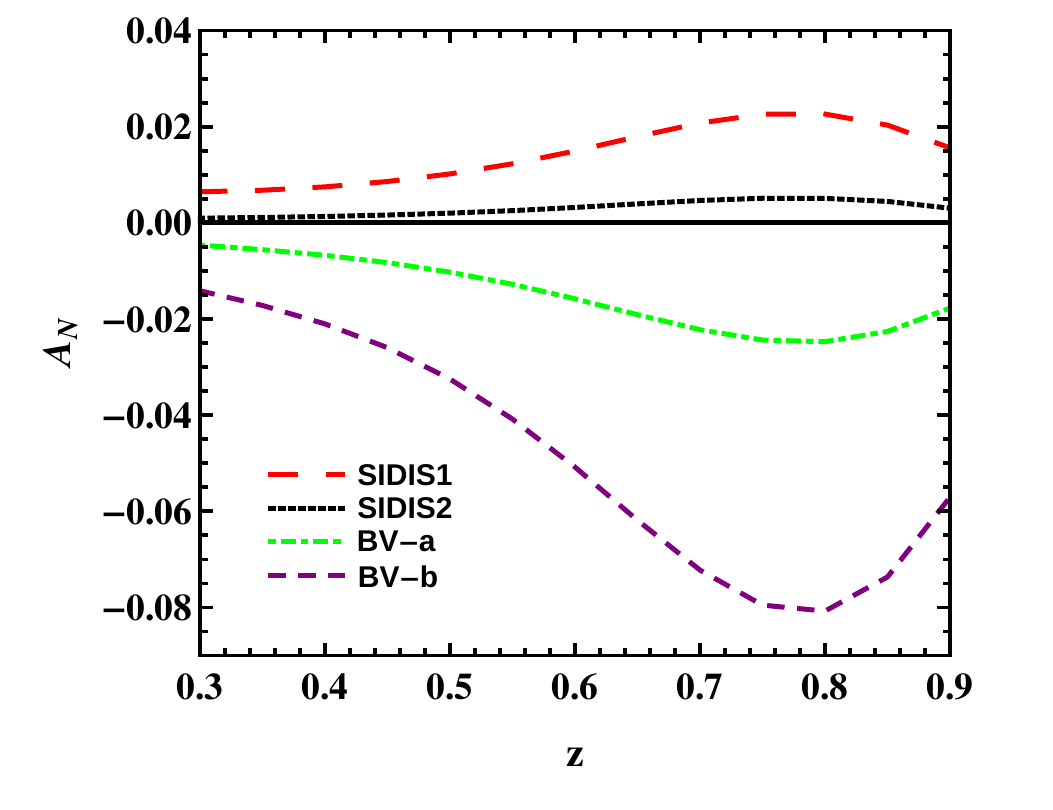}
\end{minipage}
\caption{\label{fig5}Single spin asymmetry  in $e+p^{\uparrow}\rightarrow J/\psi +X$
process as function of 
(a) $P_T$ (left panel) and  (b) $z$ (right panel) at $\sqrt{s}=17.2$ GeV (COMPASS). The integration ranges 
are $0<P_{T}\leq1$ GeV  and $0.3<z<0.9$. For convention of lines see the text.}
\end{figure}
\begin{figure}[H]
\begin{minipage}[c]{0.99\textwidth}
\small{(a)}\includegraphics[width=8cm,height=6.5cm,clip]{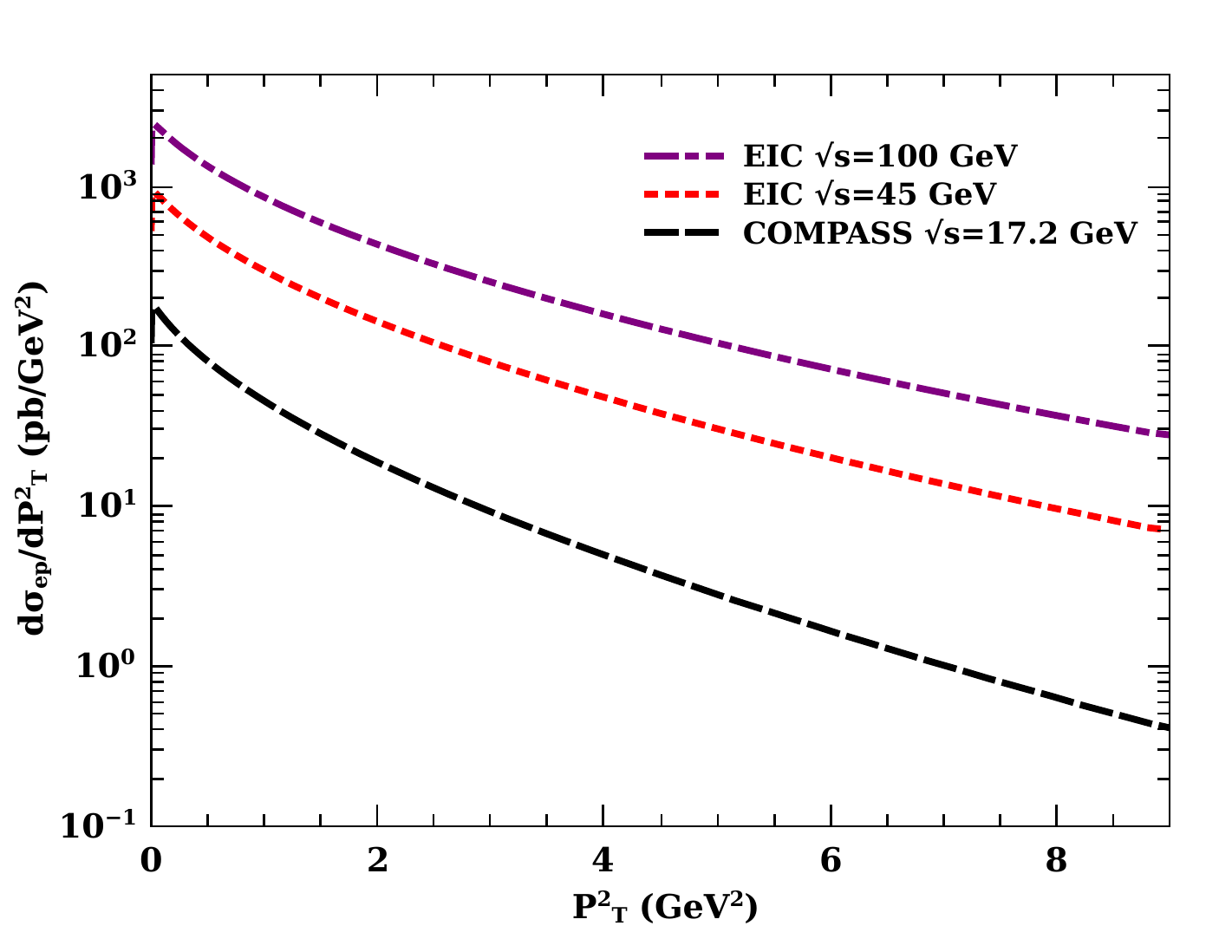}
\hspace{0.1cm}
\small{(b)}\includegraphics[width=8cm,height=6.5cm,clip]{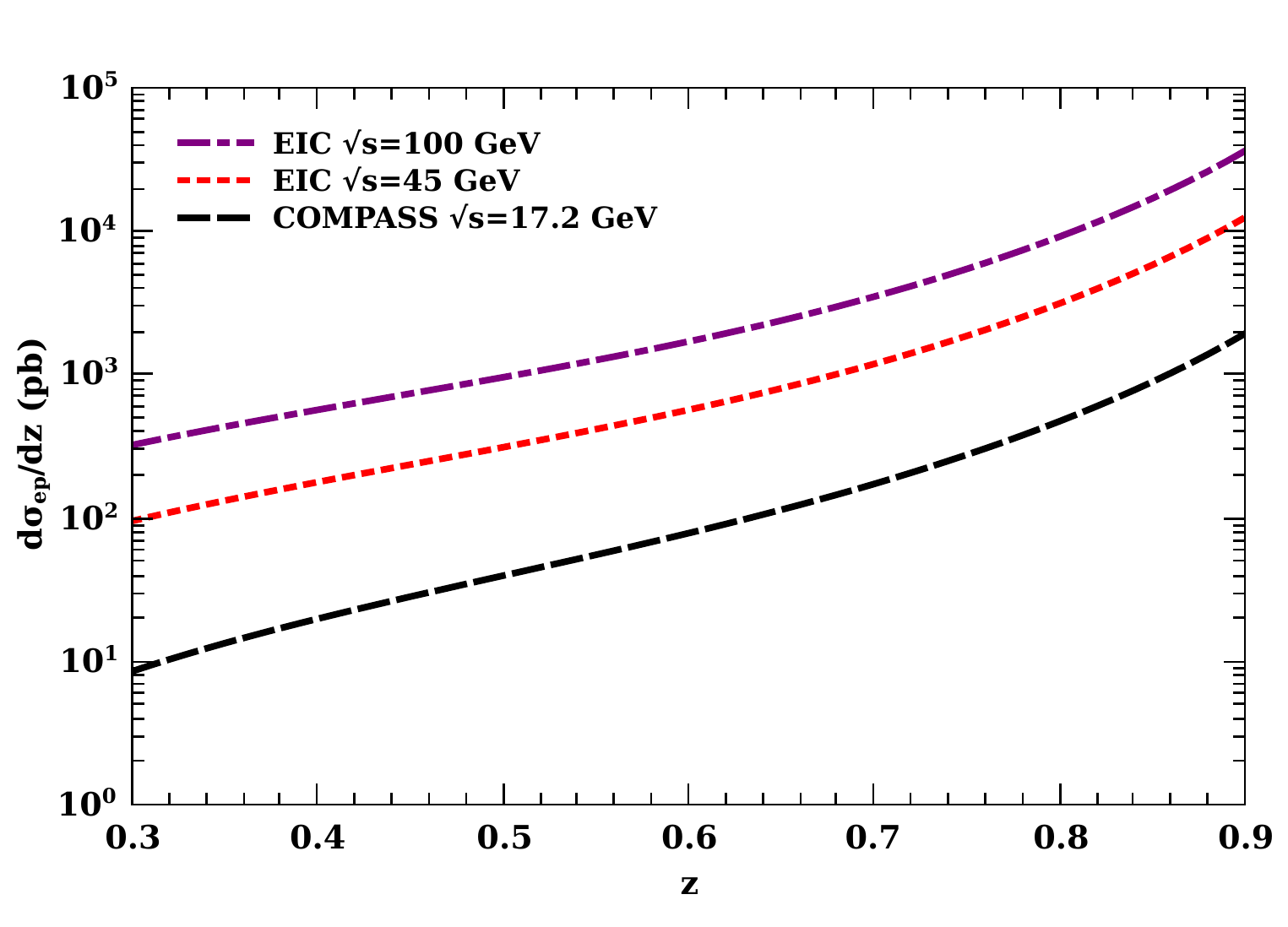}
\end{minipage}
\caption{\label{fig6} Unpolarized differential cross section in $e+p\rightarrow J/\psi +X$
process as function of (a) $P_T$ (left panel) and  (b) $z$ (right panel) at $\sqrt{s}=100, 45$ GeV (EIC) and
$\sqrt{s}=17.2$ GeV (COMPASS)  with $\langle k^2_{\perp g}\rangle=1$ GeV$^2$. The each curve is obtained 
by 
taking into account the color singlet and color octet states contribution to $J/\psi$ production. 
The integration ranges are $0<P_{T}\leq3$ GeV and $0.3<z<0.9$. LDMEs are from \cite{Zhang:2014ybe}.}
\end{figure}
\begin{figure}[H]
\begin{minipage}[c]{0.99\textwidth}
\small{(a)}\includegraphics[width=8cm,height=6.5cm,clip]{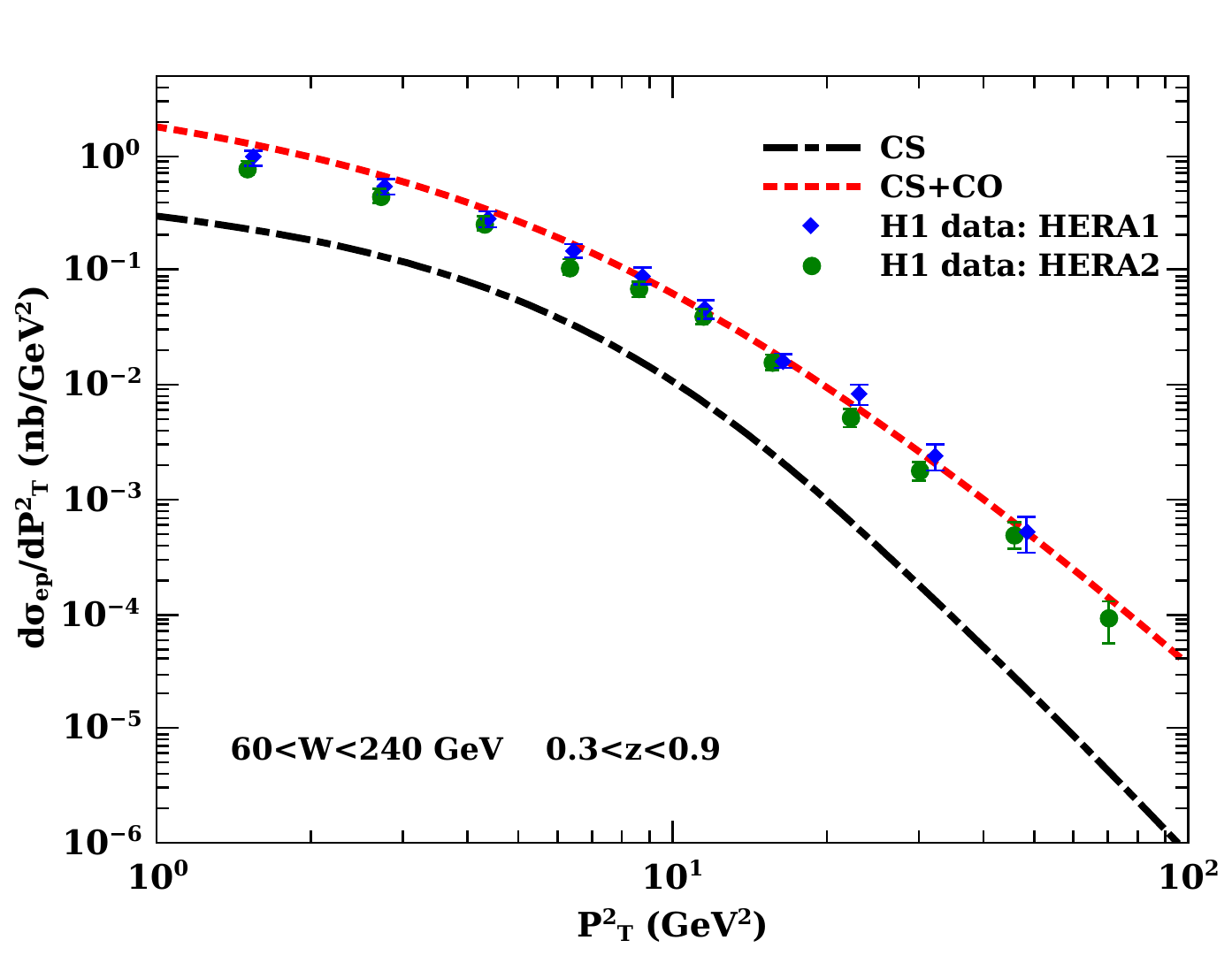}
\hspace{0.1cm}
\small{(b)}\includegraphics[width=8cm,height=6.5cm,clip]{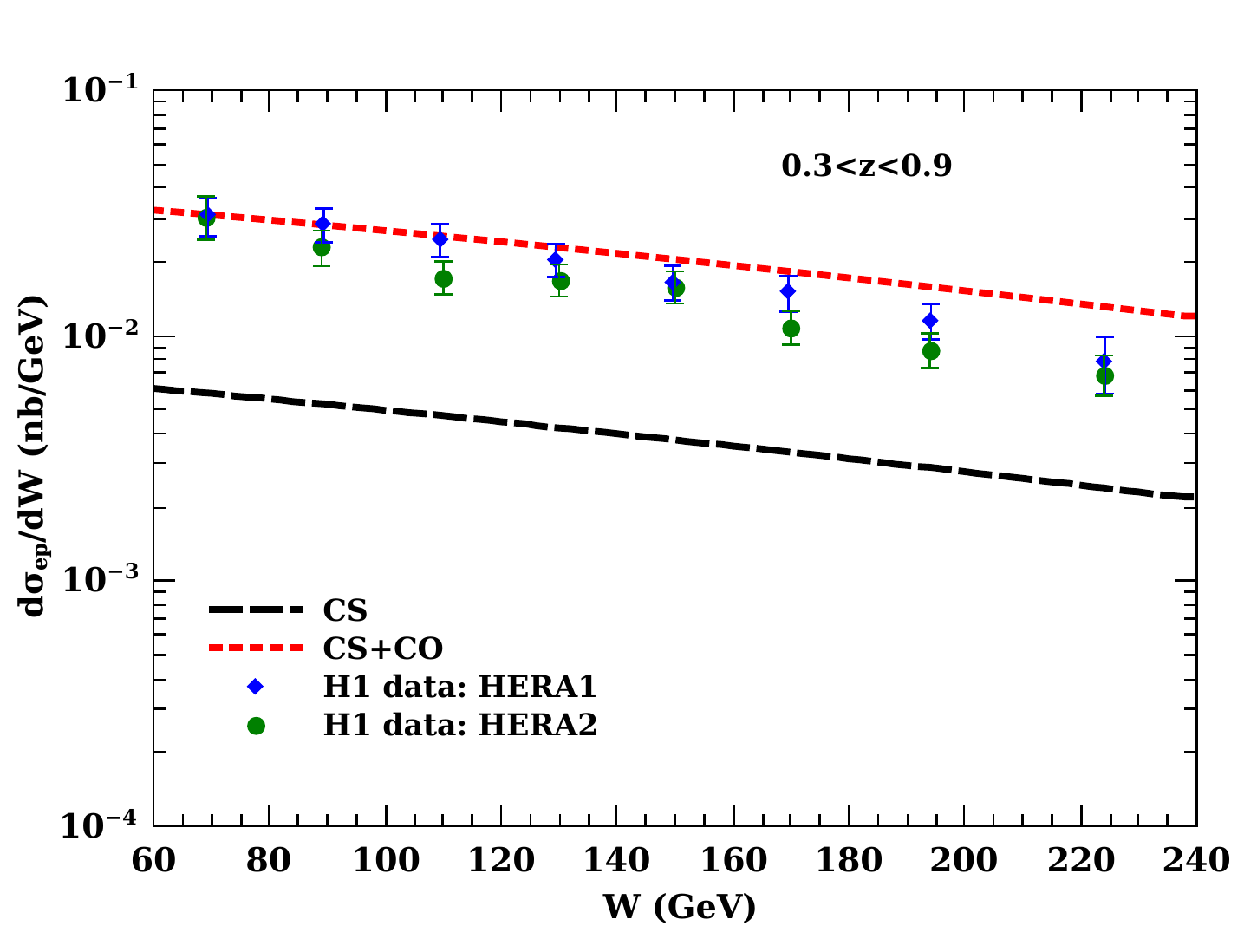}
\end{minipage}
\begin{center}
 \small{(c)}\includegraphics[width=8cm,height=6.5cm,clip]{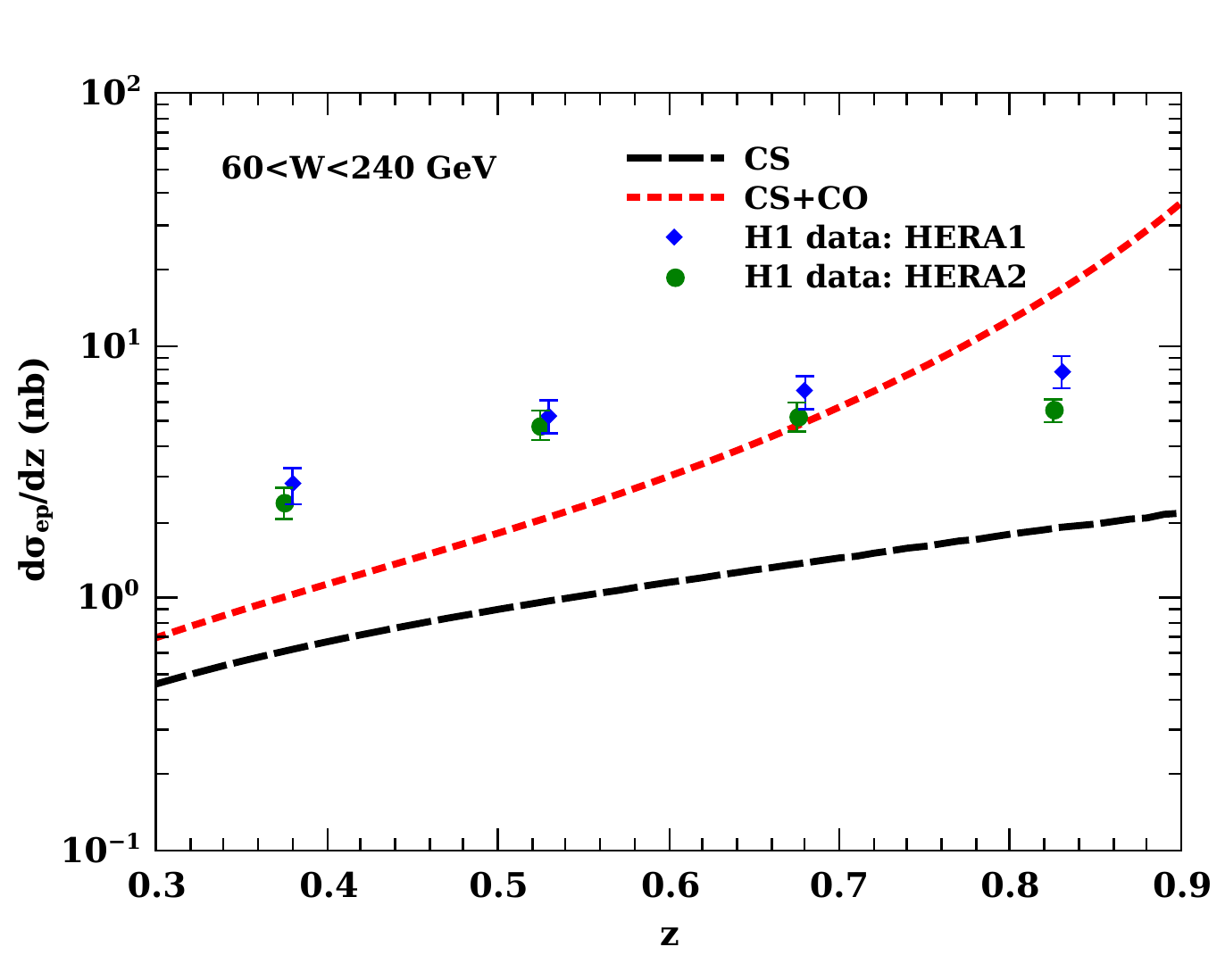}
\end{center}
\caption{\label{fig7} Unpolarized differential cross section in $e+p\rightarrow J/\psi +X$
process as function of (a) $P_T$ (left panel),  (b) $W$ (right panel) and (c) $z$ (lower pannel) at 
HERA ($\sqrt{s}=318$ GeV) with $\langle k^2_{\perp g}\rangle=1$ GeV$^2$. The H1 data  from 
\cite{Adloff:2002ex,Aaron:2010gz} and LDMEs are from \cite{Zhang:2014ybe}. The integration ranges are 
$1<P_{T}\leq 10$ GeV, $60<W<240$ GeV and $0.3<z<0.9$. The curves \textquotedblleft CS\textquotedblright  
 and \textquotedblleft CS+CO\textquotedblright  represent the consideration of $J/\psi$ production only in 
color singlet model and color singlet plus color octet model respectively.}
\end{figure}

%-----------------------------------------------------------------------------
\begin{table}[h!]
  \centering
  \caption{Numerical values of LDMEs.}
  \label{table2}
   \begin{tabular}{ccccc}
    \toprule
  \hline
  \hline
  $~~~~~~~~~~~$
&$\langle  \mathcal{O}_1^{J/\psi}(\leftidx{^3}{S}{_1})\rangle$
&$~~\langle \mathcal{O}_8^{J/\psi}(\leftidx{^3}{S}{_1})\rangle$
&$~~\langle \mathcal{O}_8^{J/\psi}(\leftidx{^1}{S}{_0})\rangle$
 &$~~\langle \mathcal{O}_8^{J/\psi}(\leftidx{^3}{P}{_0})\rangle$  \\
 $~~~~~~~~~~~$
 &$\mathrm{GeV}^3$
&$~\times 10^{-2}\mathrm{GeV}^3$
&$~\times 10^{-2}\mathrm{GeV}^3$
&$~\times 10^{-2}\mathrm{GeV}^5$\\
 \hline
\midrule
 Ref.\cite{Chao:2012iv}&1.16 & $0.3\pm 0.12$ &$ 8.9\pm 0.98$ & $1.26\pm0.47 $ \\
  Ref.\cite{Butenschoen:2011yh}&1.32 & $0.168\pm0.046$ & $3.04\pm0.35$ &$ -0.908\pm0.161$\\
Ref.\cite{Zhang:2014ybe}&$0.645\pm0.405$ & $1.0\pm0.3$ & $0.785\pm0.42$ & $3.8\pm1.1$ \\
  \hline 
  \hline
\bottomrule
  \end{tabular}
\end{table}
%--------------------------------------------------------------------------------------
We have estimated the SSA at $\sqrt{s}=100,45$ GeV (EIC) and $\sqrt{s}=17.2$ GeV (COMPASS)  
energies using Eq.\eqref{asy} by fixing the $J/\psi$ production plane as discussed in 
\cite{Anselmino:2009pn}. The SSA as a function of $P_T$ and $z$ is obtained by integrating 
$0.3<z\leq0.9$ and $0<P_T\leq1$ GeV respectively, and is shown in \figurename{\ref{fig3}-\ref{fig5}}.
The  light-cone 
momentum fraction $x_\gamma$ of quasi-real photon  is integrated over the 
range  $0<x_\gamma<1$  in \figurename{\ref{fig3}-\ref{fig6}}. The 
upper bound on the virtuality of the photon in Eq.(\ref{flux}), $Q^2_{max}=1~\mathrm{GeV}^2$ is considered in 
 \figurename{\ref{fig3}-\ref{fig6}}.
The integration w.r.t the light-cone momentum fraction of initial gluon $x_g$ in Eq.\eqref{d3} and 
\eqref{d4} is carried out  by using the Dirac delta function as discussed in appendix \ref{ap2}. The 
conventions in the \figurename{\ref{fig3}-\ref{fig5}} are the following. The obtained  asymmetry
using D'Alesio et al. \cite{alesio} fit parameters of GSF is represented by 
\textquotedblleft SIDIS1\textquotedblright and \textquotedblleft SIDIS2\textquotedblright. The 
\textquotedblleft BV-a\textquotedblright and \textquotedblleft BV-b\textquotedblright  curves are 
obtained by using Anselmino et al. \cite{Anselmino:2016uie} fit parameters as defined in Eq.\eqref{ab}.
As aforementioned, due to the final state interactions the  asymmetry is nonzero when the heavy quark pair is 
produced
in the CO state in $ep$ collision \cite{Yuan:2008vn}. Therefore, we have considered the initial heavy quark 
pair production is to be only in the CO state for calculating the numerator part of Eq.\eqref{asy}.
However, the denominator of Eq.\eqref{asy} is basically two times the unpolarized cross section and CS state 
do contribute significantly to unpolarized cross section as shown in \figurename{\ref{fig7}}. Hence, CS state
contribution of $J/\psi$ is taken into account in the denominator of asymmetry. The asymmetry is increased by 
maximum about $30\%$ if the CS state contribution is not considered in the denominator.
The SSA decreases as center-of-mass (C.M) energy increases in the kinematical range considered. \par
\begin{figure}[H]
\begin{minipage}[c]{0.99\textwidth}
\small{(a)}\includegraphics[width=8cm,height=6.5cm,clip]{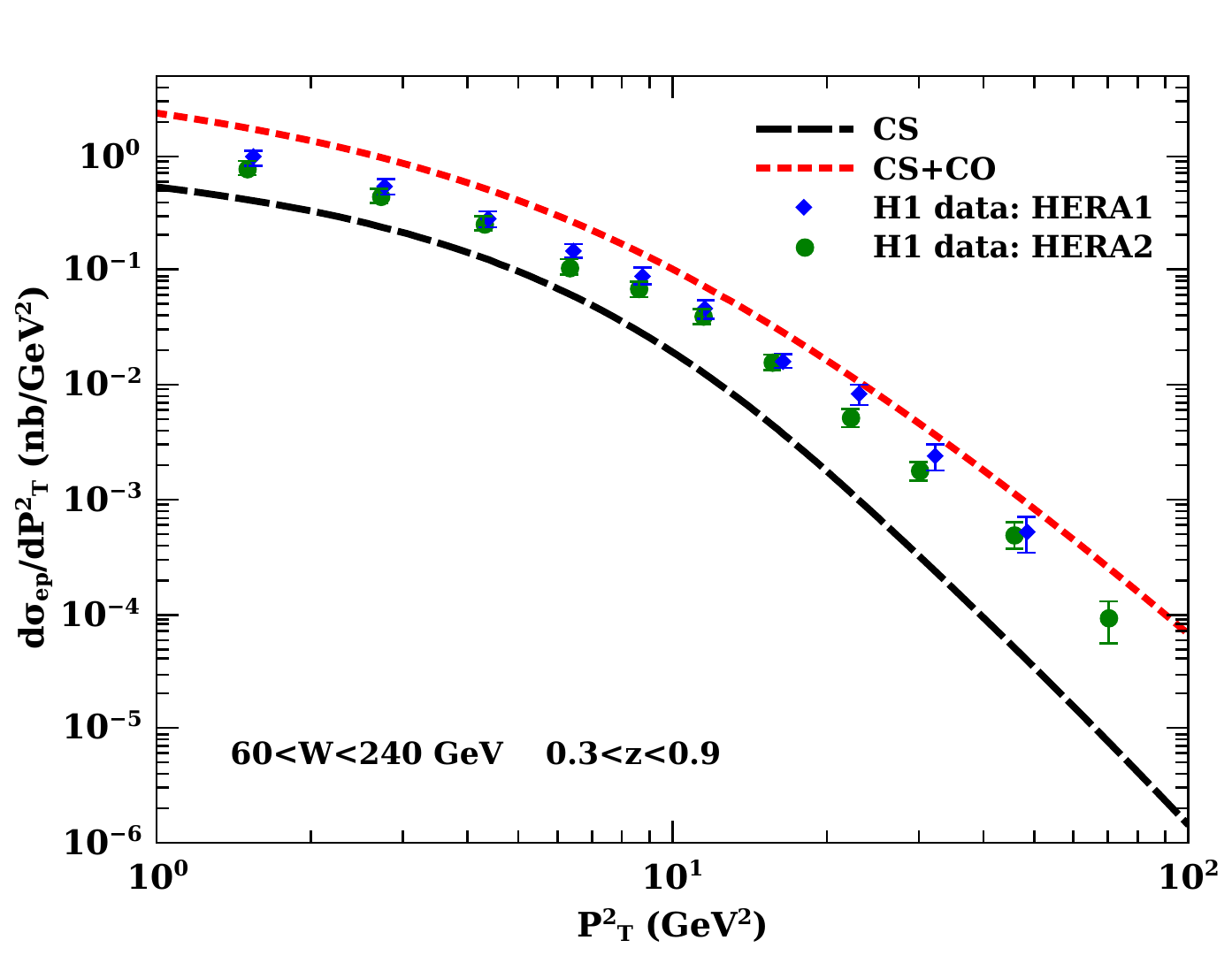}
\hspace{0.1cm}
\small{(b)}\includegraphics[width=8cm,height=6.5cm,clip]{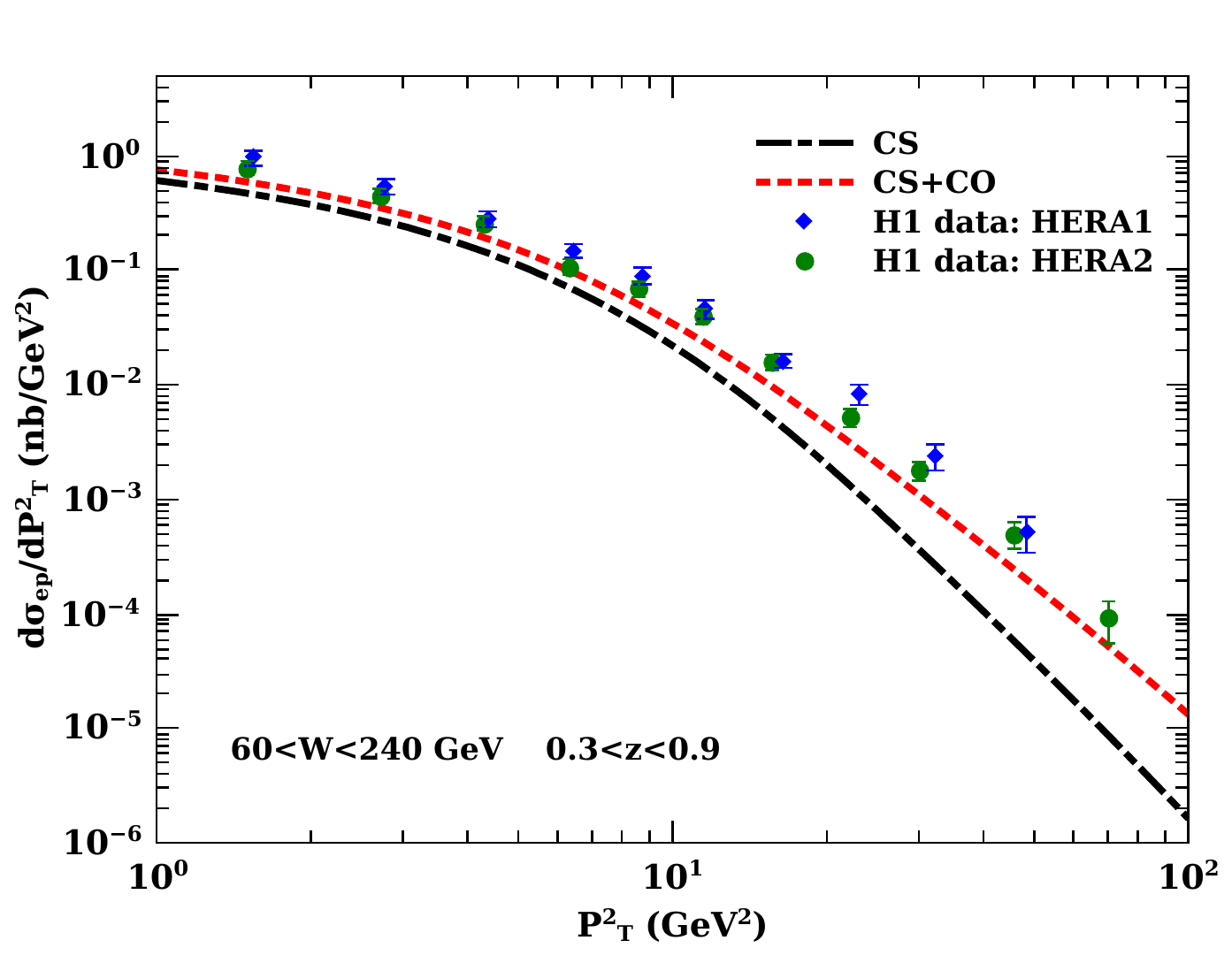}
\end{minipage}
\caption{\label{fig8} Unpolarized differential cross section in $e+p\rightarrow J/\psi +X$
process as function of  $P_T$ at HERA ($\sqrt{s}=318$ GeV) using the LDMEs  from (a) Ref. \cite{Chao:2012iv}
(left panel) and  (b) Ref. \cite{Butenschoen:2011yh} (right panel)
with $\langle k^2_{\perp g}\rangle=1$ GeV$^2$. 
The H1 data  from 
\cite{Adloff:2002ex,Aaron:2010gz}. The integration ranges are 
$1<P_{T}\leq 10$ GeV, $60<W<240$ GeV and $0.3<z<0.9$. The convention of lines is same as 
\figurename{\ref{fig7}}.}
\end{figure}

%%%%%%%%%%%%%%%%%%%%%%%%%%%%%%%%%%%%%%%%%%%%%%%%%%%%%%%%%%%%%%%%%%%%%%%%%%%%%%%%%%%%%%%%%%%%%%%%%%%%%%%%%%%%%%%%%%%%%
From  \figurename{\ref{fig3}-\ref{fig5}}, SIDIS and BV parameters are estimating positive and negative 
asymmetry respectively as a function 
of $P_T$ and $z$. However, the estimated asymmetry  using \textquotedblleft SIDIS2\textquotedblright fit is 
almost close to zero for all $\sqrt{s}$. The obtained asymmetry as a function of $P_T$ using 
\textquotedblleft BV-b\textquotedblright 
parameters is maximum about  14\% at  COMPASS $\sqrt{s}$.
Basically, asymmetry is proportional  to GSF which is considered as 
an average of $u$ and $d$ quark's $x$-dependent normalization $\mathcal{N}(x_g)$ in \textquotedblleft BV-a\textquotedblright  parameterization as defined
in Eq.\eqref{ab}. 
The sign of the asymmetry depends on relative magnitude of $N_u$ and $N_d$ and these have opposite sign which
can be observed in \tablename{~\ref{table1}}. The magnitude of $\mathcal{N}_d(x_g)$ is dominant compared to 
$\mathcal{N}_u(x_g)$  as a result the asymmetry is negative.
The LDMEs from Ref. \cite{Chao:2012iv} and \cite{Zhang:2014ybe} estimate similar asymmetry as presented in 
\figurename{\ref{fig3}-\ref{fig5}}. However, the obtained asymmetry using LDMEs of 
Ref. \cite{Butenschoen:2011yh} is one order magnitude lesser than that of \figurename{\ref{fig3}-\ref{fig5}}.
This is due to the fact that CS state contribution that apper only in the denominator is much larger than  CO 
state as shown in the right panel of \figurename{\ref{fig8}}.
 Nevertheless, the magnitude and sign of the asymmetry 
strongly depends on the modeling of GSF. Asymmetry increases slightly for higher values of Gaussian widths of 
unpolarized gluon TMD which appears in the denominator of asymmetry definition.
\par
%------------------------------------------------------------------------------------------------
In \figurename{\ref{fig6}}, the unpolarized differential cross section as a function of $P_T$ and $z$ 
using the LDMEs from Ref.\cite{Zhang:2014ybe}
at EIC and COMPASS energies is shown. The CS state, $\leftidx{^{3}}{S}{_1}^{(1)}$, 
contribution to $J/\psi$ production
is considered along with CO states to obtain the  \figurename{\ref{fig6}}.
The energy spectrum of $J/\psi$, right panel in 
\figurename{\ref{fig6}},  is restricted to $z\leq 0.9$ as  we are interested in the inelastic $J/\psi$ 
production. The
Gaussian parametrization of gluon TMD as defined in
Eq.\eqref{unp} with Gaussian width $\langle k^2_{\perp g}\rangle=1$ GeV$^2$ is considered.  For lower 
values of TMD width, i.e., $\langle k^2_{\perp g}\rangle=0.5$ GeV$^2$, the 
cross section differential in $z$ is increased by 10\%  at low z region. 
Whereas the differential cross section as a function of $P_T$ is increased by 4.5\% in the low 
$P_T$ region. The $\leftidx{^{3}}{P}{_J}^{(8a)}$ state contribution to $J/\psi$ production is 
significantly large compared to $\leftidx{^{3}}{S}{_1}^{(8a)}$ and $\leftidx{^{1}}{S}{_0}^{(8a)}$ 
states for the  LDMEs of Ref.\cite{Zhang:2014ybe}.\par
%------------------------------------------------------------------------------------------
The obtained unpolarized differential cross section of $J/\psi$ using the LDMEs of Ref.\cite{Zhang:2014ybe}  
is compared with H1 data \cite{Adloff:2002ex,Aaron:2010gz} in 
\figurename{\ref{fig7}}. The theoretical results are calculated within the same kinematical region of H1 data,
i.e., $\sqrt{s}=318$ GeV, $P^2_T>1$ GeV$^2$, $60<W<240$ GeV, $0.3<z<0.9$ and $Q^2_{max}=2.5$ GeV$^2$. The 
C.M energy of the photon-proton system is $W$ and $W^2=(P+q)^2\approx x_\gamma s$, where $s=(P+l)^2$ is the 
C.M energy square of the proton-lepton system. The $P_T$ and $W$ spectra obtained by considering the 
$J/\psi$ production in CS state along with the CO states are in good agreement with data. 
However, the CS contribution to the $J/\psi$ production is below the data. In 
\figurename{\ref{fig7}}, the $d\sigma/dz$ distribution is not well described by both CS and CO contributions 
of $J/\psi$. From 
\figurename{\ref{fig7}}, it is obvious that the CO states contribution is dominated for higher $z$ 
values.\par
\begin{figure}[H]
\begin{minipage}[c]{0.99\textwidth}
\small{(a)}\includegraphics[width=8cm,height=6.5cm,clip]{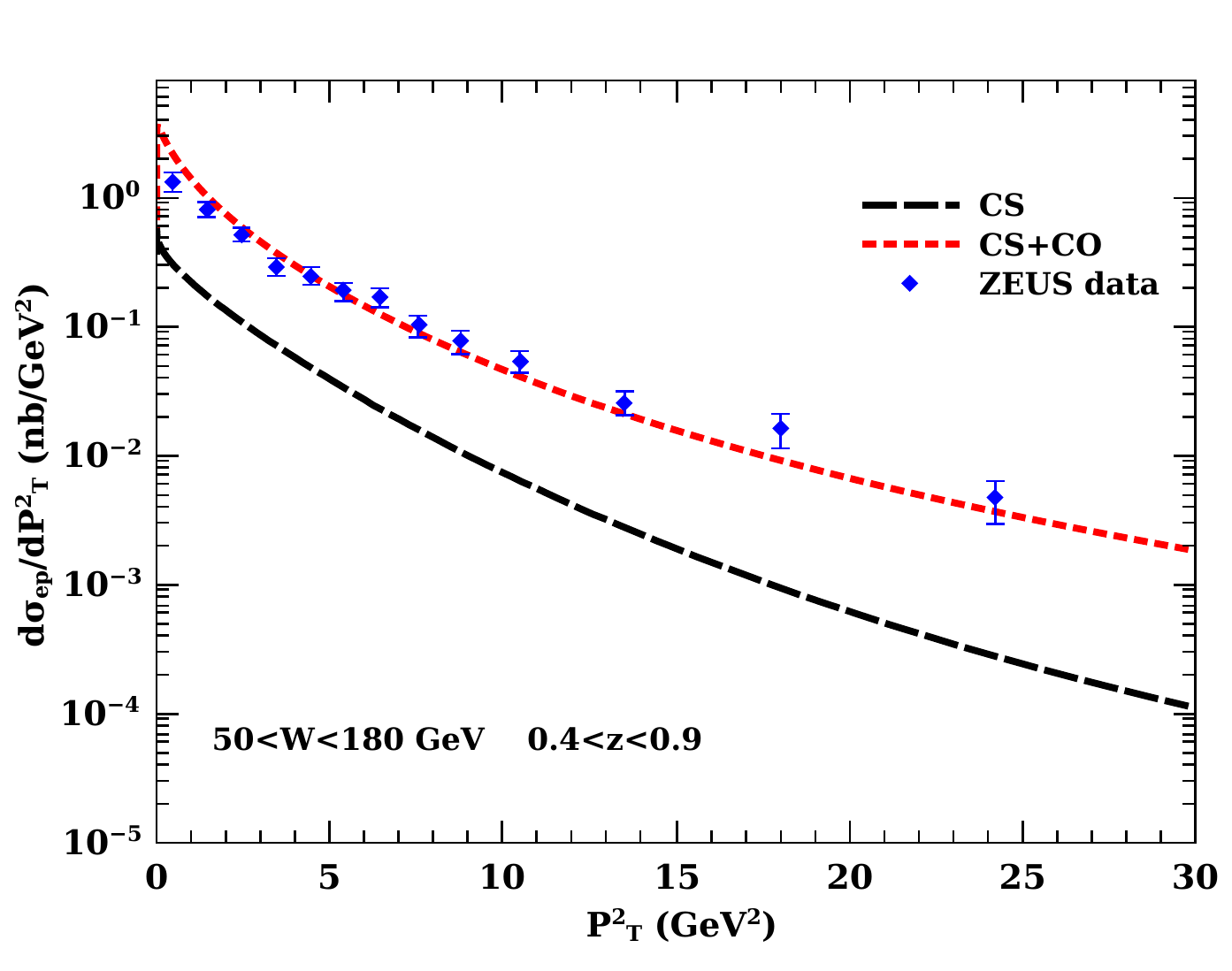}
\hspace{0.1cm}
\small{(b)}\includegraphics[width=8cm,height=6.5cm,clip]{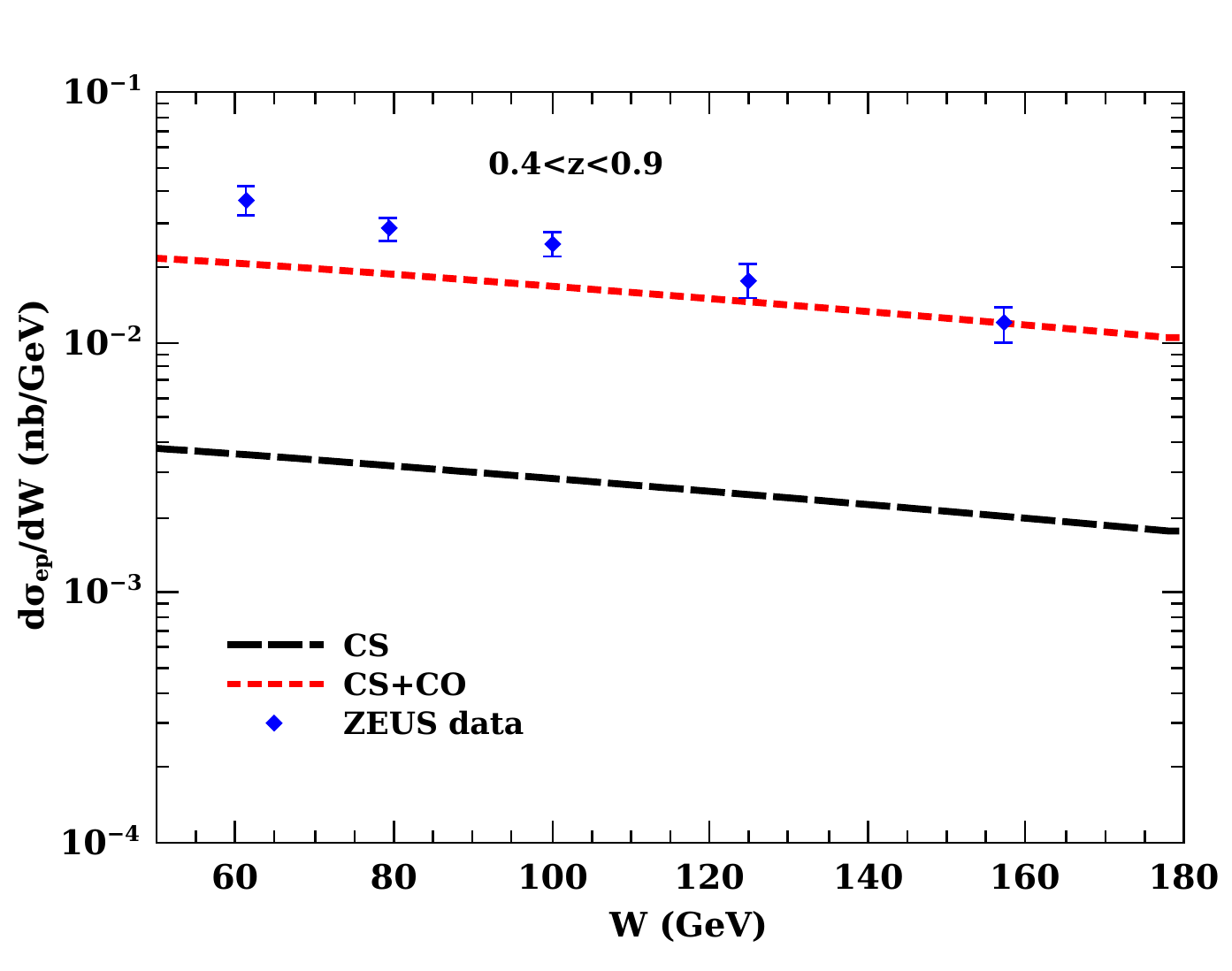}
\end{minipage}
\begin{center}
 \small{(c)}\includegraphics[width=8cm,height=6.5cm,clip]{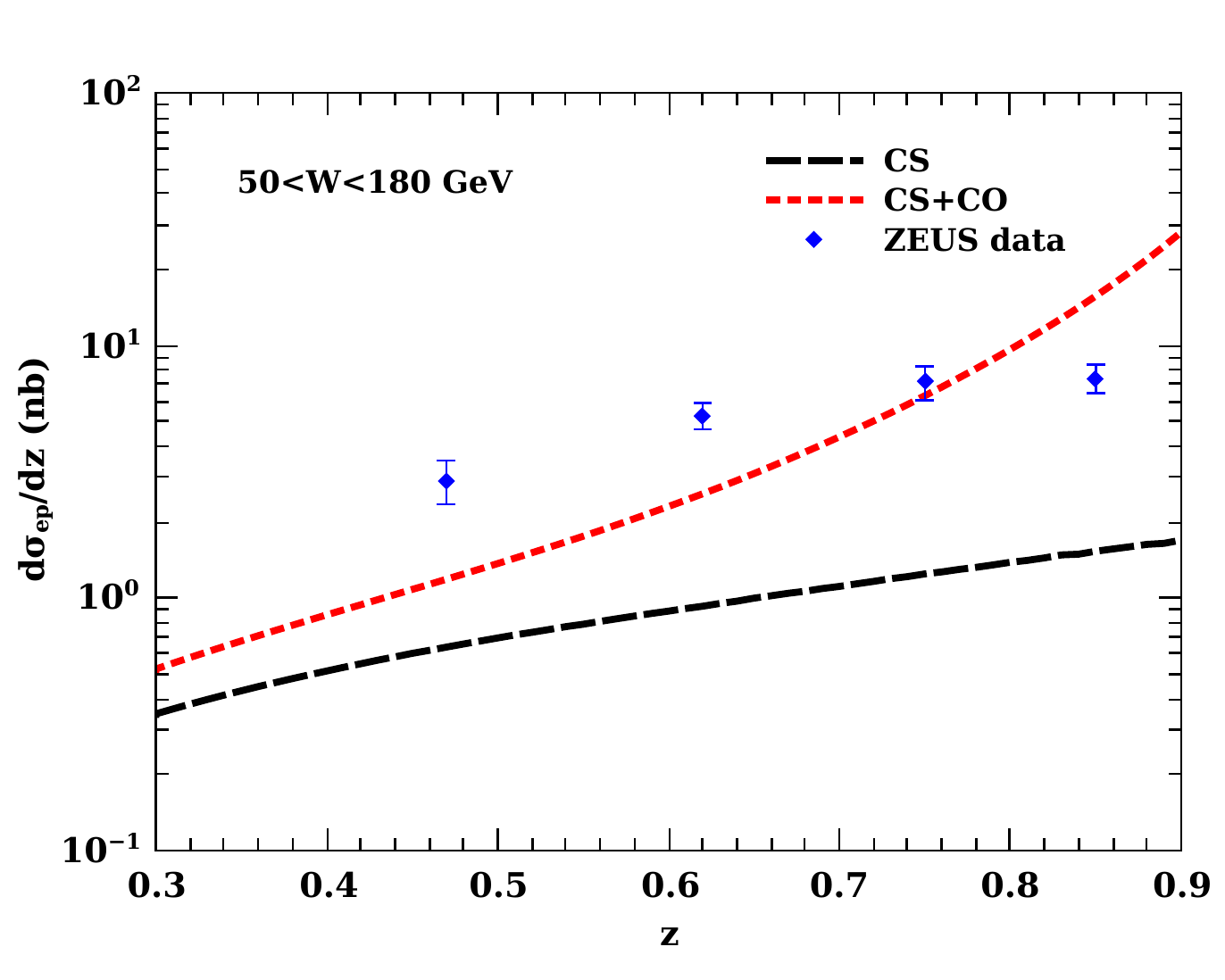}
\end{center}
\caption{\label{fig9} Unpolarized differential cross section in $e+p \rightarrow  J/\psi +X$
process as function of (a) $P_T$ (left panel),  (b) $W$ (right panel) and (c) $z$ (lower pannel) at 
HERA ($\sqrt{s}=300$ GeV) with $\langle k^2_{\perp g}\rangle=1$ GeV$^2$. The ZEUS data  from 
\cite{Chekanov:2002at} and LDMEs are from \cite{Zhang:2014ybe}. The integration ranges are 
$1<P_{T}\leq 5$ GeV, $50<W<180$ GeV and $0.4<z<0.9$. The convention of lines is same as 
\figurename{\ref{fig7}}.}
\end{figure}
\begin{figure}[H]
\begin{minipage}[c]{0.99\textwidth}
\small{(a)}\includegraphics[width=8cm,height=6.5cm,clip]{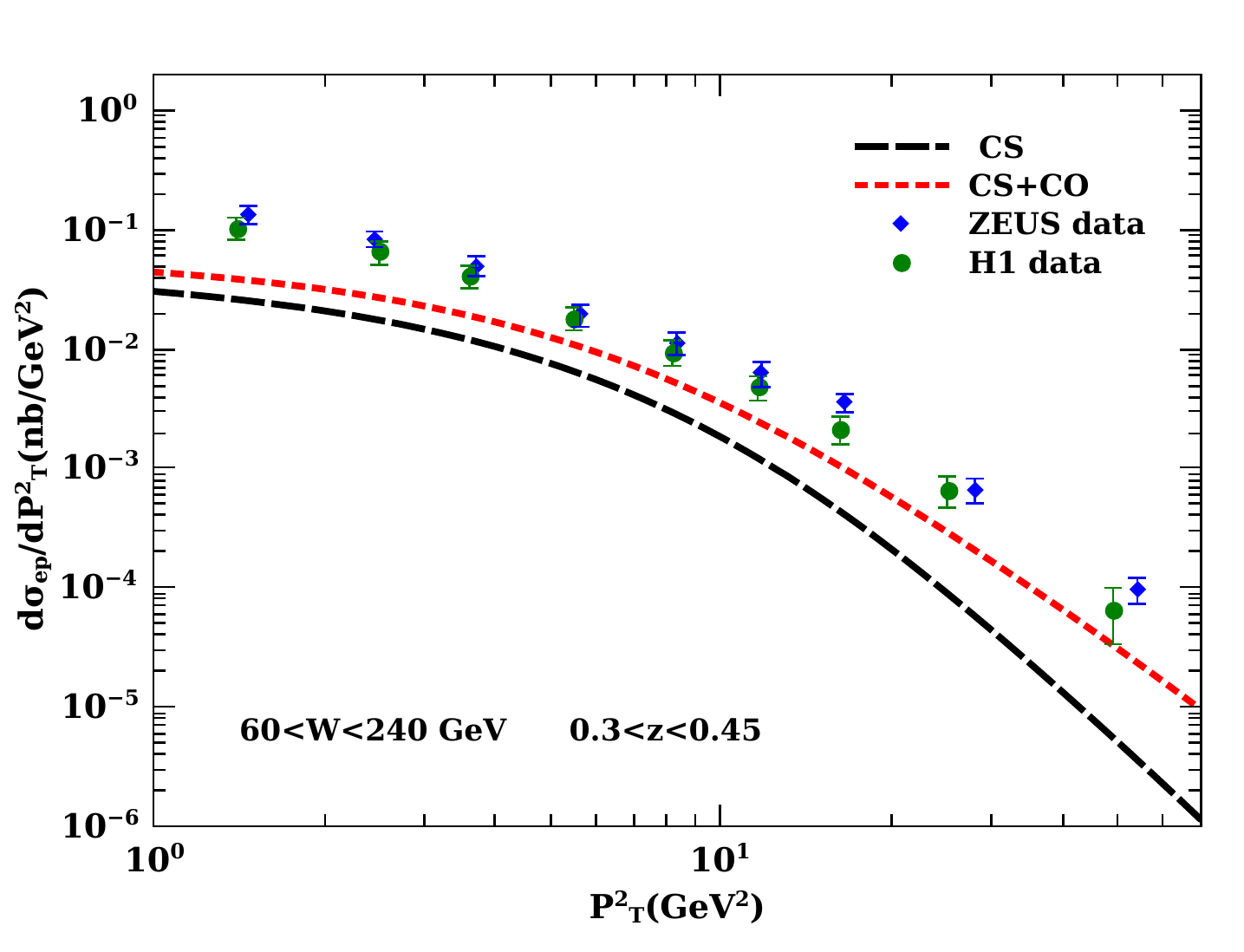}
\hspace{0.1cm}
\small{(b)}\includegraphics[width=8cm,height=6.5cm,clip]{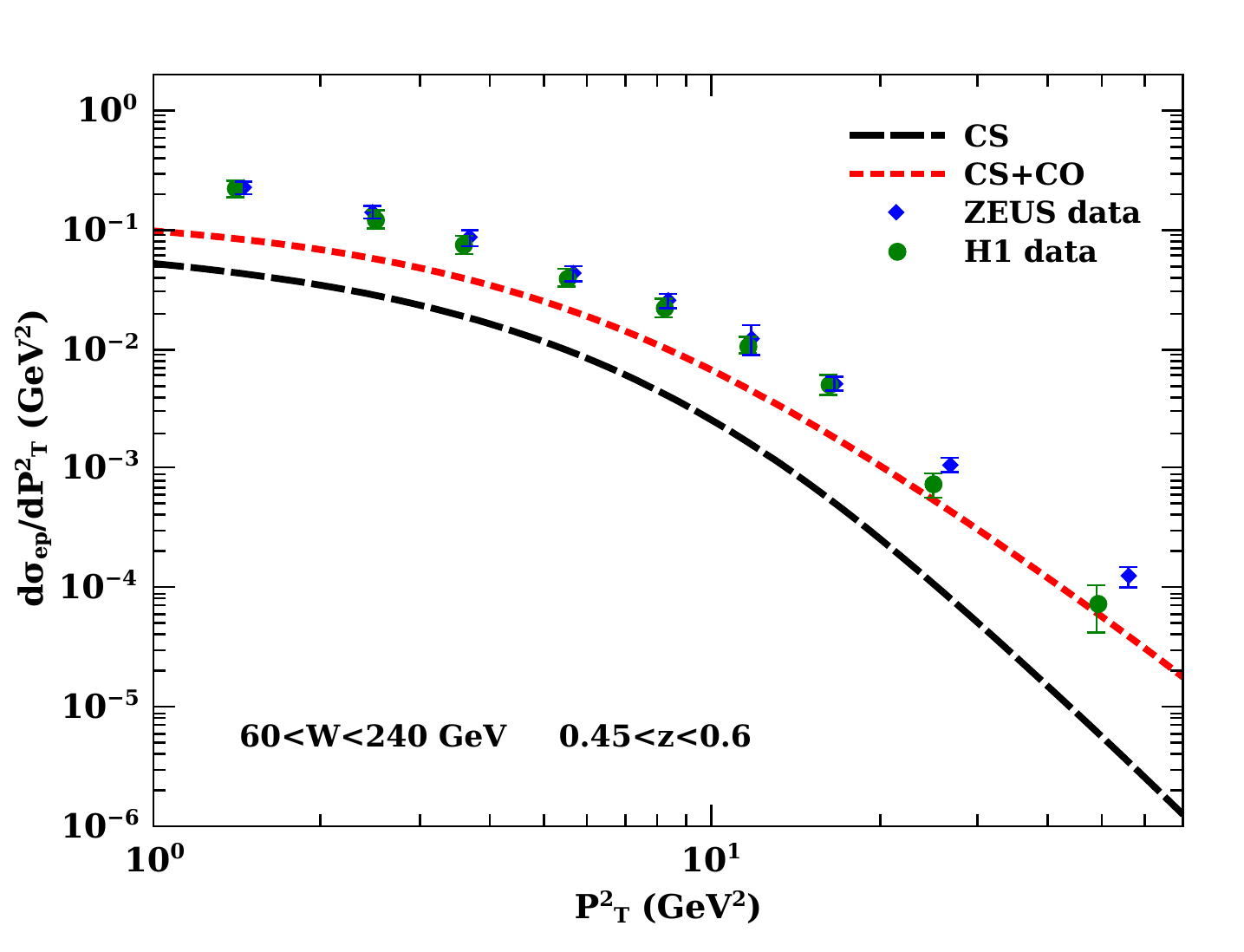}
\end{minipage}
\begin{minipage}[c]{0.99\textwidth}
\small{(c)}\includegraphics[width=8cm,height=6.5cm,clip]{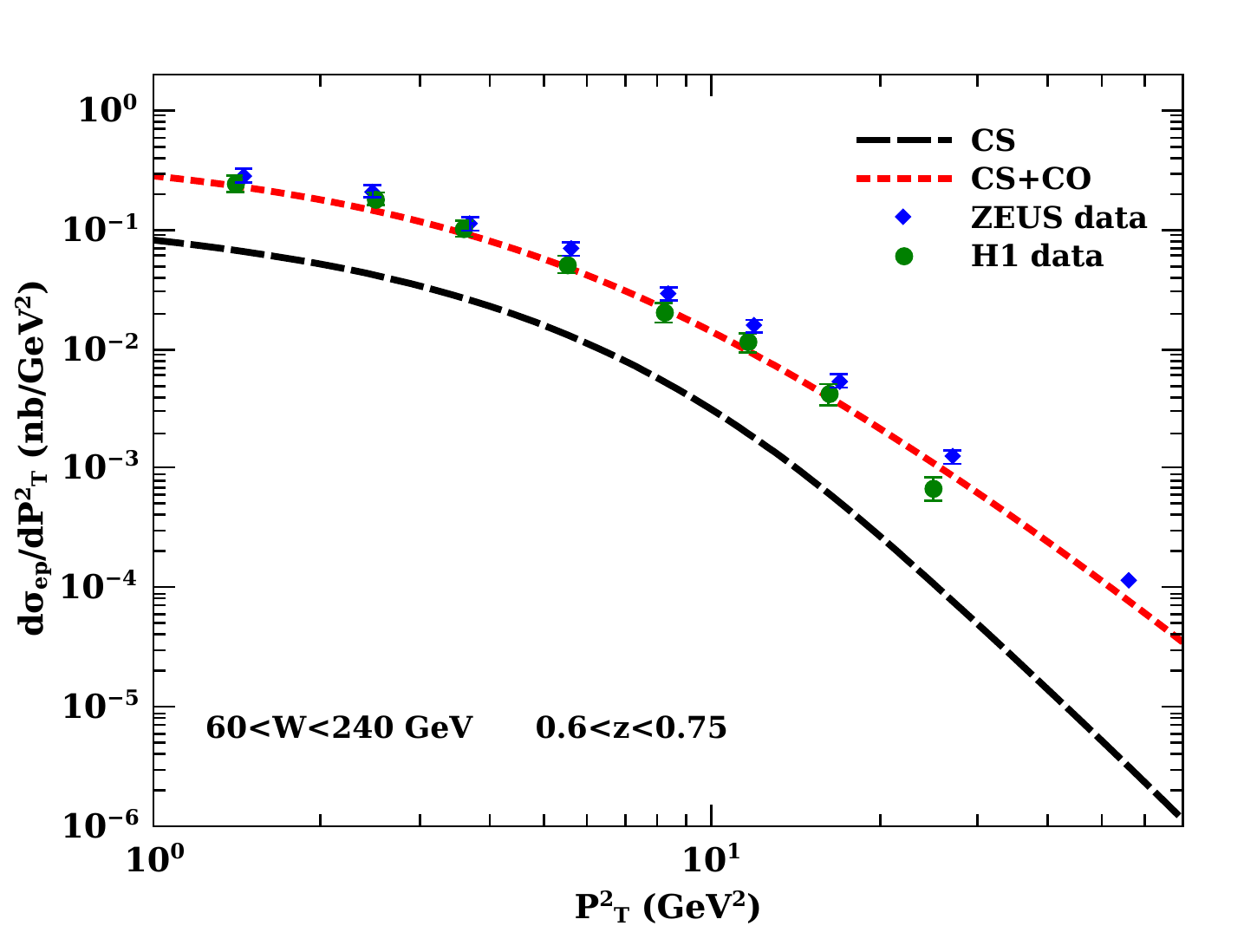}
\hspace{0.1cm}
\small{(d)}\includegraphics[width=8cm,height=6.5cm,clip]{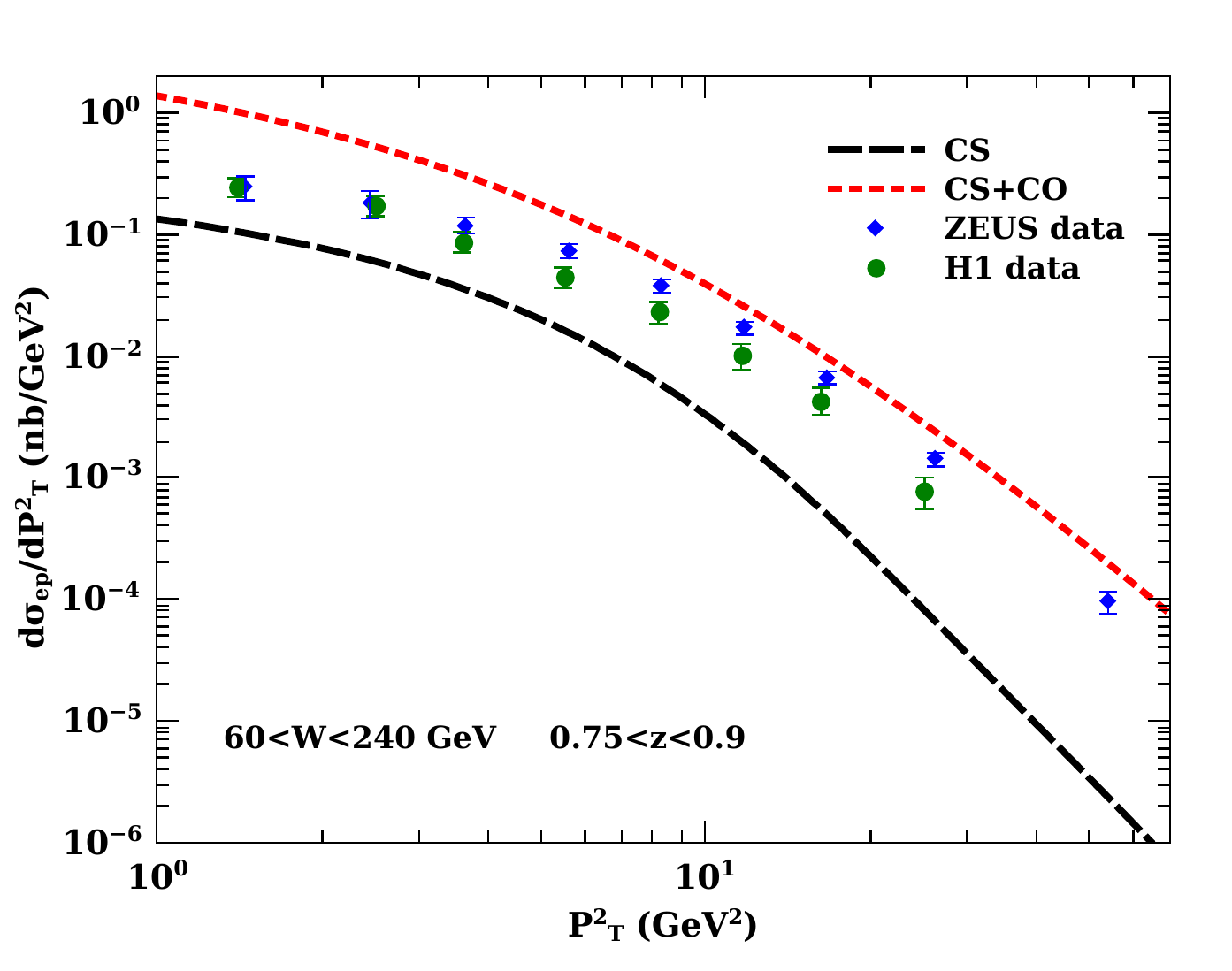}
\end{minipage}

\caption{\label{fig10} Unpolarized differential cross section in $e+p  \rightarrow J/\psi +X$
process as function of $P_T$ for each $z$ bin (a) $0.3<z<0.45$,  (b) $0.45<z<0.6$,  (c) $0.6<z<0.75$ and (d) 
$0.75<z<0.9$ at HERA ($\sqrt{s}=318$ GeV) with $\langle k^2_{\perp g}\rangle=1$ GeV$^2$. The  H1, ZEUS 
data  from \cite{Aaron:2010gz,Abramowicz:2012dh} and LDMEs are from \cite{Zhang:2014ybe}. The integration 
range 
of $W$ is $60<W<240$ GeV.  The convention of lines is same as
\figurename{\ref{fig7}}. }
\end{figure}

%------------------------------------------------
\begin{table}[h!]
  \centering
   \caption{$\chi^2$/d.o.f   for the LDMEs of 
Ref.\cite{Chao:2012iv,Butenschoen:2011yh,Zhang:2014ybe}.}
    \begin{tabular}{cccc}
    \toprule
  \hline
  \hline
  data
&LDMEs of \cite{Chao:2012iv}$~~~$
&LDMEs of \cite{Butenschoen:2011yh}$~~~$
&LDMEs of \cite{Zhang:2014ybe}  \\
 \hline
\midrule
 H1 data \cite{Aaron:2010gz} & 62.129 &3.83 & 7.92 \\
  ZEUS data \cite{Chekanov:2002at} &12.56 & 9.12 & 2.541\\
  \hline 
  \hline
\bottomrule
  \end{tabular}
   \label{table3}
\end{table}
%------------------------------------------------------------------------------------
The H1 data are compared with the theoretical results obtained by using the LDMEs of Ref. 
\cite{Chao:2012iv} and  \cite{Butenschoen:2011yh}, which are presented in \figurename{\ref{fig8}}.
The LDMEs of \cite{Chao:2012iv} over estimate the result as shown in the left panel of 
\figurename{\ref{fig8}}. Whereas Ref. \cite{Butenschoen:2011yh} LDMEs predict the result very close to the 
data, which is illustrated in the right 
panel 
of \figurename{\ref{fig8}}. The same behavior is also noticed for $z$ and $W$ spectra which are not shown.
To assess the agreement between the data and theoretical results, 
$\chi^2$/d.o.f is calculated for three sets of LDMEs from the $P_T$ spectrum of \figurename{\ref{fig7}},
\figurename{\ref{fig8}} and \figurename{\ref{fig9}} at a fixed  $\langle k^2_{\perp g}\rangle=1$ GeV$^2$, 
which is tabulated in \tablename{~\ref{table3}}. The $\chi^2$/d.o.f for $\langle k^2_{\perp g}\rangle=1$ 
GeV$^2$ is observed to be smaller than that of $\langle k^2_{\perp g}\rangle=0.5$ GeV$^2$ and $\langle 
k^2_{\perp 
g}\rangle=0.25$ GeV$^2$ for three sets of LDMEs. Therefore, we have considered the unpolarized TMD Gaussian 
width to be $\langle k^2_{\perp g}\rangle=1$ GeV$^2$ in the analysis of $J/\psi$ photoproduction.
Since the $\chi^2$/d.o.f for LDMEs of \cite{Zhang:2014ybe} is 7.92 and 2.541 for H1 and ZEUS data 
respectively, only the 
LDMEs of Ref.\cite{Zhang:2014ybe} have been used in the \figurename{\ref{fig9}} and \figurename{\ref{fig10}}.
%-------------------------------------------------------------------------------------
 The ZEUS data \cite{Chekanov:2002at} are compared with theoretical results within the kinematical region 
$\sqrt{s}=300$ GeV, $50<W<180$ GeV, $0.4<z<0.9$ and $Q^2_{max}=1$ GeV$^2$, and is shown in
\figurename{\ref{fig9}}. The $W$ and $z$ spectra are obtained by integrating the $P_T$ over the range 
$1<P_T<5$ GeV. In \figurename{\ref{fig10}}, the $P_T$ spectrum for each $z$ bin is compared with H1 
\cite{Aaron:2010gz} and ZEUS \cite{Abramowicz:2012dh} data. The $P_T$ spectrum is away from the data in the 
$0.3<z<0.5$, $0.45<z<0.6$ and $0.75<z<0.9$ bins. However, the theoretical result is in good agreement with 
the 
data for the bin $0.6<z<0.75$.

%======================================================================================================
%------------------------------------------------------------------------------------------------------------

%========================================================================================================
%=======================================================================================================
%=======================================================================================================
%=======================================================================================================
%=======================================================================================================
%-----------------------------------------------------------------------------------------------------------------------------
\section{Conclusion}\label{sec5}
%==================================
We have calculated the single-spin asymmetry and unpolarized differential cross section  in the inelastic 
photoproduction of $J/\psi$ in polarized and unpolarized  $ep$ collision respectively, 
where the scattered electron with small angle produces low virtuality photons. The NLO subprocess for $J/\psi$ production
is the photon-gluon fusion process $\gamma+g\rightarrow J/\psi+g$. Within the NRQCD based COM framework, the 
color octet states  $\leftidx{^{3}}{S}{_1}^{(8)}$, $\leftidx{^{1}}{S}{_0}^{(8)}$ and 
$\leftidx{^{3}}{P}{_{J(0,1,2)}}^{(8)}$ contribution to $J/\psi$ production is calculated. Sizable asymmetry
is obtained as a function of $P_T$ and $z$ in the kinematical range $0<P_T\leq 1$ GeV and $0.3< z\leq 0.9$ 
respectively. The infrared singularity at $z=1$, arises when the final gluon becomes soft, is excluded by restricting 
the analysis in the region $z\leq 0.9$. The resolved photoproduction contribution is removed by considering $z> 0.3$.
We also presented the unpolarized differential cross section of inelastic $J/\psi$ photoproduction as a function of 
$P_T$, $z$ and $W$, and is found to be in good agreement with the H1 and ZEUS data.
The sizable  asymmetry indicates that the inelastic photoproduction of $J/\psi$
in $ep^\uparrow$ collision is a useful process to probe the gluon Sivers function over a wide kinematical region accessible to the future electron-ion collider (EIC).

%=================================
\section*{Acknowledgment}
We would like to thank Jean-Philippe Lansberg for fruitful discussion during his stay at IIT Bombay.  
%--------------------------------------------------------------------------
\appendix
%\section*{Appendices}
% \begin{center}
 \section{S\lowercase{quare of the amplitude for} $\gamma + g\rightarrow J/\psi +g$ 
\lowercase{process}}\label{ap1}
%\end{center}
The summation over only the transverse polarizations of the initial and final on-shell gluons and photon is 
achieved
by invoking  \cite{Baier:1983va}
\be
\sum_{\lambda_a=1}^2\varepsilon^{\lambda_a}_\mu(k)\varepsilon^{\ast\lambda_a}_{\mu^\prime}(k)=-g_{
\mu\mu^\prime}+
\frac{k_\mu n_{\mu^\prime}+k_{\mu^\prime}n_\mu}{k.n}-\frac{k_\mu k_{\mu^\prime}}{(k.n)^2}
\ee
with $n^\mu=\frac{P^\mu_h}{M}$.
We define the following variables for computation purpose
\be
s_1=\hat{s}-M^2,~~t_1=\hat{t}-M^2,~~u_1=\hat{u}-M^2.
\ee
 FORM package \cite{Kuipers:2012rf} is used to obtain the square of the amplitude. The CS, 
 $\leftidx{^{3}}{S}{_1}^{(1)}$ state amplitude calculation is similar to CO, $\leftidx{^{3}}{S}{_1}^{(8a)}$, 
except a change in the color factor.
The amplitude square of  $\leftidx{^{3}}{S}{_1}^{(1,8)}$, 
$\leftidx{^{1}}{S}{_0}^{(8)}$, $\leftidx{^{3}}{P}{_0}^{(8)}$, $\leftidx{^{3}}{P}{_1}^{(8)}$ and 
$\leftidx{^{3}}{P}{_2}^{(8)}$
states is given below
\begin{eqnarray}\label{13s1}
\begin{aligned}
|\mathcal{M}[\leftidx{^{3}}{S}{_1}^{(1)}]|^2=&{}\frac{2\pi^3e_c^2\alpha^2_s\alpha}{27M}\langle 0\mid 
\mathcal{O}_1^{J/\psi}(\leftidx{^3}{S}{_1})\mid 0\rangle
\frac{512 M^2} {s_1^2 t_1^2 u_1^2}\\
&\times\big\{s_1^2(s_1+M^2)^2+u_1^2(u_1+M^2)^2+t_1^2(t_1+M^2)^2\big\}
\end{aligned}
\end{eqnarray}
\begin{eqnarray}\label{83s1}
\begin{aligned}
|\mathcal{M}[\leftidx{^{3}}{S}{_1}^{(8)}]|^2=&{}\frac{5\pi^3e_c^2\alpha^2_s\alpha}{36M}\langle 0\mid 
\mathcal{O}_8^{J/\psi}(\leftidx{^3}{S}{_1})\mid 0\rangle
\frac{512 M^2} {s_1^2 t_1^2 u_1^2}\\
&\times\big\{s_1^2(s_1+M^2)^2+u_1^2(u_1+M^2)^2+t_1^2(t_1+M^2)^2\big\}
\end{aligned}
\end{eqnarray}
\begin{equation}\label{1s0}
\begin{split}
|\mathcal{M}[\leftidx{^{1}}{S}{_0}^{(8)}]|^2=&\frac{3\pi^3e_c^2\alpha^2_s\alpha}{4M}\langle 0\mid 
\mathcal{O}_8^{J/\psi}(\leftidx{^1}{S}{_0})\mid 0\rangle 
\frac{128}{s_1^2 t_1^2 u_1^2 \left(M^2+u_1\right)^2}
\Big\{8 M^{14}+4 M^{12}\\
&\times(4(s_1+t_1)
+7 u_1)+2 M^{10} \left(8
   s_1^2+17 u_1 (s_1+t_1)+12 s_1 t_1+8 t_1^2+19
   u_1^2\right)\\
 &  +2 M^8 \big(7 s_1^3
   +4 u_1 \big(s_1^2+5 s_1
   t_1   +t_1^2\big)+5 s_1^2 t_1+5 s_1 t_1^2+6 u_1^2
   (s_1+t_1)+7 t_1^3\\
   &+13 u_1^3\big) +2 M^6 \big(2 s_1^4+4
   u_1 \left(s_1^3+t_1^3\right)+s_1^3 t_1
   +u_1^2 \big(-17
   s_1^2+7 s_1 t_1-17 t_1^2\big)\\
   &+2 s_1^2 t_1^2+s_1t_1^3-8 u_1^3 (s_1+t_1)+2 t_1^4+5 u_1^4\big)+2 M^4
   \big(6 u_1^3 \big(-3 s_1^2
   +s_1 t_1-3 t_1^2\big)\\
   &-6   u_1^2 (s_1+t_1)\left(s_1^2+t_1^2\right)+u_1
   (s_1-t_1)^2 \left(s_1^2+4 s_1 t_1+t_1^2\right)-6
   u_1^4 (s_1+t_1)\\
&-3 s_1 t_1
   (s_1-t_1)^2
   (s_1+t_1)+u_1^5\big)+M^2 \big(-2 u_1^4 \left(5 s_1^2-11
   s_1 t_1+5 t_1^2\right)+u_1^2\\
   &\times\big(-2 s_1^4+s_1^3
   t_1-5 s_1^2 t_1^2+s_1 t_1^3-2 t_1^4\big)-2 u_1^5
   (s_1+t_1)-6 u_1^3 (s_1-t_1)^2 (s_1+t_1)\\
   &+s_1 t_1 u_1 (s_1-t_1)^2 (s_1+t_1)-s_1 t_1
   (s_1-t_1)^2
   (2 s_1+t_1) (s_1+2 t_1)\big)+s_1
   t_1 u_1 \\
   &\big(3 u_1^2 \left(s_1^2+s_1
   t_1+t_1^2\right)
   +(s_1-t_1)^2 \left(s_1^2+s_1
   t_1+t_1^2\right)+8 u_1^3
   (s_1+t_1)+8
   u_1^4\big)\Big\}
\end{split}
\end{equation}
   \begin{equation}\label{3p0}
\begin{split}
    |\mathcal{M}[\leftidx{^{3}}{P}{_0}^{(8)}]|^2=&\frac{\pi^3e_c^2\alpha^2_s\alpha}{4M}\langle 0\mid 
\mathcal{O}_8^{J/\psi}(\leftidx{^3}{P}{_0})\mid 0\rangle
\frac{512}{M^2 s_1^4 t_1^4
   u_1^4 \left(M^2+u_1\right)^2}\big\{ 32 s_1 t_1 u_1 M^{20}\\
   &+16 \left(-5 s_1^2   t_1^2-\left(s_1^2-8 t_1 s_1+t_1^2\right) 
    u_1^2\right)
   M^{18}+16 u_1 \big(s_1 t_1 \left(2 s_1^2-13 t_1 s_1+2
   t_1^2\right)\\
   &-2 \left(s_1^2-6 t_1 s_1+t_1^2\right)
   u_1^2\big) M^{16}+8 \big(2 \left(s_1^2+8 t_1
   s_1+t_1^2\right)   u_1^4+s_1 t_1 \big(2 s_1^2-17
   t_1 s_1\\
   &+2 t_1^2\big) u_1^2
   -2 s_1^2 t_1^2 \left(3
   s_1^2-7 t_1 s_1+3 t_1^2\right)\big) M^{14}+8 u_1 \big(2
   \left(5 s_1^2+2 t_1 s_1+5 t_1^2\right) u_1^4\\
   &-\left(2   s_1^4+15 t_1 s_1^3-19 t_1^2 s_1^2+15 t_1^3
   s_1+2   t_1^4\right) u_1^2-s_1 t_1 \big(s_1^4+8 t_1
   s_1^3   -22 t_1^2 s_1^2\\
  & +8 t_1^3 s_1+t_1^4\big)\big)
   M^{12}+4 \big(16 \left(s_1^2+t_1^2\right) u_1^6-\left(3 s_1^2+16
   t_1 s_1+3 t_1^2\right) \big(4 s_1^2-7 t_1 s_1\\
   &+4   t_1^2\big) u_1^4+s_1 t_1 \left(3 s_1^4+4 t_1
   s_1^3-2 t_1^2 s_1^2+4 t_1^3 s_1+3 t_1^4\right)
   u_1^2+2 s_1^2 t_1^2 \big(2 s_1^4\\
   &+7 t_1 s_1^3-10
   t_1^2 s_1^2+7 t_1^3 s_1+2 t_1^4\big)\big) M^{10}+4
   u_1 \big(4 \left(s_1^2+t_1^2\right) u_1^6-\big(12 s_1^4\\
   &+19   t_1 s_1^3-73 t_1^2 s_1^2+19 t_1^3 s_1+12
   t_1^4\big) u_1^4+s_1 t_1 \big(11 s_1^4+6 t_1
   s_1^3-26 t_1^2 s_1^2\\
  & +6 t_1^3 s_1+11 t_1^4\big)
   u_1^2-s_1^2 t_1^2 (s_1+t_1)^4\big) M^8+2 \big(-2 \big(4
   s_1^4+t_1 s_1^3-32 t_1^2 s_1^2\\
   &+t_1^3 s_1+4
   t_1^4\big) u_1^6+s_1 t_1 \left(10 s_1^4-19 t_1
   s_1^3-15 t_1^2 s_1^2-19 t_1^3 s_1+10 t_1^4\right)
   u_1^4\\
   &-s_1^2 t_1^2 \left(11 s_1^4+3 t_1 s_1^3-22
   t_1^2 s_1^2+3 t_1^3 s_1+11 t_1^4\right) u_1^2-2
   s_1^3 (s_1-t_1)^2 t_1^3 \\
   &\times(2 s_1+t_1) (s_1+2
   t_1)\big) M^6+2 s_1 t_1 u_1 \big(2 \left(s_1^2+7 t_1
   s_1+t_1^2\right) u_1^6-\big(2 s_1^4   +11 t_1 s_1^3\\
   &-12   t_1^2 s_1^2+11 t_1^3 s_1+2 t_1^4\big) u_1^4+s_1
   t_1 \left(s_1^4+10 t_1 s_1^3+10 t_1^2 s_1^2+10
   t_1^3 s_1+t_1^4\right) u_1^2\\
   &+6 s_1^2 t_1^2
   \left(s_1^2-t_1^2\right)^2\big) M^4+s_1^2 t_1^2 u_1^2
   \big(2 \left(4 s_1^2+11 t_1 s_1+4 t_1^2\right) u_1^4+\big(4
   s_1^4-16 t_1 s_1^3\\
   &-19 t_1^2 s_1^2-16 t_1^3 s_1+4
   t_1^4\big) u_1^2-3 s_1 (s_1-t_1)^2 t_1 \left(2
   s_1^2+3 t_1 s_1+2 t_1^2\right)\big) M^2+s_1^3 t_1^3\\
   &\times   \left(s_1^2+t_1 s_1+t_1^2\right) u_1^3
   \left((s_1-t_1)^2+3 u_1^2\right)\Big\}
   \end{split}
\end{equation}

  \begin{equation}\label{3p1}
\begin{split}
 |\mathcal{M}[\leftidx{^{3}}{P}{_1}^{(8)}]|^2=&\frac{\pi^3e_c^2\alpha^2_s\alpha}{8M}\langle 0\mid 
\mathcal{O}_8^{J/\psi}(\leftidx{^3}{P}{_1})\mid 0\rangle
\frac{2048}{m^2 s_1^4 t_1^4 u_1^4\left(m^2+u_1\right)^2}
\Big\{8 s_1 t_1 u_1 m^{20}+4 \big(5 s_1^2   t_1^2\\
&   +(s_1+t_1)^2 u_1^2\big) m^{18}+4 u_1 \left(2
   \left(s_1^2-4 t_1 s_1+t_1^2\right) u_1^2+s_1 t_1
   \left(5 s_1^2+8 t_1 s_1+5 t_1^2\right)\right)\\
&\times   m^{16}+2 \big(-2   \left(s_1^2+16 t_1 s_1+t_1^2\right) u_1^4+s_1 t_1
   \left(12 s_1^2+23 t_1 s_1+12 t_1^2\right) u_1^2+2 s_1^2t_1^2\\
&  \times \left(3 s_1^2-7 t_1 s_1+3 t_1^2\right)\big) m^{14}+2
   u_1 \big(-10 (s_1+t_1)^2 u_1^4+\big(2 s_1^4-16 t_1
   s_1^3+71 t_1^2 s_1^2\\
&   -16 t_1^3 s_1+2 t_1^4\big)
   u_1^2+s_1 t_1 \left(-2 s_1^4+3 t_1 s_1^3-22 t_1^2
   s_1^2+3 t_1^3 s_1-2 t_1^4\right)\big) m^{12}\\
&  +\big(-\left(16
   s_1^2+7 t_1 s_1+16 t_1^2\right) u_1^6+2 \left(6 s_1^4-31
   t_1 s_1^3+109 t_1^2 s_1^2-31 t_1^3 s_1+6
   t_1^4\right) u_1^4\\
&-s_1 t_1 \left(3 s_1^4+16 t_1
   s_1^3+20 t_1^2 s_1^2+16 t_1^3 s_1+3 t_1^4\right)
   u_1^2-2 s_1^2 t_1^2 \big(2 s_1^4+7 t_1 s_1^3-10
   t_1^2 s_1^2\\
&+7 t_1^3 s_1+2 t_1^4\big)\big) m^{10}+u_1
   \big(\left(-4 s_1^2+3 t_1 s_1-4 t_1^2\right) u_1^6+\big(12
   s_1^4-28 t_1 s_1^3+159 t_1^2 s_1^2\\
&-28 t_1^3 s_1+12 t_1^4\big) u_1^4+s_1 t_1 \left(5 s_1^4-16 t_1
   s_1^3-45 t_1^2 s_1^2-16 t_1^3 s_1+5 t_1^4\right)
   u_1^2-s_1^2 t_1^2\\
& \times\left(3 s_1^4+5 t_1 s_1^3-12
   t_1^2 s_1^2+5 t_1^3 s_1+3 t_1^4\right)\big) m^8+\big(3
   s_1 t_1 u_1^8+\left(4 s_1^4+55 t_1^2 s_1^2+4
   t_1^4\right)\\
&\times   u_1^6 +s_1 t_1 \left(3 s_1^4-16 t_1
   s_1^3-87 t_1^2 s_1^2-16 t_1^3 s_1+3 t_1^4\right)
   u_1^4+s_1 t_1 \big(2 s_1^6-t_1 s_1^5+21 t_1^2s_1^4\\
&   -15 t_1^3 s_1^3+21 t_1^4 s_1^2-t_1^5 s_1+2
   t_1^6\big) u_1^2+s_1^3 (s_1-t_1)^2 t_1^3 (2
   s_1+t_1) (s_1+2 t_1)\big) m^6\\
& +s_1 t_1 u_1\big(u_1^8+\left(2 s_1^2+3 t_1 s_1+2 t_1^2\right)
   u_1^6-\left(s_1^4+12 t_1 s_1^3+59 t_1^2 s_1^2+12
   t_1^3 s_1+t_1^4\right) u_1^4\\
& +\left(2 s_1^6-7 t_1
   s_1^5+24 t_1^2 s_1^4-7 t_1^3 s_1^3+24 t_1^4
   s_1^2-7 t_1^5 s_1+2 t_1^6\right) u_1^2-s_1^2
   (s_1-t_1)^2\\
&   \times t_1^2 \left(s_1^2+t_1
   s_1+t_1^2\right)\big) m^4-s_1^2 t_1^2 u_1^2 \big(3
   u_1^6+\left(2 s_1^2+13 t_1 s_1+2 t_1^2\right)   u_1^4\\
&+\left(5 s_1^4-13 t_1 s_1^3-7 t_1^2 s_1^2-13
   t_1^3 s_1+5 t_1^4\right) u_1^2+s_1^2 (s_1-t_1)^2
   t_1^2\big) m^2\\
   &+s_1^3 t_1^3 \left(s_1^2+t_1
   s_1+t_1^2\right) u_1^3 \left((s_1-t_1)^2+3
   u_1^2\right)\Big\}
   \end{split}
\end{equation}
 \begin{equation}\label{3p2}
\begin{split}
 |\mathcal{M}[\leftidx{^{3}}{P}{_2}^{(8)}]|^2=&\frac{3\pi^3e_c^2\alpha^2_s\alpha}{20M}\langle 0\mid 
\mathcal{O}_8^{J/\psi}(\leftidx{^3}{P}{_2})\mid 0\rangle
\frac{1024} {3 M^2 s_1^4 t_1^4
   u_1^4 \left(M^2+u_1\right)^2} \Big\{104 s_1 t_1 u_1 M^{20}+4\\
&\times\left(-5 s_1^2   t_1^2-\left(s_1^2-86 t_1 s_1+t_1^2\right)
u_1^2\right)M^{18}+4 u_1 \big(\left(-2 s_1^2+99 t_1 s_1-2 t_1^2\right)
   u_1^2\\
  & +2 s_1 t_1 \left(13 s_1^2-23 t_1 s_1+13
   t_1^2\right)\big) M^{16}+2 \big(2 \left(s_1^2+47 t_1
   s_1+t_1^2\right)u_1^4+s_1 t_1\\
   &\times\left(122 s_1^2-107   t_1 s_1+122 t_1^2\right) u_1^2-2 s_1^2 t_1^2 \left(3
   s_1^2-7 t_1 s_1+3 t_1^2\right)\big) M^{14}+2 u_1 \\
   &\times\big(10\left(s_1^2+4 t_1 s_1+t_1^2\right) u_1^4+\left(-2
   s_1^4+63 t_1 s_1^3+133 t_1^2 s_1^2+63 t_1^3 s_1-2
   t_1^4\right) u_1^2\\
   &+s_1 t_1 \left(23 s_1^4-77 t_1
   s_1^3+52 t_1^2 s_1^2-77 t_1^3 s_1+23 t_1^4\right)\big)
   M^{12}+\big(\big(16 s_1^2+99 t_1 s_1\\
   &+16 t_1^2\big) u_1^6-2\left(6 s_1^4+59 t_1 s_1^3-305 t_1^2 s_1^2+59 t_1^3
   s_1+6 t_1^4\right) u_1^4+s_1 t_1 \big(171 s_1^4\\
   &-476   t_1 s_1^3+220 t_1^2 s_1^2-476 t_1^3 s_1+171
   t_1^4\big) u_1^2+2 s_1^2 t_1^2 \big(2 s_1^4+7 t_1
   s_1^3-10 t_1^2 s_1^2\\
   &+7 t_1^3 s_1+2 t_1^4\big)\big)
   M^{10}+u_1 \big(\left(4 s_1^2+69 t_1 s_1+4 t_1^2\right)
   u_1^6-\big(12 s_1^4+154 t_1 s_1^3\\
   &-355 t_1^2 s_1^2+154
   t_1^3 s_1+12 t_1^4\big) u_1^4+s_1 t_1 \big(227
   s_1^4-612 t_1 s_1^3+295 t_1^2 s_1^2\\
   &-612 t_1^3
   s_1+227 t_1^4\big) u_1^2-s_1 t_1 \big(18 s_1^6+19
   t_1 s_1^5+19 t_1^2 s_1^4-60 t_1^3 s_1^3+19 t_1^4
   s_1^2\\
   &+19 t_1^5 s_1+18 t_1^6\big)\big) M^8+\big(21 s_1
   t_1 u_1^8-\big(4 s_1^4+52 t_1 s_1^3-53 t_1^2
   s_1^2+52 t_1^3 s_1\\
   &+4 t_1^4\big) u_1^6+s_1 t_1   \big(125 s_1^4-374 t_1 s_1^3+219 t_1^2 s_1^2-374 t_1^3
   s_1+125 t_1^4\big) u_1^4+s_1 t_1 \\
   &\times\big(-30 s_1^6-13
   t_1 s_1^5+39 t_1^2 s_1^4+83 t_1^3 s_1^3+39 t_1^4
   s_1^2-13 t_1^5 s_1-30 t_1^6\big) u_1^2-s_1^3\\
   &\times(s_1-t_1)^2 t_1^3(2 s_1+t_1) (s_1+2 t_1)\big)
   M^6+s_1 t_1 u_1 \big(3 u_1^8-\left(2 s_1^2+17 t_1
   s_1+2 t_1^2\right) \\
   &\times u_1^6+\big(23 s_1^4-100 t_1 s_1^3+81
   t_1^2 s_1^2-100 t_1^3 s_1+23 t_1^4\big)
   u_1^4+\big(-12 s_1^6+17 t_1 s_1^5\\
   &+80 t_1^2 s_1^4+11   t_1^3 s_1^3+80 t_1^4 s_1^2+17 t_1^5 s_1-12
   t_1^6\big) u_1^2+3 s_1^2 (s_1-t_1)^2 t_1^2 \big(3
   s_1^2+7 t_1 s_1\\
   &+3 t_1^2\big)\big) M^4+s_1^2 t_1^2
   u_1^2 \big(-9 u_1^6+\left(-4 s_1^2+13 t_1 s_1-4
   t_1^2\right) u_1^4+\big(7 s_1^4+23 t_1 s_1^3\\
   &-13 t_1^2   s_1^2+23 t_1^3 s_1+7 t_1^4\big) u_1^2+3 s_1^2   (s_1-t_1)^2 t_1^2\big) M^2+s_1^3 t_1^3
   \left(s_1^2+t_1 s_1+t_1^2\right) u_1^3\\
&\times   \left((s_1-t_1)^2+3 u_1^2\right)\Big\}
  \end{split}
\end{equation}

 %==================================================================
 \section{K\lowercase{inematics}}\label{ap2}
 We consider the frame in which the proton and electron are moving along -$z$ and +$z$-axises respectively and 
their four momenta are given by
\be
P=\frac{\sqrt{s}}{2}(1,0,0,-1),~~l=\frac{\sqrt{s}}{2}(1,0,0,1).
\ee
The C.M energy of electron-proton system is $s=(P+l)^2$.
The above four momenta in light-cone coordinate system can be written as 
\be
P^\mu=\sqrt{\frac s2}n_-^\mu,~~ l^\mu=\sqrt{\frac s2}n_+^\mu,
\ee
where $n_+$ and $n_-$ are two light-like vectors with $n_+.n_-=1$ and $n_+^2=n^2_-=0$. 
\be
n_+^\mu=(1,0,{\bm 0}),~~~n_-^\mu=(0,1,{\bm 0}).
\ee
We assume that the quasi-real photon is collinear to the electron.
The quasi-real photon and gluon four momenta are given by
\be
q^\mu=x_\gamma\sqrt{\frac s2}n_+^\mu,
\ee
\be
k=\frac{k^2_{\perp g}}{2x_g\sqrt{\frac s2}}n_+^\mu  +x_g\sqrt{\frac s2}n_-^\mu + {\bm k}^\mu_\perp \approx x_g\sqrt{\frac s2}n_-^\mu + {\bm k}^\mu_\perp ,
\ee
where $x_\gamma=\frac{q^+}{l^+}$ and $x_g=\frac{k^-}{P^-}$ are the light-cone momentum fractions. 
 The four momentum of the $J/\psi$ is given by
\be
P_h^\mu=zx_\gamma\sqrt{\frac s2}n_+^\mu+\frac{M^2+P^2_{T}}{2zx_\gamma\sqrt{\frac s2}}n_-^\mu+{\bm P}_{T}^\mu.
\ee
The inelastic variable is defined as $z=\frac{P.P_h}{P.q}=\frac{P_h^+}{q^+}$.
By using the above relations, we can write down the expressions of  Mandelstam variables as below
\be\label{sh}
\hat{s}=(k+q)^2=2k.q=sx_gx_\gamma,
\ee
\be\label{th}
\hat{t}&=&(k-P_h)^2=M^2-2k.P_h\nonumber\\
&=&M^2-zsx_gx_\gamma+2k_{\perp g}P_{T}\cos(\phi-\phi_h),
\ee
\be\label{uh}
\hat{u}&=&(q-P_h)^2=M^2-2q.P_h\nonumber\\
&=&M^2-\frac{M^2+P^2_{T}}{z}.
\ee
Here M being the mass of  $J/\psi$. The $\phi$ and $\phi_h$ are the azimuthal angles of the gluon and 
$J/\psi$ transverse momentum vector respectively. $\phi_h=0$ for estimating the asymmetry since the 
production of $J/\psi$ is considered to be in the xz plane as shown in \figurename{\ref{fig1}}.
The delta function in Eq.\eqref{d1} can be used to find the solution 
of $x_g$.
From Eq.\eqref{sh}, \eqref{th} and \eqref{uh}, the  delta function can be written as follows 
\be\label{soft}
\delta(\hat{s}+\hat{t}+\hat{u}-M^2)&=&\delta\left(sx_gx_\gamma +M^2-zsx_gx_\gamma+2k_{\perp g}P_{T}\cos(\phi-\phi_h)+M^2-\frac{M^2+P^2_{T}}{z}
-M^2\right)\nonumber\\
&=&\delta\left(sx_gx_\gamma(1-z)+2k_{\perp g}P_{T}\cos(\phi-\phi_h)-\frac{M^2+P^2_{T}}{z} +M^2\right)\nonumber\\
&=&\frac{1}{sx_\gamma (1-z)}\delta\left(x_g-a_1\right),
\ee
where $a_1$ is defined as 
\be
a_1=\frac{M^2+P^2_{T}-zM^2-2zk_{\perp g}P_{T}\cos(\phi-\phi_h)}{sx_\gamma z(1-z)}.
\ee

The phase space integration of $J/\psi$ can be written as 
\be
\frac{d^3{\bm P}_h}{E_h}=\frac{1}{z}dzd^2{\bm P}_{T}.
\ee
In line with Ref.\cite{Berger:1980ni}, we impose the following kinematical cuts on Mandelstam variables
\be
M^2\leq \hat{s}\leq s,~~~0\geq\hat{t}\geq -(\hat{s}-M^2),~~~0\geq\hat{u}\geq -(\hat{s}-M^2).
\ee
%-------------------------------------------------------------------------------
\bibliographystyle{apsrev}
\bibliography{refer}

\begin{thebibliography}{69}
\expandafter\ifx\csname natexlab\endcsname\relax\def\natexlab#1{#1}\fi
\expandafter\ifx\csname bibnamefont\endcsname\relax
  \def\bibnamefont#1{#1}\fi
\expandafter\ifx\csname bibfnamefont\endcsname\relax
  \def\bibfnamefont#1{#1}\fi
\expandafter\ifx\csname citenamefont\endcsname\relax
  \def\citenamefont#1{#1}\fi
\expandafter\ifx\csname url\endcsname\relax
  \def\url#1{\texttt{#1}}\fi
\expandafter\ifx\csname urlprefix\endcsname\relax\def\urlprefix{URL }\fi
\providecommand{\bibinfo}[2]{#2}
\providecommand{\eprint}[2][]{\url{#2}}

\bibitem[{\citenamefont{Sivers}(1990)}]{Sivers:1989cc}
\bibinfo{author}{\bibfnamefont{D.~W.} \bibnamefont{Sivers}},
  \bibinfo{journal}{Phys. Rev.} \textbf{\bibinfo{volume}{D41}},
  \bibinfo{pages}{83} (\bibinfo{year}{1990}).

\bibitem[{\citenamefont{Sivers}(1991)}]{Sivers:1990fh}
\bibinfo{author}{\bibfnamefont{D.~W.} \bibnamefont{Sivers}},
  \bibinfo{journal}{Phys. Rev.} \textbf{\bibinfo{volume}{D43}},
  \bibinfo{pages}{261} (\bibinfo{year}{1991}).

\bibitem[{\citenamefont{Airapetian et~al.}(2005)}]{Airapetian:2004tw}
\bibinfo{author}{\bibfnamefont{A.}~\bibnamefont{Airapetian}}
  \bibnamefont{et~al.} (\bibinfo{collaboration}{HERMES}),
  \bibinfo{journal}{Phys. Rev. Lett.} \textbf{\bibinfo{volume}{94}},
  \bibinfo{pages}{012002} (\bibinfo{year}{2005}), \eprint{hep-ex/0408013}.

\bibitem[{\citenamefont{Airapetian et~al.}(2009)}]{Airapetian:2009ae}
\bibinfo{author}{\bibfnamefont{A.}~\bibnamefont{Airapetian}}
  \bibnamefont{et~al.} (\bibinfo{collaboration}{HERMES}),
  \bibinfo{journal}{Phys. Rev. Lett.} \textbf{\bibinfo{volume}{103}},
  \bibinfo{pages}{152002} (\bibinfo{year}{2009}), \eprint{0906.3918}.

\bibitem[{\citenamefont{Airapetian et~al.}(2014)}]{Airapetian:2013bim}
\bibinfo{author}{\bibfnamefont{A.}~\bibnamefont{Airapetian}}
  \bibnamefont{et~al.} (\bibinfo{collaboration}{HERMES}),
  \bibinfo{journal}{Phys. Lett.} \textbf{\bibinfo{volume}{B728}},
  \bibinfo{pages}{183} (\bibinfo{year}{2014}), \eprint{1310.5070}.

\bibitem[{\citenamefont{Adolph et~al.}(2012)}]{Adolph:2012sp}
\bibinfo{author}{\bibfnamefont{C.}~\bibnamefont{Adolph}} \bibnamefont{et~al.}
  (\bibinfo{collaboration}{COMPASS}), \bibinfo{journal}{Phys. Lett.}
  \textbf{\bibinfo{volume}{B717}}, \bibinfo{pages}{383} (\bibinfo{year}{2012}),
  \eprint{1205.5122}.

\bibitem[{\citenamefont{Adolph et~al.}(2014)}]{Adolph:2014fjw}
\bibinfo{author}{\bibfnamefont{C.}~\bibnamefont{Adolph}} \bibnamefont{et~al.}
  (\bibinfo{collaboration}{COMPASS}), \bibinfo{journal}{Phys. Lett.}
  \textbf{\bibinfo{volume}{B736}}, \bibinfo{pages}{124} (\bibinfo{year}{2014}),
  \eprint{1401.7873}.

\bibitem[{\citenamefont{Adolph et~al.}(2017)}]{Adolph:2017pgv}
\bibinfo{author}{\bibfnamefont{C.}~\bibnamefont{Adolph}} \bibnamefont{et~al.}
  (\bibinfo{collaboration}{COMPASS}), \bibinfo{journal}{Phys. Lett.}
  \textbf{\bibinfo{volume}{B772}}, \bibinfo{pages}{854} (\bibinfo{year}{2017}),
  \eprint{1701.02453}.

\bibitem[{\citenamefont{Aghasyan et~al.}(2017)}]{Aghasyan:2017jop}
\bibinfo{author}{\bibfnamefont{M.}~\bibnamefont{Aghasyan}} \bibnamefont{et~al.}
  (\bibinfo{collaboration}{COMPASS}), \bibinfo{journal}{Phys. Rev. Lett.}
  \textbf{\bibinfo{volume}{119}}, \bibinfo{pages}{112002}
  (\bibinfo{year}{2017}), \eprint{1704.00488}.

\bibitem[{\citenamefont{Qian et~al.}(2011)}]{Qian:2011py}
\bibinfo{author}{\bibfnamefont{X.}~\bibnamefont{Qian}} \bibnamefont{et~al.}
  (\bibinfo{collaboration}{Jefferson Lab Hall A}), \bibinfo{journal}{Phys. Rev.
  Lett.} \textbf{\bibinfo{volume}{107}}, \bibinfo{pages}{072003}
  (\bibinfo{year}{2011}), \eprint{1106.0363}.

\bibitem[{\citenamefont{Zhao et~al.}(2014)}]{Zhao:2014qvx}
\bibinfo{author}{\bibfnamefont{Y.~X.} \bibnamefont{Zhao}} \bibnamefont{et~al.}
  (\bibinfo{collaboration}{Jefferson Lab Hall A}), \bibinfo{journal}{Phys.
  Rev.} \textbf{\bibinfo{volume}{C90}}, \bibinfo{pages}{055201}
  (\bibinfo{year}{2014}), \eprint{1404.7204}.

\bibitem[{\citenamefont{Adare et~al.}(2010)}]{Adare:2010bd}
\bibinfo{author}{\bibfnamefont{A.}~\bibnamefont{Adare}} \bibnamefont{et~al.}
  (\bibinfo{collaboration}{PHENIX}), \bibinfo{journal}{Phys. Rev.}
  \textbf{\bibinfo{volume}{D82}}, \bibinfo{pages}{112008}
  (\bibinfo{year}{2010}), \bibinfo{note}{[Erratum: Phys.
  Rev.D86,099904(2012)]}, \eprint{1009.4864}.

\bibitem[{\citenamefont{Adamczyk et~al.}(2016)}]{Adamczyk:2015gyk}
\bibinfo{author}{\bibfnamefont{L.}~\bibnamefont{Adamczyk}} \bibnamefont{et~al.}
  (\bibinfo{collaboration}{STAR}), \bibinfo{journal}{Phys. Rev. Lett.}
  \textbf{\bibinfo{volume}{116}}, \bibinfo{pages}{132301}
  (\bibinfo{year}{2016}), \eprint{1511.06003}.

\bibitem[{\citenamefont{Collins}(2002)}]{Collins:2002kn}
\bibinfo{author}{\bibfnamefont{J.~C.} \bibnamefont{Collins}},
  \bibinfo{journal}{Phys. Lett.} \textbf{\bibinfo{volume}{B536}},
  \bibinfo{pages}{43} (\bibinfo{year}{2002}), \eprint{hep-ph/0204004}.

\bibitem[{\citenamefont{Boer et~al.}(2003)\citenamefont{Boer, Mulders, and
  Pijlman}}]{Boer:2003cm}
\bibinfo{author}{\bibfnamefont{D.}~\bibnamefont{Boer}},
  \bibinfo{author}{\bibfnamefont{P.~J.} \bibnamefont{Mulders}},
  \bibnamefont{and} \bibinfo{author}{\bibfnamefont{F.}~\bibnamefont{Pijlman}},
  \bibinfo{journal}{Nucl. Phys.} \textbf{\bibinfo{volume}{B667}},
  \bibinfo{pages}{201} (\bibinfo{year}{2003}), \eprint{hep-ph/0303034}.

\bibitem[{\citenamefont{Anselmino
  et~al.}(2017{\natexlab{a}})\citenamefont{Anselmino, Boglione, D'Alesio,
  Murgia, and Prokudin}}]{Anselmino:2016uie}
\bibinfo{author}{\bibfnamefont{M.}~\bibnamefont{Anselmino}},
  \bibinfo{author}{\bibfnamefont{M.}~\bibnamefont{Boglione}},
  \bibinfo{author}{\bibfnamefont{U.}~\bibnamefont{D'Alesio}},
  \bibinfo{author}{\bibfnamefont{F.}~\bibnamefont{Murgia}}, \bibnamefont{and}
  \bibinfo{author}{\bibfnamefont{A.}~\bibnamefont{Prokudin}},
  \bibinfo{journal}{JHEP} \textbf{\bibinfo{volume}{04}}, \bibinfo{pages}{046}
  (\bibinfo{year}{2017}{\natexlab{a}}), \eprint{1612.06413}.

\bibitem[{\citenamefont{Mulders and Rodrigues}(2001)}]{Mulders:2000sh}
\bibinfo{author}{\bibfnamefont{P.~J.} \bibnamefont{Mulders}} \bibnamefont{and}
  \bibinfo{author}{\bibfnamefont{J.}~\bibnamefont{Rodrigues}},
  \bibinfo{journal}{Phys. Rev.} \textbf{\bibinfo{volume}{D63}},
  \bibinfo{pages}{094021} (\bibinfo{year}{2001}), \eprint{hep-ph/0009343}.

\bibitem[{\citenamefont{Buffing et~al.}(2013)\citenamefont{Buffing, Mukherjee,
  and Mulders}}]{Buffing:2013kca}
\bibinfo{author}{\bibfnamefont{M.~G.~A.} \bibnamefont{Buffing}},
  \bibinfo{author}{\bibfnamefont{A.}~\bibnamefont{Mukherjee}},
  \bibnamefont{and} \bibinfo{author}{\bibfnamefont{P.~J.}
  \bibnamefont{Mulders}}, \bibinfo{journal}{Phys. Rev.}
  \textbf{\bibinfo{volume}{D88}}, \bibinfo{pages}{054027}
  (\bibinfo{year}{2013}), \eprint{1306.5897}.

\bibitem[{\citenamefont{Godbole et~al.}(2012)\citenamefont{Godbole, Misra,
  Mukherjee, and Rawoot}}]{Godbole:2012bx}
\bibinfo{author}{\bibfnamefont{R.~M.} \bibnamefont{Godbole}},
  \bibinfo{author}{\bibfnamefont{A.}~\bibnamefont{Misra}},
  \bibinfo{author}{\bibfnamefont{A.}~\bibnamefont{Mukherjee}},
  \bibnamefont{and} \bibinfo{author}{\bibfnamefont{V.~S.}
  \bibnamefont{Rawoot}}, \bibinfo{journal}{Phys. Rev.}
  \textbf{\bibinfo{volume}{D85}}, \bibinfo{pages}{094013}
  (\bibinfo{year}{2012}), \eprint{1201.1066}.

\bibitem[{\citenamefont{Mukherjee and
  Rajesh}(2017{\natexlab{a}})}]{Mukherjee:2016qxa}
\bibinfo{author}{\bibfnamefont{A.}~\bibnamefont{Mukherjee}} \bibnamefont{and}
  \bibinfo{author}{\bibfnamefont{S.}~\bibnamefont{Rajesh}},
  \bibinfo{journal}{Eur. Phys. J.} \textbf{\bibinfo{volume}{C77}},
  \bibinfo{pages}{854} (\bibinfo{year}{2017}{\natexlab{a}}),
  \eprint{1609.05596}.

\bibitem[{\citenamefont{Boer}(2017)}]{Boer:2016bfj}
\bibinfo{author}{\bibfnamefont{D.}~\bibnamefont{Boer}}, \bibinfo{journal}{Few
  Body Syst.} \textbf{\bibinfo{volume}{58}}, \bibinfo{pages}{32}
  (\bibinfo{year}{2017}), \eprint{1611.06089}.

\bibitem[{\citenamefont{Anselmino
  et~al.}(2017{\natexlab{b}})\citenamefont{Anselmino, Barone, and
  Boglione}}]{Anselmino:2016fhz}
\bibinfo{author}{\bibfnamefont{M.}~\bibnamefont{Anselmino}},
  \bibinfo{author}{\bibfnamefont{V.}~\bibnamefont{Barone}}, \bibnamefont{and}
  \bibinfo{author}{\bibfnamefont{M.}~\bibnamefont{Boglione}},
  \bibinfo{journal}{Phys. Lett.} \textbf{\bibinfo{volume}{B770}},
  \bibinfo{pages}{302} (\bibinfo{year}{2017}{\natexlab{b}}),
  \eprint{1607.00275}.

\bibitem[{\citenamefont{Boer et~al.}(2016)\citenamefont{Boer, Mulders, Pisano,
  and Zhou}}]{Boer:2016fqd}
\bibinfo{author}{\bibfnamefont{D.}~\bibnamefont{Boer}},
  \bibinfo{author}{\bibfnamefont{P.~J.} \bibnamefont{Mulders}},
  \bibinfo{author}{\bibfnamefont{C.}~\bibnamefont{Pisano}}, \bibnamefont{and}
  \bibinfo{author}{\bibfnamefont{J.}~\bibnamefont{Zhou}},
  \bibinfo{journal}{JHEP} \textbf{\bibinfo{volume}{08}}, \bibinfo{pages}{001}
  (\bibinfo{year}{2016}), \eprint{1605.07934}.

\bibitem[{\citenamefont{Anselmino et~al.}(2004)\citenamefont{Anselmino,
  Boglione, D'Alesio, Leader, and Murgia}}]{Anselmino:2004nk}
\bibinfo{author}{\bibfnamefont{M.}~\bibnamefont{Anselmino}},
  \bibinfo{author}{\bibfnamefont{M.}~\bibnamefont{Boglione}},
  \bibinfo{author}{\bibfnamefont{U.}~\bibnamefont{D'Alesio}},
  \bibinfo{author}{\bibfnamefont{E.}~\bibnamefont{Leader}}, \bibnamefont{and}
  \bibinfo{author}{\bibfnamefont{F.}~\bibnamefont{Murgia}},
  \bibinfo{journal}{Phys. Rev.} \textbf{\bibinfo{volume}{D70}},
  \bibinfo{pages}{074025} (\bibinfo{year}{2004}), \eprint{hep-ph/0407100}.

\bibitem[{\citenamefont{D'Alesio
  et~al.}(2017{\natexlab{a}})\citenamefont{D'Alesio, Murgia, Pisano, and
  Taels}}]{DAlesio:2017rzj}
\bibinfo{author}{\bibfnamefont{U.}~\bibnamefont{D'Alesio}},
  \bibinfo{author}{\bibfnamefont{F.}~\bibnamefont{Murgia}},
  \bibinfo{author}{\bibfnamefont{C.}~\bibnamefont{Pisano}}, \bibnamefont{and}
  \bibinfo{author}{\bibfnamefont{P.}~\bibnamefont{Taels}},
  \bibinfo{journal}{Phys. Rev.} \textbf{\bibinfo{volume}{D96}},
  \bibinfo{pages}{036011} (\bibinfo{year}{2017}{\natexlab{a}}),
  \eprint{1705.04169}.

\bibitem[{\citenamefont{Mukherjee and Rajesh}(2016)}]{Mukherjee:2015smo}
\bibinfo{author}{\bibfnamefont{A.}~\bibnamefont{Mukherjee}} \bibnamefont{and}
  \bibinfo{author}{\bibfnamefont{S.}~\bibnamefont{Rajesh}},
  \bibinfo{journal}{Phys. Rev.} \textbf{\bibinfo{volume}{D93}},
  \bibinfo{pages}{054018} (\bibinfo{year}{2016}), \eprint{1511.04319}.

\bibitem[{\citenamefont{Mukherjee and
  Rajesh}(2017{\natexlab{b}})}]{Mukherjee:2016cjw}
\bibinfo{author}{\bibfnamefont{A.}~\bibnamefont{Mukherjee}} \bibnamefont{and}
  \bibinfo{author}{\bibfnamefont{S.}~\bibnamefont{Rajesh}},
  \bibinfo{journal}{Phys. Rev.} \textbf{\bibinfo{volume}{D95}},
  \bibinfo{pages}{034039} (\bibinfo{year}{2017}{\natexlab{b}}),
  \eprint{1611.05974}.

\bibitem[{\citenamefont{Bodwin et~al.}(1995)\citenamefont{Bodwin, Braaten, and
  Lepage}}]{Bodwin:1994jh}
\bibinfo{author}{\bibfnamefont{G.~T.} \bibnamefont{Bodwin}},
  \bibinfo{author}{\bibfnamefont{E.}~\bibnamefont{Braaten}}, \bibnamefont{and}
  \bibinfo{author}{\bibfnamefont{G.~P.} \bibnamefont{Lepage}},
  \bibinfo{journal}{Phys. Rev.} \textbf{\bibinfo{volume}{D51}},
  \bibinfo{pages}{1125} (\bibinfo{year}{1995}), \bibinfo{note}{[Erratum: Phys.
  Rev.D55,5853(1997)]}, \eprint{hep-ph/9407339}.

\bibitem[{\citenamefont{Carlson and Suaya}(1976)}]{Carlson:1976cd}
\bibinfo{author}{\bibfnamefont{C.~E.} \bibnamefont{Carlson}} \bibnamefont{and}
  \bibinfo{author}{\bibfnamefont{R.}~\bibnamefont{Suaya}},
  \bibinfo{journal}{Phys. Rev.} \textbf{\bibinfo{volume}{D14}},
  \bibinfo{pages}{3115} (\bibinfo{year}{1976}).

\bibitem[{\citenamefont{Berger and Jones}(1981)}]{Berger:1980ni}
\bibinfo{author}{\bibfnamefont{E.~L.} \bibnamefont{Berger}} \bibnamefont{and}
  \bibinfo{author}{\bibfnamefont{D.~L.} \bibnamefont{Jones}},
  \bibinfo{journal}{Phys. Rev.} \textbf{\bibinfo{volume}{D23}},
  \bibinfo{pages}{1521} (\bibinfo{year}{1981}).

\bibitem[{\citenamefont{Baier and Ruckl}(1981)}]{Baier:1981uk}
\bibinfo{author}{\bibfnamefont{R.}~\bibnamefont{Baier}} \bibnamefont{and}
  \bibinfo{author}{\bibfnamefont{R.}~\bibnamefont{Ruckl}},
  \bibinfo{journal}{Phys. Lett.} \textbf{\bibinfo{volume}{102B}},
  \bibinfo{pages}{364} (\bibinfo{year}{1981}).

\bibitem[{\citenamefont{Baier and Ruckl}(1982)}]{Baier:1981zz}
\bibinfo{author}{\bibfnamefont{R.}~\bibnamefont{Baier}} \bibnamefont{and}
  \bibinfo{author}{\bibfnamefont{R.}~\bibnamefont{Ruckl}},
  \bibinfo{journal}{Nucl. Phys.} \textbf{\bibinfo{volume}{B201}},
  \bibinfo{pages}{1} (\bibinfo{year}{1982}).

\bibitem[{\citenamefont{Braaten and Fleming}(1995)}]{Braaten:1994vv}
\bibinfo{author}{\bibfnamefont{E.}~\bibnamefont{Braaten}} \bibnamefont{and}
  \bibinfo{author}{\bibfnamefont{S.}~\bibnamefont{Fleming}},
  \bibinfo{journal}{Phys. Rev. Lett.} \textbf{\bibinfo{volume}{74}},
  \bibinfo{pages}{3327} (\bibinfo{year}{1995}), \eprint{hep-ph/9411365}.

\bibitem[{\citenamefont{Cho and Leibovich}(1996{\natexlab{a}})}]{Cho:1995vh}
\bibinfo{author}{\bibfnamefont{P.~L.} \bibnamefont{Cho}} \bibnamefont{and}
  \bibinfo{author}{\bibfnamefont{A.~K.} \bibnamefont{Leibovich}},
  \bibinfo{journal}{Phys. Rev.} \textbf{\bibinfo{volume}{D53}},
  \bibinfo{pages}{150} (\bibinfo{year}{1996}{\natexlab{a}}),
  \eprint{hep-ph/9505329}.

\bibitem[{\citenamefont{Cho and Leibovich}(1996{\natexlab{b}})}]{Cho:1995ce}
\bibinfo{author}{\bibfnamefont{P.~L.} \bibnamefont{Cho}} \bibnamefont{and}
  \bibinfo{author}{\bibfnamefont{A.~K.} \bibnamefont{Leibovich}},
  \bibinfo{journal}{Phys. Rev.} \textbf{\bibinfo{volume}{D53}},
  \bibinfo{pages}{6203} (\bibinfo{year}{1996}{\natexlab{b}}),
  \eprint{hep-ph/9511315}.

\bibitem[{\citenamefont{Lepage et~al.}(1992)\citenamefont{Lepage, Magnea,
  Nakhleh, Magnea, and Hornbostel}}]{Lepage:1992tx}
\bibinfo{author}{\bibfnamefont{G.~P.} \bibnamefont{Lepage}},
  \bibinfo{author}{\bibfnamefont{L.}~\bibnamefont{Magnea}},
  \bibinfo{author}{\bibfnamefont{C.}~\bibnamefont{Nakhleh}},
  \bibinfo{author}{\bibfnamefont{U.}~\bibnamefont{Magnea}}, \bibnamefont{and}
  \bibinfo{author}{\bibfnamefont{K.}~\bibnamefont{Hornbostel}},
  \bibinfo{journal}{Phys. Rev.} \textbf{\bibinfo{volume}{D46}},
  \bibinfo{pages}{4052} (\bibinfo{year}{1992}), \eprint{hep-lat/9205007}.

\bibitem[{\citenamefont{Abe et~al.}(1997)}]{Abe:1997jz}
\bibinfo{author}{\bibfnamefont{F.}~\bibnamefont{Abe}} \bibnamefont{et~al.}
  (\bibinfo{collaboration}{CDF}), \bibinfo{journal}{Phys. Rev. Lett.}
  \textbf{\bibinfo{volume}{79}}, \bibinfo{pages}{572} (\bibinfo{year}{1997}).

\bibitem[{\citenamefont{Acosta et~al.}(2005)}]{Acosta:2004yw}
\bibinfo{author}{\bibfnamefont{D.}~\bibnamefont{Acosta}} \bibnamefont{et~al.}
  (\bibinfo{collaboration}{CDF}), \bibinfo{journal}{Phys. Rev.}
  \textbf{\bibinfo{volume}{D71}}, \bibinfo{pages}{032001}
  (\bibinfo{year}{2005}), \eprint{hep-ex/0412071}.

\bibitem[{\citenamefont{Adloff et~al.}(2002)}]{Adloff:2002ex}
\bibinfo{author}{\bibfnamefont{C.}~\bibnamefont{Adloff}} \bibnamefont{et~al.}
  (\bibinfo{collaboration}{H1}), \bibinfo{journal}{Eur. Phys. J.}
  \textbf{\bibinfo{volume}{C25}}, \bibinfo{pages}{25} (\bibinfo{year}{2002}),
  \eprint{hep-ex/0205064}.

\bibitem[{\citenamefont{Aaron et~al.}(2010)}]{Aaron:2010gz}
\bibinfo{author}{\bibfnamefont{F.~D.} \bibnamefont{Aaron}} \bibnamefont{et~al.}
  (\bibinfo{collaboration}{H1}), \bibinfo{journal}{Eur. Phys. J.}
  \textbf{\bibinfo{volume}{C68}}, \bibinfo{pages}{401} (\bibinfo{year}{2010}),
  \eprint{1002.0234}.

\bibitem[{\citenamefont{Chekanov et~al.}(2003)}]{Chekanov:2002at}
\bibinfo{author}{\bibfnamefont{S.}~\bibnamefont{Chekanov}} \bibnamefont{et~al.}
  (\bibinfo{collaboration}{ZEUS}), \bibinfo{journal}{Eur. Phys. J.}
  \textbf{\bibinfo{volume}{C27}}, \bibinfo{pages}{173} (\bibinfo{year}{2003}),
  \eprint{hep-ex/0211011}.

\bibitem[{\citenamefont{Abramowicz et~al.}(2013)}]{Abramowicz:2012dh}
\bibinfo{author}{\bibfnamefont{H.}~\bibnamefont{Abramowicz}}
  \bibnamefont{et~al.} (\bibinfo{collaboration}{ZEUS}), \bibinfo{journal}{JHEP}
  \textbf{\bibinfo{volume}{02}}, \bibinfo{pages}{071} (\bibinfo{year}{2013}),
  \eprint{1211.6946}.

\bibitem[{\citenamefont{Butenschoen and
  Kniehl}(2011{\natexlab{a}})}]{Butenschoen:2010rq}
\bibinfo{author}{\bibfnamefont{M.}~\bibnamefont{Butenschoen}} \bibnamefont{and}
  \bibinfo{author}{\bibfnamefont{B.~A.} \bibnamefont{Kniehl}},
  \bibinfo{journal}{Phys. Rev. Lett.} \textbf{\bibinfo{volume}{106}},
  \bibinfo{pages}{022003} (\bibinfo{year}{2011}{\natexlab{a}}),
  \eprint{1009.5662}.

\bibitem[{\citenamefont{Ma et~al.}(2011)\citenamefont{Ma, Wang, and
  Chao}}]{Ma:2010jj}
\bibinfo{author}{\bibfnamefont{Y.-Q.} \bibnamefont{Ma}},
  \bibinfo{author}{\bibfnamefont{K.}~\bibnamefont{Wang}}, \bibnamefont{and}
  \bibinfo{author}{\bibfnamefont{K.-T.} \bibnamefont{Chao}},
  \bibinfo{journal}{Phys. Rev.} \textbf{\bibinfo{volume}{D84}},
  \bibinfo{pages}{114001} (\bibinfo{year}{2011}), \eprint{1012.1030}.

\bibitem[{\citenamefont{Chao et~al.}(2012)\citenamefont{Chao, Ma, Shao, Wang,
  and Zhang}}]{Chao:2012iv}
\bibinfo{author}{\bibfnamefont{K.-T.} \bibnamefont{Chao}},
  \bibinfo{author}{\bibfnamefont{Y.-Q.} \bibnamefont{Ma}},
  \bibinfo{author}{\bibfnamefont{H.-S.} \bibnamefont{Shao}},
  \bibinfo{author}{\bibfnamefont{K.}~\bibnamefont{Wang}}, \bibnamefont{and}
  \bibinfo{author}{\bibfnamefont{Y.-J.} \bibnamefont{Zhang}},
  \bibinfo{journal}{Phys. Rev. Lett.} \textbf{\bibinfo{volume}{108}},
  \bibinfo{pages}{242004} (\bibinfo{year}{2012}), \eprint{1201.2675}.

\bibitem[{\citenamefont{Butenschoen and
  Kniehl}(2011{\natexlab{b}})}]{Butenschoen:2011yh}
\bibinfo{author}{\bibfnamefont{M.}~\bibnamefont{Butenschoen}} \bibnamefont{and}
  \bibinfo{author}{\bibfnamefont{B.~A.} \bibnamefont{Kniehl}},
  \bibinfo{journal}{Phys. Rev.} \textbf{\bibinfo{volume}{D84}},
  \bibinfo{pages}{051501} (\bibinfo{year}{2011}{\natexlab{b}}),
  \eprint{1105.0820}.

\bibitem[{\citenamefont{Zhang et~al.}(2015)\citenamefont{Zhang, Sun, Sang, and
  Li}}]{Zhang:2014ybe}
\bibinfo{author}{\bibfnamefont{H.-F.} \bibnamefont{Zhang}},
  \bibinfo{author}{\bibfnamefont{Z.}~\bibnamefont{Sun}},
  \bibinfo{author}{\bibfnamefont{W.-L.} \bibnamefont{Sang}}, \bibnamefont{and}
  \bibinfo{author}{\bibfnamefont{R.}~\bibnamefont{Li}}, \bibinfo{journal}{Phys.
  Rev. Lett.} \textbf{\bibinfo{volume}{114}}, \bibinfo{pages}{092006}
  (\bibinfo{year}{2015}), \eprint{1412.0508}.

\bibitem[{\citenamefont{Butenschoen and Kniehl}(2012)}]{Butenschoen:2012px}
\bibinfo{author}{\bibfnamefont{M.}~\bibnamefont{Butenschoen}} \bibnamefont{and}
  \bibinfo{author}{\bibfnamefont{B.~A.} \bibnamefont{Kniehl}},
  \bibinfo{journal}{Phys. Rev. Lett.} \textbf{\bibinfo{volume}{108}},
  \bibinfo{pages}{172002} (\bibinfo{year}{2012}), \eprint{1201.1872}.

\bibitem[{\citenamefont{Yuan}(2008)}]{Yuan:2008vn}
\bibinfo{author}{\bibfnamefont{F.}~\bibnamefont{Yuan}}, \bibinfo{journal}{Phys.
  Rev.} \textbf{\bibinfo{volume}{D78}}, \bibinfo{pages}{014024}
  (\bibinfo{year}{2008}), \eprint{0801.4357}.

\bibitem[{\citenamefont{Matoušek}(2016)}]{Matousek:2016xbl}
\bibinfo{author}{\bibfnamefont{J.}~\bibnamefont{Matoušek}}
  (\bibinfo{collaboration}{COMPASS}), \bibinfo{journal}{J. Phys. Conf. Ser.}
  \textbf{\bibinfo{volume}{678}}, \bibinfo{pages}{012050}
  (\bibinfo{year}{2016}).

\bibitem[{\citenamefont{Kniehl and Kramer}(1999)}]{Kniehl:1998qy}
\bibinfo{author}{\bibfnamefont{B.~A.} \bibnamefont{Kniehl}} \bibnamefont{and}
  \bibinfo{author}{\bibfnamefont{G.}~\bibnamefont{Kramer}},
  \bibinfo{journal}{Eur. Phys. J.} \textbf{\bibinfo{volume}{C6}},
  \bibinfo{pages}{493} (\bibinfo{year}{1999}), \eprint{hep-ph/9803256}.

\bibitem[{\citenamefont{Ryskin}(1993)}]{Ryskin:1992ui}
\bibinfo{author}{\bibfnamefont{M.~G.} \bibnamefont{Ryskin}},
  \bibinfo{journal}{Z. Phys.} \textbf{\bibinfo{volume}{C57}},
  \bibinfo{pages}{89} (\bibinfo{year}{1993}).

\bibitem[{\citenamefont{Butenschoen and Kniehl}(2010)}]{Butenschoen:2009zy}
\bibinfo{author}{\bibfnamefont{M.}~\bibnamefont{Butenschoen}} \bibnamefont{and}
  \bibinfo{author}{\bibfnamefont{B.~A.} \bibnamefont{Kniehl}},
  \bibinfo{journal}{Phys. Rev. Lett.} \textbf{\bibinfo{volume}{104}},
  \bibinfo{pages}{072001} (\bibinfo{year}{2010}), \eprint{0909.2798}.

\bibitem[{\citenamefont{Li and Liu}(1998)}]{Li:1996jk}
\bibinfo{author}{\bibfnamefont{Y.-d.} \bibnamefont{Li}} \bibnamefont{and}
  \bibinfo{author}{\bibfnamefont{L.-s.} \bibnamefont{Liu}},
  \bibinfo{journal}{Commun. Theor. Phys.} \textbf{\bibinfo{volume}{29}},
  \bibinfo{pages}{99} (\bibinfo{year}{1998}).

\bibitem[{\citenamefont{Artoisenet et~al.}(2009)\citenamefont{Artoisenet,
  Campbell, Maltoni, and Tramontano}}]{Artoisenet:2009xh}
\bibinfo{author}{\bibfnamefont{P.}~\bibnamefont{Artoisenet}},
  \bibinfo{author}{\bibfnamefont{J.~M.} \bibnamefont{Campbell}},
  \bibinfo{author}{\bibfnamefont{F.}~\bibnamefont{Maltoni}}, \bibnamefont{and}
  \bibinfo{author}{\bibfnamefont{F.}~\bibnamefont{Tramontano}},
  \bibinfo{journal}{Phys. Rev. Lett.} \textbf{\bibinfo{volume}{102}},
  \bibinfo{pages}{142001} (\bibinfo{year}{2009}), \eprint{0901.4352}.

\bibitem[{\citenamefont{Kramer}(1996)}]{Kramer:1995nb}
\bibinfo{author}{\bibfnamefont{M.}~\bibnamefont{Kramer}},
  \bibinfo{journal}{Nucl. Phys.} \textbf{\bibinfo{volume}{B459}},
  \bibinfo{pages}{3} (\bibinfo{year}{1996}), \eprint{hep-ph/9508409}.

\bibitem[{\citenamefont{Ko et~al.}(1996)\citenamefont{Ko, Lee, and
  Song}}]{Ko:1996xw}
\bibinfo{author}{\bibfnamefont{P.}~\bibnamefont{Ko}},
  \bibinfo{author}{\bibfnamefont{J.}~\bibnamefont{Lee}}, \bibnamefont{and}
  \bibinfo{author}{\bibfnamefont{H.~S.} \bibnamefont{Song}},
  \bibinfo{journal}{Phys. Rev.} \textbf{\bibinfo{volume}{D54}},
  \bibinfo{pages}{4312} (\bibinfo{year}{1996}), \bibinfo{note}{[Erratum: Phys.
  Rev.D60,119902(1999)]}, \eprint{hep-ph/9602223}.

\bibitem[{\citenamefont{D'Alesio
  et~al.}(2017{\natexlab{b}})\citenamefont{D'Alesio, Flore, and
  Murgia}}]{DAlesio:2017nrd}
\bibinfo{author}{\bibfnamefont{U.}~\bibnamefont{D'Alesio}},
  \bibinfo{author}{\bibfnamefont{C.}~\bibnamefont{Flore}}, \bibnamefont{and}
  \bibinfo{author}{\bibfnamefont{F.}~\bibnamefont{Murgia}},
  \bibinfo{journal}{Phys. Rev.} \textbf{\bibinfo{volume}{D95}},
  \bibinfo{pages}{094002} (\bibinfo{year}{2017}{\natexlab{b}}),
  \eprint{1701.01148}.

\bibitem[{\citenamefont{Frixione et~al.}(1993)\citenamefont{Frixione, Mangano,
  Nason, and Ridolfi}}]{Frixione:1993yw}
\bibinfo{author}{\bibfnamefont{S.}~\bibnamefont{Frixione}},
  \bibinfo{author}{\bibfnamefont{M.~L.} \bibnamefont{Mangano}},
  \bibinfo{author}{\bibfnamefont{P.}~\bibnamefont{Nason}}, \bibnamefont{and}
  \bibinfo{author}{\bibfnamefont{G.}~\bibnamefont{Ridolfi}},
  \bibinfo{journal}{Phys. Lett.} \textbf{\bibinfo{volume}{B319}},
  \bibinfo{pages}{339} (\bibinfo{year}{1993}), \eprint{hep-ph/9310350}.

\bibitem[{\citenamefont{D'Alesio et~al.}(2015)\citenamefont{D'Alesio, Murgia,
  and Pisano}}]{alesio}
\bibinfo{author}{\bibfnamefont{U.}~\bibnamefont{D'Alesio}},
  \bibinfo{author}{\bibfnamefont{F.}~\bibnamefont{Murgia}}, \bibnamefont{and}
  \bibinfo{author}{\bibfnamefont{C.}~\bibnamefont{Pisano}},
  \bibinfo{journal}{JHEP} \textbf{\bibinfo{volume}{09}}, \bibinfo{pages}{119}
  (\bibinfo{year}{2015}), \eprint{1506.03078}.

\bibitem[{\citenamefont{Adare et~al.}(2014)}]{rhic1}
\bibinfo{author}{\bibfnamefont{A.}~\bibnamefont{Adare}} \bibnamefont{et~al.}
  (\bibinfo{collaboration}{PHENIX}), \bibinfo{journal}{Phys. Rev.}
  \textbf{\bibinfo{volume}{D90}}, \bibinfo{pages}{012006}
  (\bibinfo{year}{2014}), \eprint{1312.1995}.

\bibitem[{\citenamefont{Boer and Vogelsang}(2004)}]{Boer:2003tx}
\bibinfo{author}{\bibfnamefont{D.}~\bibnamefont{Boer}} \bibnamefont{and}
  \bibinfo{author}{\bibfnamefont{W.}~\bibnamefont{Vogelsang}},
  \bibinfo{journal}{Phys. Rev.} \textbf{\bibinfo{volume}{D69}},
  \bibinfo{pages}{094025} (\bibinfo{year}{2004}), \eprint{hep-ph/0312320}.

\bibitem[{\citenamefont{Baier and Ruckl}(1983)}]{Baier:1983va}
\bibinfo{author}{\bibfnamefont{R.}~\bibnamefont{Baier}} \bibnamefont{and}
  \bibinfo{author}{\bibfnamefont{R.}~\bibnamefont{Ruckl}}, \bibinfo{journal}{Z.
  Phys.} \textbf{\bibinfo{volume}{C19}}, \bibinfo{pages}{251}
  (\bibinfo{year}{1983}).

\bibitem[{\citenamefont{Boer and Pisano}(2012)}]{Boer:2012bt}
\bibinfo{author}{\bibfnamefont{D.}~\bibnamefont{Boer}} \bibnamefont{and}
  \bibinfo{author}{\bibfnamefont{C.}~\bibnamefont{Pisano}},
  \bibinfo{journal}{Phys. Rev.} \textbf{\bibinfo{volume}{D86}},
  \bibinfo{pages}{094007} (\bibinfo{year}{2012}), \eprint{1208.3642}.

\bibitem[{\citenamefont{Kuhn et~al.}(1979)\citenamefont{Kuhn, Kaplan, and
  Safiani}}]{Kuhn:1979bb}
\bibinfo{author}{\bibfnamefont{J.~H.} \bibnamefont{Kuhn}},
  \bibinfo{author}{\bibfnamefont{J.}~\bibnamefont{Kaplan}}, \bibnamefont{and}
  \bibinfo{author}{\bibfnamefont{E.~G.~O.} \bibnamefont{Safiani}},
  \bibinfo{journal}{Nucl. Phys.} \textbf{\bibinfo{volume}{B157}},
  \bibinfo{pages}{125} (\bibinfo{year}{1979}).

\bibitem[{\citenamefont{Guberina et~al.}(1980)\citenamefont{Guberina, Kuhn,
  Peccei, and Ruckl}}]{Guberina:1980dc}
\bibinfo{author}{\bibfnamefont{B.}~\bibnamefont{Guberina}},
  \bibinfo{author}{\bibfnamefont{J.~H.} \bibnamefont{Kuhn}},
  \bibinfo{author}{\bibfnamefont{R.~D.} \bibnamefont{Peccei}},
  \bibnamefont{and} \bibinfo{author}{\bibfnamefont{R.}~\bibnamefont{Ruckl}},
  \bibinfo{journal}{Nucl. Phys.} \textbf{\bibinfo{volume}{B174}},
  \bibinfo{pages}{317} (\bibinfo{year}{1980}).

\bibitem[{\citenamefont{Martin et~al.}(2009)\citenamefont{Martin, Stirling,
  Thorne, and Watt}}]{Martin:2009iq}
\bibinfo{author}{\bibfnamefont{A.~D.} \bibnamefont{Martin}},
  \bibinfo{author}{\bibfnamefont{W.~J.} \bibnamefont{Stirling}},
  \bibinfo{author}{\bibfnamefont{R.~S.} \bibnamefont{Thorne}},
  \bibnamefont{and} \bibinfo{author}{\bibfnamefont{G.}~\bibnamefont{Watt}},
  \bibinfo{journal}{Eur. Phys. J.} \textbf{\bibinfo{volume}{C63}},
  \bibinfo{pages}{189} (\bibinfo{year}{2009}), \eprint{0901.0002}.

\bibitem[{\citenamefont{Kuipers et~al.}(2013)\citenamefont{Kuipers, Ueda,
  Vermaseren, and Vollinga}}]{Kuipers:2012rf}
\bibinfo{author}{\bibfnamefont{J.}~\bibnamefont{Kuipers}},
  \bibinfo{author}{\bibfnamefont{T.}~\bibnamefont{Ueda}},
  \bibinfo{author}{\bibfnamefont{J.~A.~M.} \bibnamefont{Vermaseren}},
  \bibnamefont{and} \bibinfo{author}{\bibfnamefont{J.}~\bibnamefont{Vollinga}},
  \bibinfo{journal}{Comput. Phys. Commun.} \textbf{\bibinfo{volume}{184}},
  \bibinfo{pages}{1453} (\bibinfo{year}{2013}), \eprint{1203.6543}.

\bibitem[{\citenamefont{Anselmino et~al.}(2010)\citenamefont{Anselmino,
  Boglione, D'Alesio, Melis, Murgia, and Prokudin}}]{Anselmino:2009pn}
\bibinfo{author}{\bibfnamefont{M.}~\bibnamefont{Anselmino}},
  \bibinfo{author}{\bibfnamefont{M.}~\bibnamefont{Boglione}},
  \bibinfo{author}{\bibfnamefont{U.}~\bibnamefont{D'Alesio}},
  \bibinfo{author}{\bibfnamefont{S.}~\bibnamefont{Melis}},
  \bibinfo{author}{\bibfnamefont{F.}~\bibnamefont{Murgia}}, \bibnamefont{and}
  \bibinfo{author}{\bibfnamefont{A.}~\bibnamefont{Prokudin}},
  \bibinfo{journal}{Phys. Rev.} \textbf{\bibinfo{volume}{D81}},
  \bibinfo{pages}{034007} (\bibinfo{year}{2010}), \eprint{0911.1744}.

\end{thebibliography}

\end{document}